\documentclass[a4paper,12pt]{article}
\pdfoutput=1
\usepackage{epsfig,graphicx,xcolor,amsbsy,amssymb,latexsym,amsfonts,amsmath,,tcolorbox,setspace}
\usepackage{pstricks}
\usepackage{color}
\usepackage{soul}
\usepackage[numbers,square,comma, compress]{natbib}
\usepackage{placeins}
\usepackage{wrapfig,framed,caption}
\usepackage[makeroom]{cancel}

\usepackage{wasysym}

\usepackage{multirow}

\definecolor{shadecolor}{gray}{0.925}

\usepackage{eurosym}
\usepackage[a4paper]{geometry}
\usepackage{hyperref}
\usepackage{graphicx}
\usepackage{tikz}
\usetikzlibrary{decorations.pathreplacing,decorations.markings,snakes}
\usetikzlibrary{decorations.pathmorphing}
\usetikzlibrary{patterns}
\usetikzlibrary{calc}
\usepackage{xcolor}
\usepackage{amsmath,amssymb,amsfonts,pstricks,setspace}
\usepackage[enableskew]{youngtab}

\numberwithin{equation}{section}

\usepackage{wrapfig}

\newcommand{\bea}{\begin{eqnarray}\displaystyle}
\newcommand{\eea}{\end{eqnarray}}

\newcommand{\shan}{\mathfrak{S}}

\newcommand{\hma}[1]{{\color{cyan!70!black}{#1}}}
\newcommand{\hmb}[1]{{\color{orange!70!black}{#1}}}

\title{
{\bf Information Clustering and Pathogen Evolution}\\[10pt]}

\author{\large \textsc{Baptiste~Filoche\footnote{\tt b.filoche@ip2i.in2p3.fr}}~~and~~\textsc{Stefan~Hohenegger\footnote{\tt s.hohenegger@ipnl.in2p3.fr}}
}

\begin{document}

\maketitle
\thispagestyle{empty}

\maketitle
\thispagestyle{empty}
\begin{center}
\renewcommand{\thefootnote}{\fnsymbol{footnote}}\vspace{-0.5cm}
${}^{\footnotemark[1]\,\footnotemark[2]}$ Univ Lyon, Univ Claude Bernard Lyon 1, CNRS/IN2P3, IP2I Lyon, UMR 5822, F-69622, Villeurbanne, France\\[0.2cm] 
${}^{\footnotemark[2]}$ Dept. of Physics E. Pancini, Università di Napoli Federico II, via Cintia, 80126 Napoli, Italy\\[1.25cm]
\end{center}

\begin{abstract}
Recent outbreaks of infectious diseases have been monitored closely from an epidemiological and microbiological perspective. Extracting from this wealth of data the information that is relevant for the evolution of the pathogen and predict the further dynamic of the epidemic is a difficult task. We therefore consider clusterings of these data to condense this information. We interpret the relative abundance of (genetic) variants of the pathogen as a time-dependent probability distribution and consider clusterings that keep the Fisher information (approximately) invariant, in order to ensure that they capture the dynamics of the pandemic. By first studying analytic models, we show that this condition groups variants together that interact in a similar fashion with the population and show comparable adaptation to the epidemiological situation. Moreover, we demonstrate that the same clustering can be achieved by grouping variants together according to the time-derivative of their information, which is defined as a function of the probabilities alone. A computationally simple clustering based on the probability distribution therefore allows us to probe interactions of different variants of the pathogen with its environment. To validate these findings, we consider data of 551.459 amino acid sequences of the spike protein of SARS-CoV-2 in France over the course of almost 4 years. We demonstrate that our proposed clustering enables us to identify and track point mutations that allow variants to become dominant and identify temporal correlations among such mutations. We identify indicators which point out dangerous variants, with a potential to grow to large probabilities. We show that we can accurately predict the temporal dynamics of such variants by using a universal model discussed in previous work.
\end{abstract}

\newpage

\tableofcontents

\section{Introduction}
The epidemiological modelling of the spread of infectious diseases in a population has a long and successful history. Indeed, numerous mathematical models have been developed throughout the last century to describe a large variety of different pathogens and populations of different sizes. At the heart of most approaches is a (theoretical) modelling of basic epidemiological processes, such as the transmission of the pathogen, the incubation period and the recovery/mortality of infected patients, taking into account additional biological (\emph{e.g.} mutations of the pathogen), demographical (\emph{e.g.} age and gender of infectious individuals), geographic (\emph{e.g.} seasonal or climatic effects) or socio-economic (\emph{e.g.} mobility of the population) factors as well as possible vaccinations and non-pharmaceutical interventions (\emph{e.g.} social distancing and lock-downs) to protect the population (see \emph{e.g.} \cite{PERC20171,WANG20151,WANG20161,HETHCOTErev,BaileyBook} for reviews). Concretely, these considerations lead to deterministic models (\emph{e.g.} compartmental models \cite{Hamer,HamerLect1,HamerLect2,HamerLect3,Ross1911,Ross1916,RossHudson1916II,RossHudson1916III,McKendrick1912,McKendrick1914,McKendrick1926,Kermack:1927} and differential equations \cite{DellaMorte:2020wlc,DellaMorte:2020qry,cacciapaglia2020evidence,MeRG,HealthPass,Filoche:2024xka}) or stochastic descriptions (\emph{e.g.} percolation and lattice models \cite{Cardy_1985,Grassberger1983,Pruessner,Doi1,Doi2,Peliti,Domb}), which are characterised by certain parameter spaces (see \cite{Essam,Stauffer} for reviews). The latter allow for effective modelling of different sized populations \cite{ABC,SHSannino} and can therefore be adapted to a large variety of different outbreaks.

More recent outbreaks of infectious diseases, notably the Covid-19 pandemic, along with the mass-scale collection of epidemiological and virological data have lead to new forms of tracking and predicting the dynamics of pathogens \cite{MLvariants,WasteWaterReview,KHAILANY2020100682,10.3389/fmicb.2020.01800}. Indeed, through testing and sequencing efforts in countries around the globe, not only the spread, but also the genetic evolution of SARS-CoV-2 has been very well monitored (see \emph{e.g.}~\cite{ReviewDevelop,SARS1year,Evolution,Spectrum,AntigenicDrift,v15010167,Serotypes,HU20233003,microorganisms12030467}). Such data are not only available on a national level (\emph{e.g.}~\cite{SwedishPortal,dbvar}), but large scale collection efforts such as GISAID \cite{Gisaid1,Gisaid2,Gisaid3} or NextStrain \cite{NextStrain} provide extensive data on protein sequences of samples (and their mutations) collected world-wide over the period of the entire pandemic. However, this large amount of collected material also poses the practical problem, to extract the information that is relevant from an epidemiological perspective. Concretely, in the case of protein sequences of SARS-CoV-2, to identify the mutations that allow the virus to propagate more efficiently through the population. Besides the virological interest to better understand how the virus (or by extension also other pathogens) adapts to its environment, this question is also fundamental for providing a better protection of the population in the form of adapted vaccines. In the case of SARS-CoV-2 a lot of literature has been accumulated that describes new and dangerous variants, so-called variants of concern (VOC) as well as the functionality of particular (point-)mutations that occur in such variants. In most cases, such insights have been obtained through in-vitro experiments, \emph{e.g.}~\cite{HOFFMANN20212384,BAKHSHANDEH2021104831,PHAN2020104260,AntigenicDrift,Mishra,Mishra2}. 

In this paper we propose an efficient way of extracting relevant information from such (sequencing) data by using methods from \emph{information theory} \cite{Fisher,Hotelling,Rao,Jeffreys,Lauritzen,amari2000methods}. Following our previous work \cite{Filoche:2024xka}, we interpret epidemiological data as time-dependent probability distributions. Concretely, at a given time $t\in\mathbb{R}$ we study the fraction of individuals in a population that are infected with a given variant of a pathogen relative to all infected individuals. In the case that there are $n\in \mathbb{N}$ different variants circulating in the population, this allows us to define the probabilities $p_{i=1,\ldots,n}(t)$.\footnote{In other words, $p_i(t)$ is the probability to pick among all infected individuals at time $t$ one who is carrying the $i$th variant of the pathogen.} Differences between these probability distributions as a function of time are captured by the Fisher information metric, which was used in \cite{Filoche:2024xka} to provide an efficient and universal description of epidemiological (and population dynamical) processes. In this paper we propose that such an information theoretic description also allows to identify information that is relevant to describe the time-evolution of the different variants: indeed, a way to reduce the information about the system is to consider a \emph{clustering} of the $n$ variants. By this we understand a grouping of $n$ variants into $\ell\leq n$ clusters $\mathbb{A}_{1,\ldots,\ell}$,\footnote{From a mathematical perspective this corresponds to a surjective map $f:\,\{1,\ldots,n\}\to \{\mathbb{A}_1,\ldots,\mathbb{A}_\ell\}$ (see Section~\ref{Sect:ClusteringTheory} for a precise definition and a concise notation).} where we can (unambiguously) assign each cluster $\mathbb{A}_a$ a probability $q_{a=1,\ldots,\ell}$, as a function of the original $p_{i}$. In general, the $\{q_a\}$ contain less information than the original $\{p_i\}$, however, not necessarily all information that is lost in the clustering is also relevant to describe the time evolution of the system. Indeed, if the Fisher information metric computed from both probability distributions is the same, then the clustering constitutes a \emph{sufficient statistic} and both contain the same information with regards to the time evolution \cite{FisherInt}. This criterion therefore allows us decide if the clustering has destroyed relevant information. We propose to turn this logic around and search for clusterings that keep the Fisher information (approximately) invariant. We expect that such a clustering groups together variants of the pathogen that behave similarly and have comparable properties with regard to their abilities to propagate in a population. Thus, by studying clusters of variants that are (rapidly) growing, we can compare what they have in common and what (genetic) properties and mutations allow them to propagate more efficiently through the populations than members of other clusters.

Before studying real sequencing data from GISAID in France, we first develop and test this idea in simple theoretical models. We study compartmental models~\cite{McKendrick1914,McKendrick1926,Kermack:1927,Filoche:2024xka} with multiple different classes of infectious individuals (representing the different variants) and analytically search for clusterings that keep the Fisher information metric invariant. In these models, we show that grouping the variants according to simple, identifiable quantities leads to (approximate) sufficient statistics. These quantities, which we call \emph{couplings}, are specific to the exact definition of the probability distribution and are moreover functions of the entire (closed) system described by the compartmental model. They encode, how each variant (represented through the probability $p_i(t)$) couples to the rest of the system: indeed, these couplings depend on the dynamical state of the entire system (\emph{e.g.} also on the number of susceptible or vaccinated individuals) and are therefore notably not only functions of the probability distribution itself. However, we show that the same clustering can be obtained by grouping the variants according to the time derivative of the information $\mathfrak{I}_{p_i}=-\log_2p_i$ of each probability. The time derivatives of these quantities indeed reproduce the couplings described above, up to a common additive quantity, which is identical for each variant and thus irrelevant for the clustering. Therefore, by clustering the probabilities according to $\tfrac{d}{dt}\,\mathfrak{I}_{p_i}$, which only requires information encoded by the probability distribution itself, we directly get information about the coupling of each variant to the entire system. By studying the dynamics of the probability distribution, we therefore get non-trivial information of the time-evolution of the much larger closed system. 

In the context of epidemiology, this opens a number of interesting opportunities: by clustering probability distributions of variants of pathogens according to (the time derivative of) their information we can probe quantities that dictate how these variants behave when propagating through the population. While the former is technically quite simple to compute from (epidemiological or sequencing) data, the latter is very complicated to model, since many different factors need to be taken into account (as explained above). Moreover, if data are available to define different and independent probability distributions, we can obtain access to more couplings at the same time, which allow more detailed insights into the epidemiological and temporal evolution of the pathogen. In this paper, to demonstrate the viability of this idea, we cluster sequencing data of SARS-CoV-2 in France. Concretely, we study amino acid sequences (downloaded from GISAID \cite{Gisaid1,Gisaid2,Gisaid3}) of the spike protein of SARS-CoV-2 extracted from 551.459 samples taken in France during the period of 27/01/2020-23/10/2023. In order to keep the analysis manageable, we identify variants through the spike protein sequence, which was identified in the literature as a protein carrying mutations that are important for the infectivity (and thus spread) of the virus.\footnote{Our analysis can be generalised in a straight forward fashion to include all proteins of the virus.} Furthermore, we prune the data by only considering complete sequences, thus finally dealing with 1474 distinct amino acid sequences. For each week $n$ we can associate with each variant (represented through its spike protein sequence\footnote{In the following we shall use the words 'variant' and 'sequence' to a certain degree interchangeably. Each time, we mean a virus carrying a spike protein given by a distinct amino acid sequence.}) a probability $p_i(n)$ and thus also a (time derivative of the)  information $\mathfrak{I}_{p_i}$. Using a simple k-means algorithm, we cluster the variants in each week into $\ell$ clusters, which group together those, which behave similarly. By interpreting the clusters themselves as a set of probability distributions for the amino acids at each sequence position, we can devise information theoretic means to compare clusters. In this way we can highlight differences among clusters of sequences that are growing (\emph{i.e.} that spread faster) compared to others. Concretely, our systematic analysis highlights three important aspects:
\begin{itemize}
\item[\emph{(i)}] the clustering and the subsequent comparison of the probability distributions of amino acids at each position of the protein sequences highlights the locations of strong changes. The analysis of these quantities over several weeks points to the point-mutations that allow variants to grow and thus represent competitive advantages. This works particularly well for variants of the Omicron lineage, which are known to accumulate (advantageous) mutations over periods of time \cite{ReviewDevelop}.

\item[\emph{(ii)}] The description of clusters as sets of probability distributions allows to determine certain indicators whether a newly observed variant is dangerous (\emph{i.e.} has to potential to reach large probabilities and possibly become dominant compared to the remaining variants). Indeed, by defining probabilities of variants to be part of clusters with (other) growing variants we find marked differences between variants with a strong potential for growth and those without. Moreover, once a variant with large potential has been identified, we can use the universal description proposed in \cite{Filoche:2024xka} to predict its further development: based on 4-7 weeks worth of data, we can make accurate predictions, taking into account various uncertainties. 

\item[\emph{(iii)}] By comparing dissimilarities between the probability distributions that characterise the clustering, we can deduce correlations among mutations. Concretely, we can systematically identify mutations that appear together over time. By tallying these mutations across the entire set of spike protein sequences, we can indeed show that these correlations are significant. Such relations might be useful to better understand structural properties of the various proteins.
\end{itemize}
While our results are limited to the spike protein of SARS-CoV-2, our approach can be adapted in a straightforward manner to other pathogens. Furthermore, the computations are calculationally not taxing: indeed, all computations and analyses shown in this paper have been performed using a simple laptop with a 2 GHz Quad-Core Intel i5 processor with 16 GB RAM. 

The remainder of this paper is organised as follows: In Section~\ref{Sect:ReviewInfoTheory} we review theoretical tools in information theory, which are useful for the remainder of this paper. In Section~\ref{Sect:CompartmentalComputations} we discuss compartmental models with multiple variants of a pathogen. We in particular discuss conditions for clusterings of these variants to leave the Fisher information metric invariant (sufficient statistic) and show the relation to clusterings of the time derivative of the information of each variant. In order to validate our theoretical findings in real epidemiological data, we consider in Sections~\ref{Sect:FranceSimple} and \ref{Sect:FrancLongTerm} the evolution of SARS-CoV-2 in France: in Section~\ref{Sect:FranceSimple} in order to demonstrate our methods, we fist consider a short 30-week period, which coincides with the appearance of the Omicron variant BA.5. In Section~\ref{Sect:FrancLongTerm} we extend our analysis to cover the full time period 27/01/2020-23/10/2023. We discuss in detail 8 time-periods of 15-20 weeks, which mark the appearance of new dominant variants and explain how our information theoretical tools can be used to get insights into their evolution. Section~\ref{Sect:Conclusions} contains our conclusions. Furthermore, this paper is supplemented by 3 appendices, which contain additional mathematical definitions, further results of the time evolution of mutations and a short overview over the most important SARS-CoV-2 variants discussed in the main body of the paper.

\section{Review of Information Theory}\label{Sect:ReviewInfoTheory}
In this Section, we review a number of basic concepts in information theory that are relevant to describe probability distributions and their dependence on a set of continuous parameters.

\subsection{Information and Fisher Information Metric}
We start by defining a \emph{probability distribution}~\cite{ThomasCover,amari2000methods} over a discrete set $\mathbb{V}$ as a map
\begin{align}
&p:\,\mathbb{V}\longrightarrow [0,1]\,,&&\text{such that} &&\sum_{X\in\mathbb{V}} p(X)=1\,.\label{ProbNorm}
\end{align}
Following \cite{Shannon}, we can understand these probabilities in an information theoretic context (see \cite{2004poin.book..119B,lesne_2014,Lauritzen,amari2000methods} for reviews). To this end, we interpret $\mathbb{V}$ as a set of events and define the following \emph{information} obtained by the observation of $X$ given the probability distribution $p$
\begin{align}
\mathfrak{I}_p(X)=-\log_2p(X)\,,\label{InfoDef}
\end{align}
where $\log_2$ denotes the logarithm with base $2$. The expectation value of the information $\mathfrak{I}_p$ with respect to the distribution $p$ is the \emph{Shannon entropy} \cite{Shannon,Khinchin,Faddeev} 
\begin{align}
\shan(p):=-\sum_{X\in\mathbb{V}}p(X)\,\log_2 p(X)\,,\label{ShannonEntropy}
\end{align}
which intuitively, is the 'average missing information' \cite{lesne_2014} necessary to determine $X$ provided the distribution $p$ is known. In the context where $X$ stands for the transmission of a specific message (out of all possible messages $\mathbb{V}$), $\shan(p)$ is a measure for the 'minimal number of bits' \cite{Shannon,lesne_2014} that are required to determine $X$. 

We can generalise (\ref{ProbNorm}) to depend on a set of continuous parameters $\xi^i\in(\xi^1,\ldots,\xi^d)\in \Xi\subset\mathbb{R}^d$ (with $d\in\mathbb{N}$)
\begin{align}
&p:\,\mathbb{V}\times \Xi\longrightarrow [0,1]\,,&&\text{such that} &&\sum_{X\in\mathbb{V}} p(X,\xi)=1\hspace{0.5cm}\forall \xi\in\Xi\,.\label{ProbNormPara}
\end{align}
Following \cite{Fisher,Hotelling,Rao,Jeffreys} the following \emph{Fisher information matrix}\footnote{In our conventions $\ln$ denotes the logarithm of base $e$.} is associateed with (\ref{ProbNormPara})
\begin{align}
&g_{ij}(\xi):=\sum_{X\in\mathbb{V}}(\partial_i \ln p(X,\xi))(\partial_j \ln p(X,\xi))\,p(X,\xi)=4\sum_{X\in\mathbb{V}}\left(\partial_i\sqrt{p(X,\xi)}\right)\left(\partial_j\sqrt{p(X,\xi)}\right),\label{FisherInformationMatrix}
\end{align}
where $\partial_i=\frac{\partial}{\partial\xi^i}$ with $i,j\in\{1,\ldots,d\}$. As explained in \cite{Lauritzen,amari2000methods}, $g_{ij}$ endows the family of probability distributions 
\begin{align}
\mathcal{S}=\{p(X,\xi)|X\in\mathbb{V}\text{ and }\xi\in\Xi\}\,,\label{DefStatModel}
\end{align}
(called a statistical model) with the structure of a Riemannian manifold on which $g_{ij}$ plays the role of the metric. Due to this reason, we shall refer to (\ref{FisherInformationMatrix}) also as Fisher information metric, or simply metric for short (see also \cite{amari2000methods,Filoche:2024xka}). 

In this paper we shall mostly be interested in the case $d=1$ and identify $\Xi=\mathbb{R}$ with the time variable (which we shall denote $t$) and the metric takes the form
\begin{align}
g_{tt}=\sum_{X\in\mathbb{V}}(\partial_t \ln p(X,t))^2\,p(X,t)=4\sum_{X\in\mathbb{V}}\left(\partial_t \sqrt{p(X,t)}\right)^2\,.\label{Fisher1DMetric}
\end{align}
In the one-dimensional case, this quantity is also simply called the Fisher information \cite{Fisher}.
\subsection{Sufficient Statistic and Clustering}\label{Sect:ClusteringTheory}
Let $\mathbb{W}$ be a discrete set and $f:\,\mathbb{V} \to \mathbb{W}$ a surjective map, which induces the following probability distribution on $\mathbb{W}$
\begin{align}
q:\,\mathbb{W}\times \Xi\hspace{0.2cm}&\longrightarrow \hspace{0.2cm} [0,1]\nonumber\\
(Y,\xi)\hspace{0.2cm}&\longmapsto \hspace{0.2cm}\sum_{X\in\mathbb{V}_Y}p(X,\xi)\,,\label{ClusterProbabilities}
\end{align}
where $\mathbb{V}_Y\subset\mathbb{V}$ is the pre-image of $Y\in\mathbb{W}$ under $f$. Notice that $\sum_{Y\in\mathbb{W}}q(Y,\xi)=1$ $\forall \xi\in\Xi$ due to the definition of $p$. In the same way as (\ref{FisherInformationMatrix}), we can define a metric $g_{ij}^f(\xi)$ associated with $q(Y,\xi)$. As discussed in  \cite{amari2000methods}, the difference
\begin{align}
\Delta g_{ij}(\xi):=g_{ij}(\xi)-g_{ij}^f(\xi)\,,\label{DifferenceMetric}
\end{align}
is a positive semi-definite matrix and is zero if and only if the function $f$ is a \emph{sufficient statistic}\footnote{A \emph{statistic} is a quantity (or set of quantities) that can be computed from (the values in) a given sample (\emph{i.e.} any function defined on the set $\mathbb{V}$). According to \cite{FisherInt} a statistic is \emph{sufficient} 'when no other statistic which can be calculated from the same sample provides any additional information as to the value of the parameter to be estimated'.} (for the statistical model $\mathcal{S}$ in (\ref{DefStatModel})). This difference can also be formulated as
\begin{align}
&\Delta g_{ij}(\xi)=\sum_{X\in\mathbb{V}}\left(\partial_i \ln r(X,\xi)\right)\,\left(\partial_j \ln r(X,\xi)\right)\,p(X,\xi)\,,&&\text{with} &&r(X,\xi)=\frac{p(X,\xi)}{q(f(X),\xi)}\,,
\end{align}
such that the condition of $f$ being a sufficient statistic is equivalent to
\begin{align}
&\partial_i\,r(X,\xi)=0\,,&&\forall X\in\mathbb{V}\hspace{0.2cm}\text{and}\hspace{0.2cm} \xi\in\Xi\,.\label{SufficientStatisticCondRel}
\end{align}
In this work, we shall be interested in maps $f$ that correspond to a \emph{clustering} of $\mathbb{V}$: \emph{i.e.} $\mathbb{W}$ is a set of $\ell\in\mathbb{N}$ subsets of $\mathbb{V}$
\begin{align}
&\mathbb{W}=\{\mathbb{A}_1,\ldots,\mathbb{A}_\ell\}\,,&&\text{such that} &&\mathbb{V}=\bigcup_{a=1}^\ell \mathbb{A}_a&&\text{with} &&\begin{array}{l}\mathbb{A}_a\neq \{\} \hspace{0.5cm} \forall a=1,\ldots,\ell\,,\\ \mathbb{A}_a\cap \mathbb{A}_b=\{\} \hspace{0.5cm} \forall a\neq b\,.\end{array}\label{DefClusteringGen}
\end{align}
For later use, we also introduce the \emph{variation of information} as a way to measure the distance (in the sense of an actual metric) between two clusterings: let $\mathbb{W}_A=\{\mathbb{A}_1,\ldots,\mathbb{A}_\ell\}$ and $\mathbb{W}_B=\{\mathbb{B}_1,\ldots,\mathbb{B}_k\}$ be two clusterings of the set $\mathbb{V}$  and let 
\begin{align}
p_{ab}(\xi)=\sum_{X\in \mathbb{A}_a\cap\mathbb{B}_b}p(X,\xi)\,,&&\begin{array}{l}\forall a\in\{1,\ldots,\ell\}\,,\\\forall b\in\{1,\ldots,k\}\,,\end{array}
\end{align}
then we define (see \emph{e.g.} \cite{CoverThomas}) for fixed $\xi\in\Xi$
\begin{align}
\mathcal{V}(\mathbb{W}_A,\mathbb{W}_B)=-\sum_{{a=1,\ldots,\ell}\atop{b=1,\ldots,k}}p_{ab}(\xi)\left[\ln\left(\frac{p_{ab}(\xi)}{\sum_{X\in \mathbb{A}_a}p(X,\xi)}\right)+\ln\left(\frac{p_{ab}(\xi)}{\sum_{Y\in\mathbb{B}_b}p(Y,\xi)}\right)\right]\,.\label{DefVI}
\end{align}
In the case of $\mathbb{W}_A=\mathbb{W}_B$, the respective variation of information vanishes $\mathcal{V}(\mathbb{W}_A,\mathbb{W}_B)=0$.



\section{Compartmental Models with Multiple Variants}\label{Sect:CompartmentalComputations}
In this Section we consider compartmental models (see \cite{BaileyBook,Becker,DietzSchenzle,Castillo,Dietz,Dietz2,HethcoteThousand} for reviews), mostly a S$\text{I}^n$R(S) model, and discuss how clustering, in the sense described in Section~\ref{Sect:ClusteringTheory} above, can be used to describe the spread of $n\in\mathbb{N}$ variants of a pathogen throughout a population. 
\subsection{S$\text{I}^n$R(S) Model and Clustering}\label{Sect:ClusteringGen}
\subsubsection{Dynamics and Sufficient Statistic}\label{Sect:SufficientStatistic}
We consider an isolated population of fixed size (normalised to $1$), which we divide into the following compartments:
\begin{itemize}
\item \emph{susceptible} individuals, who can become infected with any of the variants of the pathogen and whose number at time $t$ shall be denoted $S(t)$.
\item \emph{infectious} individuals, who are infected with exactly one variant of the pathogen, which they can pass on to any of the susceptible individuals. We shall denote with $I_i(t)$ (for $i\in\{1,\ldots,n\}$) the number of individuals, who are at time $t$ infected with the $i$th variant of the pathogen.
\item \emph{removed} individuals, who are currently not infected and can also not be infected with any pathogen. We denote the number of removed individuals with $R(t)$.
\end{itemize}
Individuals can pass from one compartment to another and the time evolution of the respective numbers is described by the following set of coupled first-order differential equations
\begin{align}
&\frac{dS}{dt}=-S\,\sum_{j=1}^n \gamma_j I_j+\zeta R\,,&&\frac{dR}{dt}=\sum_{j=1}^n \sigma_j I_j-\zeta R\,,&&\frac{dI_i}{dt}= (S\gamma_i -\sigma_i)I_i\hspace{0.5cm}\forall i=1,\ldots,n\,.\label{SINRS}
\end{align}
Here $\gamma_i$ is the rate at which susceptible individuals are infected with the $i$th variant of the pathogen, $\sigma_i$ the rate at which individuals infected with the $i$th variant become removed and $\zeta$ is the rate at which removed individuals may become susceptible again. 

To apply the concepts discussed in Section~\ref{Sect:ReviewInfoTheory}, we introduce the probability for an infectious individual to be infected with the $i$th variant
\begin{align}
&p_i(t)=\frac{I_i(t)}{\sum_{j=1}^n I_j(t)}\,,&&\forall i=1,\ldots,n\,.\label{PropDefsSINR}
\end{align}
Eq.~(\ref{PropDefsSINR}) is a shorthand notation for (\ref{ProbNormPara}): indeed, we consider $\mathbb{V}=\{1,\ldots,n\}$ the set of all variants of the pathogen and $t\in\mathbb{R}=\Xi$, such that $p_i(t):=p(i,t)$ (as in (\ref{PropDefsSINR})).\footnote{Due to its more compact form, we shall use (\ref{PropDefsSINR}) throughout this work.} We also remark that (\ref{PropDefsSINR}) is not the unique choice to define a probability distribution for the dynamics described by (\ref{SINRS}) and we shall indeed encounter other possibilities in Subsection~\ref{Sect:GeneralisationProbDistr}.

We next organise the set $\mathbb{V}$ of variants in terms of $\ell$ clusters, following (\ref{DefClusteringGen}):
\begin{align}
&\mathbb{V}=\{1,\ldots,n\}=\bigcup_{a=1}^\ell \mathbb{A}_a\,,&&\text{with} &&\begin{array}{l}\mathbb{A}_a\neq \{\} \hspace{0.5cm} \forall a=1,\ldots,\ell\,,\\ \mathbb{A}_a\cap \mathbb{A}_b=\{\} \hspace{0.5cm} \forall a\neq b\,,\end{array}\label{DefClustering}
\end{align}
for $1\leq \ell\leq n$. The clustering is the surjective map $f:\,\mathbb{V}\to \mathbb{W}=\{\mathbb{A}_1,\ldots,\mathbb{A}_\ell\}$, which assigns each of the $n$ variants to one of the clusters $\mathbb{A}_{1,\ldots,\ell}$. Following \cite{amari2000methods} we then define the probabilities $q(\mathbb{A}_a,t)$ in (\ref{ClusterProbabilities}) for each cluster as
\begin{align}
&q(\mathbb{A}_a,t):=\frac{\sum_{k\in \mathbb{A}_a}I_k}{\sum_{j=1}^n I_j}\,,&&\text{and} &&r_i(t):=\frac{p_i(t)}{q(f(i),t)}=\frac{I_i}{\sum_{k\in f(i)}I_k}\,.\label{ClusterProbabilities}
\end{align}
Notice that $\sum_{j\in \mathbb{A}_a} r_j=1$ $\forall a\in\{1,\ldots,\ell\}$. Given the probability distributions $\{p_i(t)\}_{i\in\mathbb{V}}$ and $\{q(\mathbb{A},t)\}_{\mathbb{A}\in\mathbb{W}}$ (as functions of time) we can define the Fisher metrics
\begin{align}
&g_{tt}=\sum_{i\in\mathbb{V}} (\partial_t \ln p_i)^2 p_i\,,&&\text{and} &&g_{tt}^{f}=\sum_{\mathbb{A}\in\mathbb{W}}(\partial_t \ln q(\mathbb{A},t))^2 q(\mathbb{A},t)\,.\label{ClusterMetric}
\end{align}
As mentioned before, the difference $\Delta g_{tt}(t):=g_{tt}(t)-g_{tt}^f(t)$ is positive semi-definite and it is zero if and only if the function $f$ is a sufficient statistic \cite{amari2000methods} for the statistical model $\mathcal{S}=\{p_i(t)|i\in\{1,\ldots,n\}\text{ and }t\in\mathbb{R}\}$. In the current context, the latter in turn is a requirement for the clustering $f$ to encode essentially the same information about the time evolution as the original probabilities $\{p_i\}_{i\in\mathbb{V}}$. The condition (\ref{SufficientStatisticCondRel}) thus imposes $\partial_t r_i=0$ $\forall i\in\mathbb{V}$ and $\forall t\in\mathbb{R}$, which concretely for the model (\ref{SINRS}) becomes
\begin{align}
\frac{dr_i}{dt}&=\frac{(\gamma_i S-\sigma_i) I_i}{\sum_{k\in f(i)}I_k}-I_i\,\frac{\sum_{j\in f(i)}(\gamma_j S-\sigma_j) I_j}{\left(\sum_{k\in f(i)}I_k\right)^2}=r_i\left[(\gamma_i S-\sigma_i)-\sum_{j\in f(i)}(\gamma_j S-\sigma_j) r_j\right]\hspace{0.5cm} \forall i\in \mathbb{V}\,.
\end{align}
The expression in the square bracket tends to zero if
\begin{align}
&(\gamma_i S-\sigma_i)\sim (\gamma_j S-\sigma_j)\,,\hspace{1cm} \forall i,j\in \mathbb{A}_a\,,\forall a=1,\ldots,\ell \,. \label{ClusteringCondition}
\end{align}
Thus, a clustering of variants with similar $(\gamma_i S-\sigma_i)$ (\emph{i.e.} such that the differences $(\gamma_i S-\sigma_i)-(\gamma_j S-\sigma_j)$ are as small as possible for $i,j\in\mathbb{A}_a$, $\forall a=1,\ldots,\ell$) constitues approximately a sufficient statistic. There are, however, two important remarks:
\begin{itemize}
\item[\emph{(i)}] in general, if $\ell<n$, (\ref{ClusteringCondition}) is not an identity for all members of a cluster and thus only approximately defines a sufficient statistic. In practice, the difference (\ref{DifferenceMetric}) may be made small, but not vanishing, such that clustering generically leads to a certain loss of information on the time evolution of the system
\item[\emph{(ii)}] the map $f$ is understood to be independent of time, whereas (\ref{ClusteringCondition}) depends on $t$. In the current approach, we therefore implicitly assume that the clustering obtained from (\ref{ClusteringCondition}) remains stable, at least for a certain time interval $T\subset \mathbb{R}$. In other words, we assume that the clusters $\{\mathbb{A}_1,\ldots,\mathbb{A}_\ell\}$ obtained from (\ref{ClusteringCondition}) remain the same $\forall t\in T$.
\end{itemize}
The relation (\ref{ClusteringCondition}) suggests how to organise a clustering procedure to minimise the loss of information regarding the time evolution of the entire system. However, it is formulated in terms of $\gamma_i,\sigma_i$ and notably $S$, which contains information that is not immediately accessible from the probability distribution $\{p_i\}_{i\in\mathbb{V}}$. Indeed, the latter only depend on the $I_i$ and not on $S$, which is a quantity describing the larger dynamical system characterised by (\ref{SINRS}). In epidemiological terms, it describes the status of the entire population, which is not entirely described by the $\{p_i\}$ alone. However, while the values $(\gamma_i S-\sigma_i)$ are not directly accessible from the probabilities, we can find a clustering of these probabilities (\ref{DefClustering}) that in fact achieves the organisation  (\ref{ClusteringCondition}). To this end, consider the time-derivatives of the information (\ref{InfoDef}) of each variant\footnote{For simplicity, we denote time-derivatives with dots, \emph{e.g.} $\dot{p}_i=\frac{dp_i}{dt}$.}
\begin{align}
\frac{d\mathfrak{I}_{p_i}}{dt}&=-\frac{d \log_2 p_i}{dt}=-\frac{1}{\ln 2}\,\frac{\dot{p}_i}{p_i}=-\frac{1}{\ln 2}\frac{\dot{I}_i}{I_i}+\frac{1}{\ln 2}\frac{\sum_{j\in \mathbb{V}}\dot{I}_j}{\sum_{j\in \mathbb{V}}I_j}=-\frac{\gamma_i S-\sigma_i}{\ln 2}+c(t)\,,\label{EnergiesSInR}
\end{align}
where $c(t)=-\frac{1}{\ln 2}\frac{\sum_{j\in \mathbb{V}}\dot{I}_j}{\sum_{j\in \mathbb{V}}I_j}$ is the same for all $i\in \mathbb{V}$. Thus, since common additive constants and multiplicative factors do not influence the clustering, grouping variants according to (differences in) $(\gamma_i S-\sigma_i)$ is equivalent to clustering according to (differences in) $\{\mathfrak{I}_{p_i}\}_{i\in \mathbb{V}}$.

\subsubsection{Numerical Example}
In order to illustrate the discussion of the previous Subsection, we shall provide here a first example by numerically solving (\ref{SINRS}) for $n=20$: an example of the the time evolution of $S$, $R$

\begin{wrapfigure}{r}{0.45\textwidth}
\vspace{-0.5cm}
\begin{center}
\includegraphics[width=7.5cm]{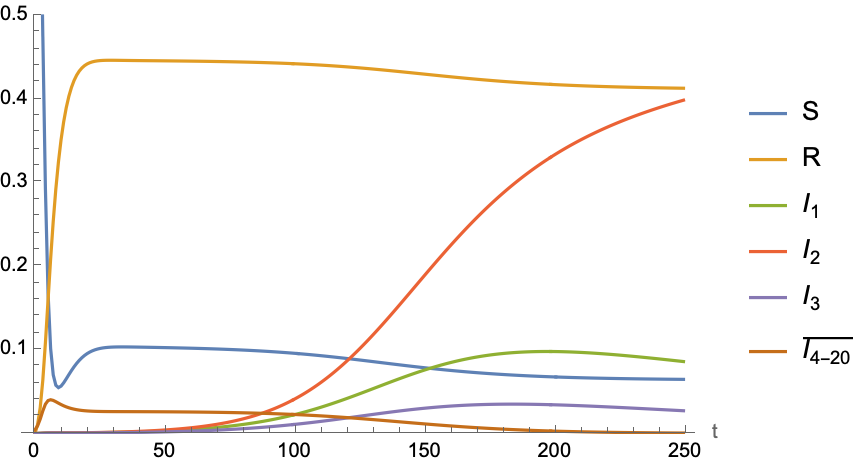}
\end{center}
\vspace{-0.5cm}
\caption{Example for the dynamics of the compartmental model (\ref{SINRS}) for $n=20$. $\overline{I}_{4-20}$ denotes the average over the variants $i=4,\ldots,20$.}
\label{Fig:SInRModelDynamics}
\end{wrapfigure}

\noindent
and the $I_n$ is shown in Figure~\ref{Fig:SInRModelDynamics}. In this example three variants (carrying labels $i=1,2,3$) are more infectious (with $\gamma_1=1.233$, $\gamma_2=1.232$, $\gamma_3=1.150$ and $\sigma_1=0.084$, $\sigma_2=0.078$, $\sigma_3=0.081$ respectively), but have lower initial values ($I_{1,2,3}(0)\sim0.00003$) compared to the remaining variants ($\gamma_i\in[0.9025,0.9975]$, $\sigma_i\in[0.095,0.105]$ and $I_i(0)\in[0.0041,0.0046]$ $\forall i=4,\ldots,20$). Furthermore $\zeta=0.1$ and $R(0)=0$. This scenario schematically simulates the appearance of three new (and more dangerous) variants, namely $i=1,2,3$, in a population where already a number of other variants are circulating. As is evident, the variant $i=2$ (red curve) becomes dominant in the population, achieving a probability of around $0.4$. 

\begin{figure}[htbp]
\begin{center}
\includegraphics[width=7.5cm]{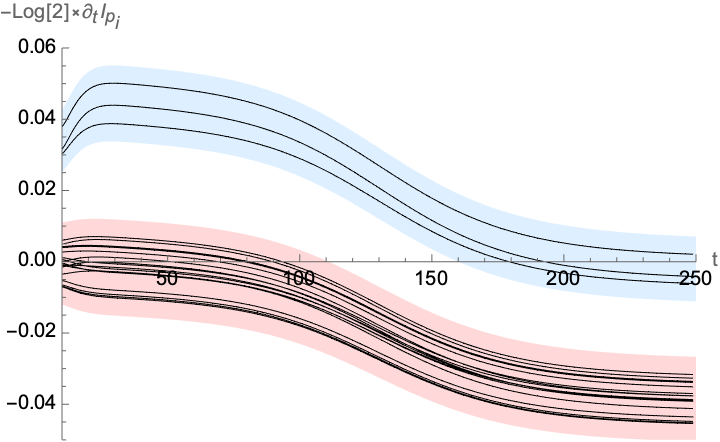}\hspace{1cm}\includegraphics[width=7.5cm]{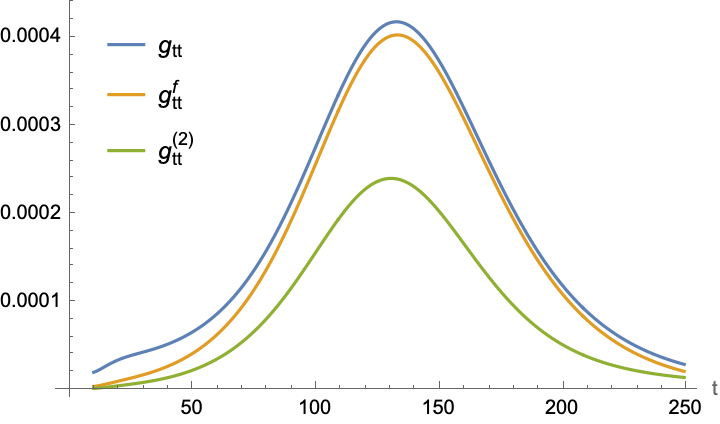}
\end{center}
\caption{Left panel: derivative of the information (\ref{EnergiesSInR}) for all variants ($i=1,\ldots,20$). The variants grouped by the blue band are $i=1,2,3$, while all remaining variants are grouped together by the red band. Right panel: comparison of the Fisher information metric computed from the probabilities $p_{i=1,\ldots,20}$ of all variant (blue curve), with the metric computed from the probabilities of the clusters (\ref{DefClustersSInR}) (orange curve) and the metric computed from the probability of the (dominant) variant $p_2$ (green curve).}
\label{Fig:ClusteringPs}
\end{figure}

The left panel of Figure~\ref{Fig:ClusteringPs} shows (up to a normalisation) the derivative of the information (\ref{EnergiesSInR}) for each of the 20 variants (shown as black lines). Due to their different natures, there exists a natural clustering into $\ell=$2 clusters of the form
\begin{align}
&\mathbb{A}_1=\{1,2,3\}\,,&&\text{and} &&\mathbb{A}_2=\{4,\ldots,20\}\,,\label{DefClustersSInR}
\end{align}
which are represented in the left panel of Figure~\ref{Fig:ClusteringPs} through coloured bands. The right panel of Figure~\ref{Fig:ClusteringPs} shows the metrics $g_{tt}$ (computed from all probabilities $p_{i=1,\ldots,20}$) and $g_{tt}^f$ (computed from the probabilities (\ref{ClusterProbabilities})) as defined in (\ref{ClusterMetric}). For comparison it also shows the metric
\begin{align}
g_{tt}^{(2)}=4\left[\left(\partial_t \sqrt{p_2(t)}\right)^2+\left(\partial_t \sqrt{1-p_2(t)}\right)^2\right]\,,
\end{align}

\noindent
based on the probability of variant 2, which becomes dominant in the considered time-frame. The plot shows that $g_{tt}>g_{tt}^f>g_{tt}^{(2)}$ in the entire time period. More importantly, the difference $g_{tt}-g_{tt}^f$ is relatively small for all $t$. Thus, while (\ref{DefClustersSInR}) is not a sufficient statistic, it still captures the time-evolution of the system (in terms of the metric) fairly well, namely better than reducing the entire dynamics to the dominant variant alone, as is evident from comparison to $g_{tt}^{(2)}$.

\subsection{Generalisations}
The approach above can be generalised in two different ways: \emph{(i)} we can consider more sophisticated types of models than (\ref{SINRS}) to reflect more elaborate epidemiological processes; \emph{(ii)} within a given model (\emph{e.g.} (\ref{SINRS})) we can consider probability distributions that are different than (\ref{PropDefsSINR}). Here, we shall briefly outline lessons that can be drawn from such generalisations.
\subsubsection{More General Compartmental Models}
The informations $\mathcal{I}_{p_i}$ in (\ref{EnergiesSInR}) encode how individual variants (measured through their probabilities $p_i$) develop in the larger system constituted by the entire population $(S,I_1,\ldots,I_n,R)$. Coarse graining the system by combining variants into clusters in such a way to preserves the information about the time-evolution, hinges upon a quantity that is specific to each variant, namely, in the case of (\ref{PropDefsSINR}), the combination $(\gamma_i S-\sigma_i)$. We expect that this quantity changes if we consider more sophisticated models than (\ref{PropDefsSINR}), thus reflecting the increased complexity of the dynamics of the entire population. For example, to get an idea, we may generalise (\ref{SINRS}) to include the additional compartment of vaccinated individuals $V$ (which are better protected against infections with all variants of the pathogen than susceptible individuals), such that the generalised dynamics is captured by the following set of differential equations
\begin{align}
&\frac{dS}{dt}=-S\,\sum_{j=1}^n \gamma_j I_j-\lambda S\,,\hspace{1cm}\frac{dV}{dt}=\lambda S+\zeta R-V\,\sum_{j=1}^n \beta_j I_j\,,\hspace{1cm}\frac{dR}{dt}=\sum_{j=1}^n \sigma_j I_j-\zeta R\,,\nonumber\\
&\frac{dI_i}{dt}= (S\gamma_i+V \beta_i -\sigma_i)I_i\hspace{0.5cm}\forall i=1,\ldots,n\,,\label{SINVRS}
\end{align}
Here $\lambda$ is the vaccination rate and $\beta_i$ the infection rates for vaccinated individuals (which for each variant of the pathogen is a priori different from the infection rate of the unvaccinated susceptible individuals) respectively. Here we assume that infection grants partial immunity, such that individuals may only return to the status of vaccinated $V$ (but not susceptible $S$). Following the discussion of Section~\ref{Sect:ClusteringTheory}, we find for the time derivative of (\ref{ClusterProbabilities})
\begin{align}
&\dot{r}_i=r_i\left[(\gamma_iS+ \beta_iV -\sigma_i)-\sum_{j\in f(i)}(\gamma_jS+\beta_jV-\sigma_j)r_j\right]\,,&&\forall i\in \mathbb{V}\,.
\end{align}
In the same way as in the absence of vaccinated individuals, clustering as in (\ref{DefClustering}) leads (approximately) to a sufficient statistic if 
\begin{align}
&(\gamma_i S+\beta_i V-\sigma_i)\sim (\gamma_j S+\beta_j V-\sigma_j)\,,\hspace{1cm} \forall i,j\in \mathbb{A}_a\,,\forall a=1,\ldots,\ell\,.\label{ClusterInfoVaccines}
\end{align}
This same quantity can still be equivalently accessed by clustering the time derivatives of the informations 
\begin{align}
&\frac{d\mathfrak{I}_{p_i}}{dt}=-\frac{\gamma_iS+ \beta_i V-\sigma_i}{\ln2}+\tilde{c}(t)\,,&&\forall i=1,\ldots,n\,.\label{InformationVaccines}
\end{align}
where $\tilde{c}(t)$ is an irrelevant (time-dependent) quantity that is the same for all variants. Notice, while it is evident from (\ref{ClusterInfoVaccines}) and (\ref{InformationVaccines}), that the clustering now takes into account more information about the more 'complex' population $\{S,V,I_1,\ldots,I_n,R\}$, the quantity (\ref{ClusterInfoVaccines}) can be fully determined from the probabilities $p_i$ through clustering. In other words, even though the quantity with respect to which the variants are clustered takes into account the more generalised external system (from the perspective of the $\{p_i\}$), parametrised by notably $S$ and $V$, it can be determined from the probability distributions alone without needing to know the full dynamics of the entire population.

This discussion can straight-forwardly be generalised to even more complicated systems, provided that the number of infectious individuals per variant (and thus the probabilities (\ref{PropDefsSINR})), can (at least approximately and over a certain period of time) be described effectively by the informations $\mathfrak{I}_{p_i}$ in the following manner
\begin{align}
&\dot{I}_i=d_i(t)\,I_i(t)\,,&&\text{with} &&d_i=-\ln 2\,\frac{d\mathfrak{I}_{p_i}}{dt}+\langle d\rangle_{p}\,,&&\forall i\in1,\ldots,n\,,\label{EffectiveDynamics}
\end{align}
where $\langle d\rangle_p$ is a time-dependent function that is (approximately) the same for all variants $i\in\{1,\ldots,n\}$. Indeed, in the two examples discussed above we have
\begin{align}
d_i=-\ln 2\,\frac{d\mathfrak{I}_{p_i}}{dt}+\langle d\rangle_p=\left\{\begin{array}{lcl}\gamma_iS-\sigma_i & \text{for} & \text{eq.~(\ref{SINRS})}\,,\\[4pt] \gamma_iS+\beta_iV-\sigma_i & \text{for} & \text{eq.~(\ref{SINVRS})}\,.\end{array}\right.\label{EffectiveCoupling}
\end{align}
Notice that $d_i$ is a general function of the (variables) of the external system (from the perspective of the variants $\{p_i\}$), namely $S$ and $(S,V)$ in the examples of (\ref{SINRS}) and (\ref{SINVRS}) respectively. In fact, it can in principle also depend on the $I_j$ themselves (as long as (\ref{EffectiveDynamics}) provides an effective description of the dynamics). Furthermore, $\langle d\rangle_p$ in (\ref{EffectiveDynamics}) can be understood as the expectation value $\langle d\rangle_p=\sum_{i=1}^n p_i\,d_i$, which follows from (\ref{EffectiveCoupling}) since $0=\sum_{i=1}^n p_i\tfrac{d\mathfrak{I}_{p_i}}{dt}$.

From (\ref{EffectiveDynamics}), we can define $d_i$ as the effective coupling of the $i$th variant to the external system. From an information theoretic perspective (\ref{EffectiveCoupling}) can be understood as a first order differential equation that governs the time-evolution of the information of the probability distribution $\{p\}_{i=1,\ldots,n}$
\begin{align}
&\frac{d\mathfrak{I}_{p_i}}{dt}=-\frac{1}{\ln 2}\left(d_i-\langle d\rangle_p\right)\,,&&\forall i=1,\ldots, n\,.\label{EffectiveEqInformation}
\end{align}
Notice in this regard that also the expectation value $\langle\cdot\rangle_p$ depends on the (choice of the) probability distribution. The clustering (\ref{DefClustering}) (with $f:\mathbb{V}\longrightarrow \{\mathbb{A}_1,\ldots,\mathbb{A}_\ell\}$ a sufficient statistic) can be seen as a similarity transformation that leaves (\ref{EffectiveEqInformation}) invariant: indeed, the time-evolution of the information of $q(\mathbb{A}_a)$ (for $a=1,\ldots,\ell$) becomes
\begin{align}
\frac{d\mathfrak{I}_{q(\mathbb{A}_a)}}{dt}=-\frac{1}{\ln 2}\frac{\dot{q}(\mathbb{A}_a,t)}{q(\mathbb{A}_a,t)}=-\frac{1}{\ln 2}\left[\frac{\sum_{i\in\mathbb{A}_a}d_i\,I_i}{\sum_{i\in\mathbb{A}_a}I_i}-\frac{\sum_{j=1}^n d_j I_j}{\sum_{j=1}^n I_j}\right]\,.
\end{align}
Assuming now $\forall a=1,\ldots,\ell$
\begin{align}
&d_i\sim d_j=:d_{\mathbb{A}_a}\,,\hspace{1cm} \forall i,j\in \mathbb{A}_a \,,\label{ClusteringConditionEffective}
\end{align}
to be the condition such that the clustering is a sufficient statistic, we obtain
\begin{align}
\frac{d\mathfrak{I}_{q(\mathbb{A}_a)}}{dt}=-\frac{1}{\ln 2}\left[d_{\mathbb{A}_a}-\sum_{b=1}^\ell d_{\mathbb{A}_b} q(\mathbb{A}_b,t)\right]=-\frac{1}{\ln 2}\left(d_{\mathbb{A}_a}-\langle d\rangle_{q}\right)\,,
\end{align}
which is structurally the same form as (\ref{EffectiveEqInformation}), except for the probability distribution $\{p_i\}$ replaced by $\{q(\mathbb{A}_a,t)\}$.
\subsubsection{Different Probability Distribution}\label{Sect:GeneralisationProbDistr}
Eq.~(\ref{PropDefsSINR}) is a simple (and natural) choice for a probability distribution in the context of the model (\ref{SINRS}) and represents the fraction of infectious individuals per variant at time $t$. There are other possible distributions on $\mathbb{V}=\{1,\ldots,n\}$ that are potentially interesting for epidemiological purposes (see \emph{e.g.} \cite{Filoche:2024xka}), for example 
\begin{align}
&p^{(1)}_i(t)=\frac{\gamma_i I_i}{\sum_{j=1}^n\gamma_j I_j}\,,&&\text{or} &&p^{(2)}_i(t)=\frac{\sigma_i I_i}{\sum_{j=1}^n\sigma_j I_j}\,.\label{ProbsExtensionNew}
\end{align}
In the context of the model (\ref{SINRS}) $\{p^{(1)}_i\}$ can be interpreted as the fraction of \emph{newly} infected individuals per variant at time $t$, while $\{p^{(2)}_i\}$ corresponds to the fraction of individuals per variant who become removed at time $t$.

Following the clustering $f:\,\{1,\ldots,n\}\longrightarrow \{\mathbb{A}_1,\ldots,\mathbb{A}_\ell\}$ introduced in Section~\ref{Sect:ClusteringGen} and using the notation (\ref{ClusterProbabilities}), we consider the time-derivative 
\begin{align}
\dot{r}_i^{(1)}=\frac{d}{dt}\left[\frac{\gamma_i I_i}{\sum_{k\in f(i)}\gamma_k I_k}\right]=r_i^{(1)}\left[(\gamma_i S-\sigma_i)-\sum_{k\in f(i)}(\gamma_k S-\sigma_k)r_k^{(1)}\right]\,,\label{rderCondition}\\
\dot{r}_i^{(2)}=\frac{d}{dt}\left[\frac{\sigma_i I_i}{\sum_{k\in f(i)}\sigma_k I_k}\right]=r_i^{(2)}\left[(\gamma_i S-\sigma_i)-\sum_{k\in f(i)}(\gamma_k S-\sigma_k)r_k^{(2)}\right]\,,
\end{align}
where we have explicitly used the dynamics (\ref{SINRS}). Thus, the condition $\dot{r}_i^{(1)}\sim 0$ or $\dot{r}_i^{(2)}\sim 0$ for the clustering to constitute a sufficient statistic for the respective statistical models $\mathcal{S}^{(1)}$ and $\mathcal{S}^{(2)}$ (see (\ref{DefStatModel})), leads to the same condition (\ref{ClusteringCondition}). 

Notice, however, that this result is due to the particular modelisation of the probabilities (\ref{ProbsExtensionNew}). For example, in the more general model described by the dynamics (\ref{SINVRS}), the fraction of newly infected is
\begin{align}
p^{(1)}_i=\frac{(\gamma_iS+ \beta_iV) I_i}{\sum_{j=1}^n(\gamma_jS+ \beta_jV)}\,,
\end{align}
which is no longer uniquely a function of the $(I_1,\ldots,I_n)$. The equivalent of (\ref{rderCondition}) then becomes {\color{red}}
\begin{align}
\dot{r}_i^{(1)}=\frac{d}{dt}\left[\frac{(\gamma_iS+\beta_iV) I_i}{\sum_{k\in f(i)}^n(\gamma_kS+ \beta_kV) I_k}\right]=r_i\left[d_i-\sum_{k\in f(i)} d_k r_k\right]\,,
\end{align}
with
\begin{align}
d_i=(\gamma_iS+ \beta_iV-\sigma_i)+\frac{S\lambda(\beta_i-\gamma_i)+\beta_i \zeta R-\sum_{j=1}^n(S\gamma_i\gamma_j+V \beta_i\beta_j)I_i}{S\gamma_i +V\beta_i}\,.
\end{align}
The condition for the clustering to constitute a sufficient statistic for the statistical model $\mathcal{S}^{(1)}$ becomes $d_i\sim d_j$ $\forall i,j\in \mathbb{A}_a$ (for all $a=1,\ldots,\ell$), which is indeed different from (\ref{ClusterInfoVaccines}). Therefore, clustering the variants of the pathogen with respect to the time evolution of the information of  different probability distributions, also clusters different functions that contain non-trivial information about the larger system made up by the entire population. Different clusterings therefore allow to probe different aspects of an (a priori) unknown system.


\section{Application to SARS-CoV-2 in France: Methods}\label{Sect:FranceSimple}
After having discussed clustering approaches in theoretical models, we shall in the remainder of this work consider an example application to real-world epidemiological data, namely the temporal evolution of the spike protein of SARS-CoV-2 in France. Indeed, this evolution is very well document in the form of amino acid sequences from samples of infected individuals during the Covid-19 pandemic, notably in the open database GISAID \cite{Gisaid1,Gisaid2,Gisaid3}. Furthermore, the effect and impact of various mutations on the transmissibility and antigenicity of different variants has been extensively studied from a virological perspective (see \emph{e.g.}~\cite{ReviewDevelop} for a nice review and references therein). In this work, in the spirit of a proof-of-concept, we show how certain aspects of this evolution can be recovered with the help of the information theoretic tools developed in the previous Section. We stress in particular, while we choose to analyse SARS-CoV-2 in this work due to the wealth of (openly) available genomic data, our approach can readily be applied to other, much less studied pathogens, allowing for similar analyses. Moreover, since our primary goal is to showcase the theoretical tools and methods developed in the previous Section (and to avoid technical subtleties with regards to the data analysis), we shall work with pruned data sets. A less restrictive analysis and a more efficient handling of the data shall be discussed elsewhere.

\subsection{Data}\label{Sect:DataTreatment}
\begin{figure}[htbp]
\begin{center}
\includegraphics[width=7.5cm]{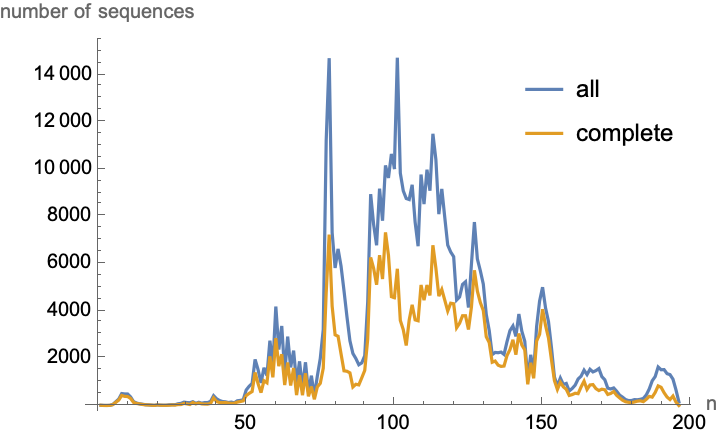}\hspace{1cm}\includegraphics[width=7.5cm]{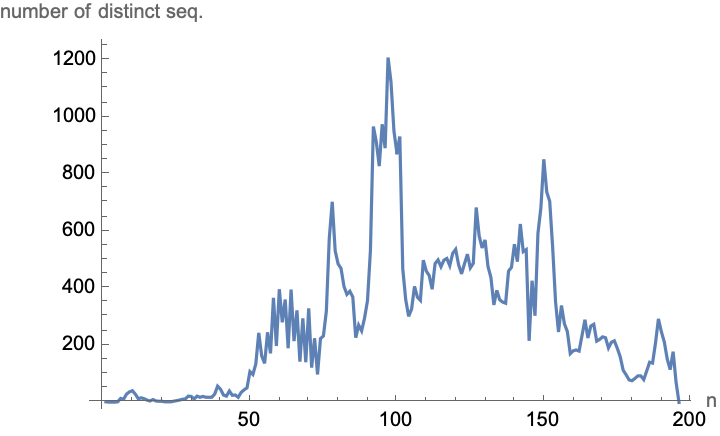}
\end{center}
\caption{Left panel: number of sequences per week in France for the entire duration of the pandemic. The blue line shows the number of all sequences, while the orange one indicates only complete ones (\emph{i.e.} without X's). Right panel: number of distinct complete sequences per week.}
\label{Fig:FranceTotalOverview}
\end{figure}

The first confirmed case of Covid-19 in France was registered on 24/01/2020 and since then roughly 39 Million infections (\cite{OurWorldIndata}, as of June 2024) with SARS-CoV-2 have been registered. Samples from these infected individuals have been genetically sequenced and in the database GISAID (accessed on 16/11/2023), 551.459 such amino acid sequences of the spike protein are available. In order to avoid subtleties relating to different strategies of reporting cases to the GISAID database by different laboratories and hospitals, we coarse-grain the sequencing data over the period of one calendar week, with week 1 corresponding to 27/01/2020. The number of sequences available each week is shown (as the blue curve) in the left panel of Figure~\ref{Fig:FranceTotalOverview}.

In each week, the GISAID database also contains a number of 'incomplete' spike protein sequences, where at least one of the amino acids could not be correctly determined and is instead replaced by an 'X'. While algorithms have been developed to infer the missing amino acids, based on statistical or probabilistic means (see \emph{e.g.} \cite{MLvariants}), in this work, for simplicity, we shall only consider complete sequences and discard incomplete ones. The number of complete sequences per week is indicated by the orange curve in the left panel of Figure~\ref{Fig:FranceTotalOverview}, while the right panel of this Figure shows the number of distinct (complete) spike-protein sequences per week. To keep the analysis tractable in the following, we shall prune the data by discarding sequences, which appear only in a very short period of time and/or accumulate a too small fraction of the sequences: in this way, we are left with a total of $N=1474$ distinct spike protein sequences, which we label by $i=1,\ldots,N$. We shall then define the \emph{probability} $p_i(n)\in[0,1]$ of the sequence $i$ in week $n$, as the number of $i$-sequences sampled during this week, divided by the number of all (complete) sequences in this week.\footnote{Using this notation, our concrete pruning criteria are to discard a sequence if it would have a non-vanishing $p_i(n)$ only for 1 or 2 weeks and/or if it would accumulate less than 0.1\% probability during the entire duration of the pandemic, \emph{i.e.} if it would result in probabilities such that $\sum_{n=1}^{196}p_i(n)<0.001$. Out of the initially 25979 distinct complete sequences, this leaves the 1474 sequences mentioned in the main text. } Notice that the probabilities are normalised such that $\sum_{i=1}^{N}p_i(n)=1$ for all weeks $n$.

 Finally, for certain computations, we shall apply a Gaussian filtering with standard deviation $\sigma$ to the probabilities $p_i(n)$ (as functions of $n$), to render them smoother and average out fluctuations in time. Since this, however, effectively smears out the information contained in the probabilities over an interval in time, we shall refrain from Gaussian filtering when discussing properties sensitive to the timing. Notice that the Gaussian filtering preserves the normalisation of the probabilities $p_i$.

\subsection{Methodology}\label{Sect:SeqMethodology}
\subsubsection{Tools to Detect Competitive Advantages}
In the following we are mainly interested in the question, whether a clustering of the sequences according to the growth of their respective  information can reveal differences (or similarities) among the analysed protein sequences. Concretely, following the examples of compartmental models in Section~\ref{Sect:CompartmentalComputations}, we shall study a clustering of the sequences according to the time-derivative of the information per sequence $\mathfrak{I}_{p_i(n)}=-\log_2 p_i(n)$. Since the $p_i(n)$ are only defined for (discrete) weeks $n$, we define a discrete approximation for its derivative. To this end, we define the shorthand notation
\begin{align}
&\mathfrak{p}_i(n):=\frac{1}{2}(p_i(n)+p_i(n+1))\,,&&\text{and} &&\mathfrak{s}_i(n):=\left\{\begin{array}{lcl}-\frac{2}{\ln2}\,\frac{p_i(n+1)-p_i(n)}{p_i(n)+p_i(n+1)} & \text{if} & \mathfrak{p}_i(n)>0\,, \\ 0 & \text{if} & \mathfrak{p}_i(n)=0\,,\end{array}\right.
\label{DiscDerivative}
\end{align}
and we take $\mathfrak{s}_i(n)\sim \frac{d\mathfrak{I}_{p_i}}{dt}$ as a discretised version of the derivative (\ref{EnergiesSInR}) of the information of the $i$th sequence in week $n$. Notice that  $\mathfrak{s}_i(n)$ yields a finite derivative for the information, even if $p_i(n)=0$ (and $p_i(n+1)>0$, \emph{i.e.} upon the occurence of a new sequence in week $n+1$). 

For a fixed week $n$, we group the $\{\mathfrak{s}_i\}_{i=1,\ldots,N}$ into $\ell\in\mathbb{N}$ clusters $\mathbb{A}_a\subset\{1,\ldots,N\}$ (with $a=1,\ldots,\ell$), according to their differences. Intuitively, one can visualise the set $\{\mathfrak{s}_i\}_{i=1,\ldots,N}$ as $N$ points along the real line that we group together based on their proximity, as schematically shown in the following Figure
\begin{center}
\begin{tikzpicture}
\draw[->] (-0.5,0) -- (9,0);
\draw[fill=green!50!white] (-0.3,-0.2) -- (1.5,-0.2) -- (1.5,0.2) -- (-0.3,0.2) -- (-0.3,-0.2);
\draw[fill=red!50!white] (2.3,-0.2) -- (4.1,-0.2) -- (4.1,0.2) -- (2.3,0.2) -- (2.3,-0.2);
\draw[fill=yellow!50!white] (6.3,-0.2) -- (8.1,-0.2) -- (8.1,0.2) -- (6.3,0.2) -- (6.3,-0.2);
\node at (9.8,0) {$\mathfrak{s}_i(n)$};
\begin{scope}[xshift=-6.5cm]
\node at (6.8,0) {$\bullet$};
\node at (7.2,0) {$\bullet$};
\node at (7.5,0) {$\bullet$};
\node at (7.8,0) {$\bullet$};
\node at (6.4,0) {$\bullet$};
\node at (7.2,0.5) {cluster $\mathbb{A}_1$};
\end{scope}
\begin{scope}[xshift=-4cm]
\node at (6.8,0) {$\bullet$};
\node at (7.1,0) {$\bullet$};
\node at (7.4,0) {$\bullet$};
\node at (7.9,0) {$\bullet$};
\node at (6.5,0) {$\bullet$};
\node at (7.3,0.5) {cluster $\mathbb{A}_2$};
\end{scope}
\node at (5.25,0.3) {\Large $\cdots$};
\node at (6.8,0) {$\bullet$};
\node at (7.2,0) {$\bullet$};
\node at (7.5,0) {$\bullet$};
\node at (7.9,0) {$\bullet$};
\node at (6.5,0) {$\bullet$};
\node at (7.3,0.5) {cluster $\mathbb{A}_\ell$};
\end{tikzpicture}
\end{center}
There are numerous different algorithms available for one-dimensional clustering (see \emph{e.g.} \cite{doi:10.1177/014662168701100401,Hubert1985ComparingP,Kaufman}), some of which require to specify $\ell$ in advance while others are capable of determining an optimal $\ell$ automatically (a review of algorithms relevant for this work is given in Appendix~\ref{App:OneDimClustering}). In this work, we shall choose an algorithm based on its capability to approximate the Fisher information metric. To this end, we first define the cluster probabilities (\ref{ClusterProbabilities})
\begin{align}
&q(\mathbb{A}_a,n):=\sum_{i\in\mathbb{A}_a}p_i(n)\,,&&\forall a=1,\ldots,\ell\,,
\end{align}
which allow to compute the (discretised) metric 
\begin{align}
g^f_{tt}(n)=\sum_{a=1}^\ell\frac{\left(q(\mathbb{A}_a,n+1)-q(\mathbb{A}_a,n)\right)^2}{\tfrac{1}{2}\left(q(\mathbb{A}_a,n)+q(\mathbb{A}_a,n+1)\right)}\,.\label{ClusterMetricDisc}
\end{align}
In most of this work, we shall use a k-means algorithm (implemented in Mathematica) with at most $\ell=6$ clusters.

After the clustering, we want to extract features of protein sequences that members of the individual clusters share in common. We are particularly interested in locating mutations that provide competitive advantages, \emph{i.e.} that allow sequences to spread faster throughout the population compared to others. The length of the spike protein sequences is not uniform. To proceed, we therefore first re-write all $N$ amino acid sequences into chains of 21 characters of common length $L$ by the insertion of gaps '--'  in a way to maximise the alignment. This can efficiently be achieved by computer tools, such as MUSCLE \cite{Edgar_2004,Edgar2}, such that we obtain aligned sequences over the alphabet
\begin{align}
\Upsilon=(\text{M},\text{F}, \text{V}, \text{L}, \text{P}, \text{S}, \text{Q}, \text{C}, \text{N}, \text{I}, \text{T}, \text{R}, \text{Y}, \text{G}, \text{D}, \text{K}, \text{H}, \text{W}, \text{A}, \text{E}, -)\,.\label{Alphabet}
\end{align}
For a given week $n$, the sequences of a given cluster $\mathbb{A}_a$ represent a vector of $L$ probability distributions over $\Upsilon$
\begin{align}
&q_\mu[\mathbb{A}_a,n]:\,\Upsilon\longrightarrow [0,1]\,,&&\text{with}&&\sum_{s\in\Upsilon}q_\mu[\mathbb{A}_a,n](s)=1\,,&&\forall \mu\in\{1,\ldots,L\}\,.\label{PropDistributionsCluster}
\end{align}
To compare these probability distributions among the clusters (in a given week $n$), we shall use the following $\alpha$-divergence 
\begin{align}
&\mathcal{D}(q_\mu[\mathbb{A}_a,n]||q_\mu[\mathbb{A}_b,n]):=\sum_{s\in\Upsilon} q_\mu[\mathbb{A}_a,n](s)\,f\left(\frac{q_\mu[\mathbb{A}_b,n](s)}{q_\mu[\mathbb{A}_a,n](s)}\right)\,,&&\text{with} &&f(u)=\frac{4}{1-\alpha^2}\left(1-u^{\frac{1+\alpha}{2}}\right)\,,\label{AlphaDiv}
\end{align}
for $\alpha\in(0,1)$ a fixed parameter. Notice that (\ref{AlphaDiv}) is not a distance but rather a dissimilarity \cite{KullbackLeibler,amari2000methods}, thus it is notably not symmetric under the exchange $a\leftrightarrow b$. In order to study competitive advantages of newly arising variants of SARS-CoV-2, we shall in the following mostly focus on scenarios when there is already an established variant circulating in the population. For a fixed week $n$, we therefore systematically choose $\mathbb{A}_a$ in (\ref{AlphaDiv}) as the largest cluster in terms of total probabilities, while for $b$ we choose clusters, whose time derivative of the information exceeds that of $\mathbb{A}_a$. Concretely, upon introducing the notation
\begin{align}
&|\mathbb{A}_c|:=\sum_{i\in\mathbb{A}_c}\mathfrak{p}_i(n)\,,&&\text{and} &&\overline{|\mathbb{A}_c|}:=\sum_{i\in\mathbb{A}_c}\mathfrak{s}_i(n)\,,&&\forall c\in\{1,\ldots,\ell\}\,,\label{DefNormsClusters}
\end{align}
with $\mathfrak{p}_i(n)$ and $\mathfrak{s}_i(n)$ defined in (\ref{DiscDerivative}), we choose
\begin{align}
&a:\hspace{0.25cm}\text{max}_{c=1,\ldots,\ell}|\mathbb{A}_c|\,,&&\text{and}
&&b:\hspace{0.25cm}\overline{|\mathbb{A}_b|}<\overline{|\mathbb{A}_a|}\,.
\end{align}
We remark that, due to the definition of $\mathfrak{I}_{p_i}$, the quantity $\overline{|\mathbb{A}_b|}$ is negative for a growing cluster.

Large divergences (\ref{AlphaDiv}) point to positions along the spike protein sequences, in which clusters that grow fast (or more correctly, which have a large absolute value for the temporal derivative of their information) differ strongly from clusters that contain sequences that are well-established in the population. Large divergences of the type (\ref{AlphaDiv}) are therefore expected to point to competitive advantages for spike proteins that grow within the population. 

Furthermore, these divergences can also be used to detect correlations among mutations: we consider mutations in two positions $\mu,\nu\in\{1,\ldots,L\}$ as correlated (in a given week $n$) if
\begin{align}
&\mathcal{D}(q_\mu[\mathbb{A}_a,n]||q_\mu[\mathbb{A}_b,n])=\mathcal{D}(q_\nu[\mathbb{A}_a,n]||q_\nu[\mathbb{A}_b,n])\,,&&\forall a,b\in 1,\ldots,\ell\,,\label{CondCorrelation}
\end{align}
\emph{i.e.} the divergences across all clusters are identical. In general, such a criterium produces numerous correlations among the data sets and to extract more meaningful ones, we shall further demand the following properties
\begin{itemize}
\item the correlation (\ref{CondCorrelation}) should remain for an extended period of time, \emph{i.e.} at least for several (consecutive) weeks
\item in addition to (\ref{CondCorrelation}), the summed divergences $\sum_{a,b=1,\ldots,\ell}\mathcal{D}(q_\mu[\mathbb{A}_a,n]||q_\mu[\mathbb{A}_b,n])$ should exceed a certain threshold. This condition guarantees that (a number of) mutations occur in the positions $\mu,\nu$ and avoids highlighting correlations between positions in which no (or only very few) mutations are observed in the data set.
\end{itemize}

\subsubsection{Towards Predicting the Spread of (Dangerous) Variants}\label{Sect:TheorySpreadDangerous}
Above we have outlined different tools that allow to pinpoint competitive advantages of individual spike proteins within their amino acid sequences. A natural question is whether knowledge of genomic features that provide (at least temporary) such advantages also allows to predict the time evolution of the spread of a certain variant (represented through its spike protein) in the population, at least at short time scales. While this is a very difficult and complex question, we shall point out features and indicators that hint towards variants with a strong potential to become dominant in the population in the near future. Moreover, once such a variant has been identified, we show how to use a recently proposed universal description of epidemiological processes \cite{Filoche:2024xka} to model the concrete time evolution.

\begin{figure}[htbp]
\begin{center}
\includegraphics[width=7.5cm]{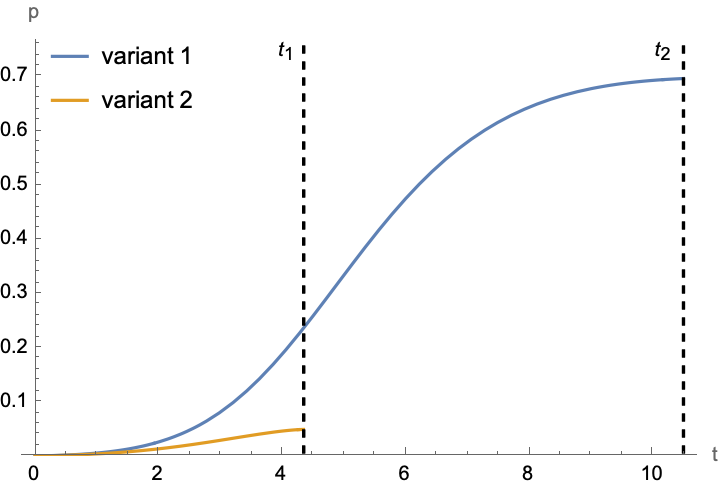}\hspace{1cm}\includegraphics[width=7.5cm]{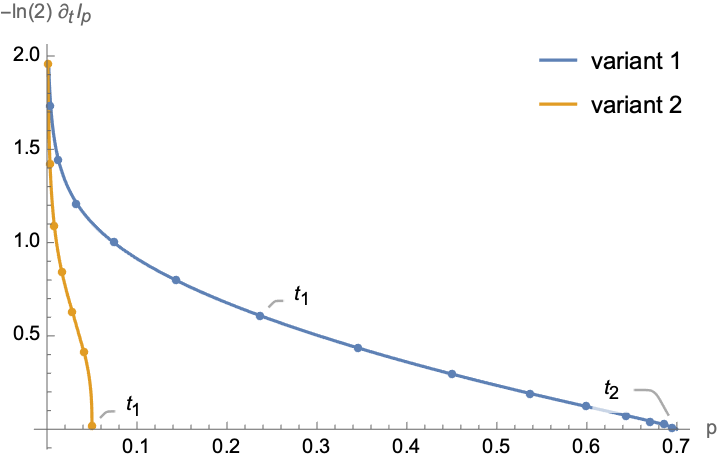}
\end{center}
\caption{Time-evolution of the probabilities (left panel) and derivative of the information $-\log 2\,\partial_t \mathcal{I}_p$ as a function of the probability (right panel) of two qualitatively different variants. For both variants, the dynamics is calculated by solving (\ref{FlowEquation}) with (\ref{ApproxMetric}) with $(a\,,b\,,c\,,p_1\,,p_2)=(1\,,0.7\,,1.5\,,0.7)$ for variant 1 and $(a\,,b\,,c\,,p_1\,,p_2)=(1\,,0.5\,,0.7\,,0\,,0.05)$ for variant 2. $t_{1,2}$ indicate the times at which variant 2 and variant 1 respectively reach the maximal probability and start declining.}
\label{Fig:SchematicAnalyticSolutions}
\end{figure}

To get a better theoretical understanding of the development of a variant, whose probability $p(t)$ is changing (in the following we assume growing) fast as a function of time, we recall the universal description of such processes proposed in \cite{Filoche:2024xka}. Indeed, there it was argued that the time evolution of $p(t)$ is very well approximated in terms of the associated Fisher information metric $g_{tt}(t)$
\begin{align}
\frac{dp}{dt}(t)=\sqrt{g_{tt}(t)\, p(t)\, (1-p(t))}\,.\label{FlowEquation}
\end{align}
Furthermore, in a regime where $p$ is a monotonic (growing) function of $t$ it was argued in \cite{Filoche:2024xka} that the Fisher metric can be approximated as the following function of $p$
\begin{align}
&g_{tt}(p)\sim a\,(p-p_1)^b\,(p_2-p)^c\,,&&\text{for} &&0\leq p_1\leq p\leq p_2\leq 1\,,\label{ApproxMetric}
\end{align} 
where $p_{1,2}$ are zeroes of the metric and $a,b,c\in\mathbb{R}$ a set of real parameters. Solving (\ref{FlowEquation}) with (\ref{ApproxMetric}) allows to model the time-evolution of the probability $p$. In Figure~\ref{Fig:SchematicAnalyticSolutions}, two cases are shown schematically, representing the probabilities of two different variants. Variant 1 (blue curve) reaches a maximal probability of 0.7, while variant 2 (orange curve) only reaches a maximum of 0.05. The former variant thus becomes dominant and should therefore be considered dangerous from an epidemiological perspective.

As we shall discuss in Section~\ref{Sect:CaseStudyFrance} (see \emph{e.g.} Figures~\ref{Fig:VshapeSeq} and \ref{Fig:PropsCluster}), the plots in Figure~\ref{Fig:SchematicAnalyticSolutions} resemble the behaviour of actual variants of SARS-CoV-2 and reveal properties of variants that have the potential to grow to very large probabilities: as seen from Figure~\ref{Fig:SchematicAnalyticSolutions}, the blue curve maintains a positive derivative $\partial_t p$ significantly longer than the orange one. In the following Subsection~\ref{Sect:CaseStudyFrance} as well as in Section~\ref{Sect:FrancLongTerm}, we shall compare this property with the probability of a given variant, represented through its protein sequence, to be part of a cluster with positive growth. We can define such a probability through the genomic properties of the variant itself. Indeed, denote the $i$th variant (with $i\in1,\ldots,N$) through the following sequence of letters in the alphabet $\Upsilon$
\begin{align}
&\mathbb{S}_i=(\mathbb{S}_{i,1},\ldots,\mathbb{S}_{i,L})\,,&&\text{with} &&\mathbb{S}_{i,\mu}\in\Upsilon\hspace{0.2cm} \forall \mu\in\{1,\ldots,L\}\,,\label{ProbabilityOccupy}
\end{align} 
and, using (\ref{PropDistributionsCluster}), interpret the cluster $\mathbb{A}_a(n)$ in week $n$ as a collection of $L$ probability distributions $q_\mu[\mathbb{A}_a,n]$ (for $\mu=1,\ldots,L$). We then interpret the product
\begin{align}
\mathfrak{h}(i,a,n):=\prod_{\mu=1}^Lq_\mu[\mathbb{A}_a,n](\mathbb{S}_{i,\mu})\in[0,1]\,,\label{ProbabilityGenetic}
\end{align}
as the probability for the $i$th sequence $\mathbb{S}_i$ to be an element of the cluster $\mathbb{A}_a(n)$ in week $n$, purely based on the genomic information encoded in the sequence. Notice, if $\mathfrak{p}_i(n)\neq 0$ (\emph{i.e.} if the variant is circulating in the population in week $n$), then $\mathfrak{h}(i,a,n)$ is non-zero at least for $i\in\mathbb{A}_a$. However, $\mathfrak{h}(i,b,n)\neq 0$ is possible also for other clusters $\mathbb{A}_b(n)$. As we shall see, sequences which have non-vanishing probabilities (\ref{ProbabilityGenetic}) in (several) clusters with positive growth tend to also obtain large probabilities.



\begin{figure}[hb]
\begin{center}
\includegraphics[width=7.5cm]{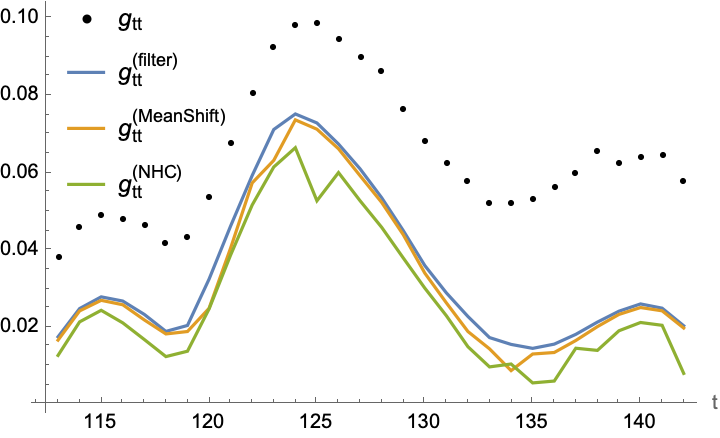}\hspace{1cm}\includegraphics[width=7.5cm]{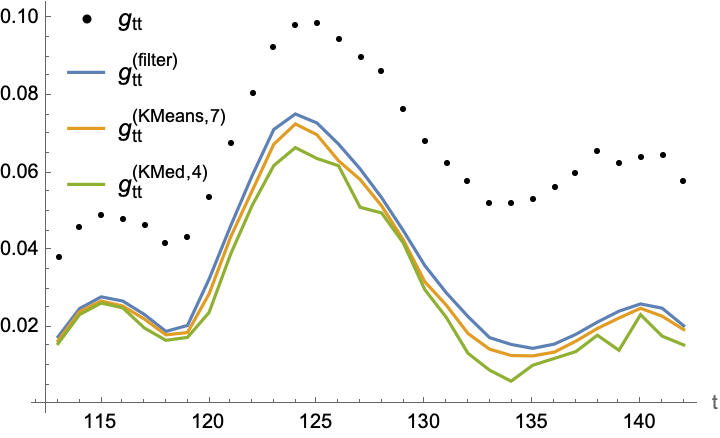}
\end{center}
\caption{Fisher information metric after clustering (pruned and filtered) sequence probabilities according to the time-derivative of their information using different algorithms (performed with Mathematica): the left panel showcases algorithms without specifying the number of clusters, while the right panel uses a predefined number of them. In both cases, the black dots indicate the Fisher metric of the pruned (but unfiltered) data, while the blue line indicates the Fisher metric computed from the Gaussian filtered ($\sigma=4$) data: as explained in  \cite{amari2000methods} all metrics computed from clustered probabilities are below this curve. The superscript of each metric indicates the algorithm used in the clustering and (if applicable) the pre-determined number of clusters.}
\label{Fig:FranceClustersMod}
\end{figure}

\subsection{Example Period: Spring/Summer 2022}\label{Sect:CaseStudyFrance}
Before moving on to analyse the data of the entire pandemic in France in subsequent Subsections, we shall first exhibit our methodology in more detail on a smaller time-scale. To this end, we consider spike-protein sequencing data from France in the weeks 109-138 (\emph{i.e.} from 21/02/2022 -- 12/09/2022), which contain a relatively low number of incomplete sequences. After pruning the data as described above, we are left with $N=119$ distinct spike protein sequences in this time-period, to which we shall apply a Gaussian filtering with $\sigma=4$ {weeks. The length of these amino acid sequences varies in the interval $[1267,1273]$ and with the tool MUSCLE \cite{Edgar_2004,Edgar2} we obtain aligned sequences of length $L=1279$ over the alphabet (\ref{Alphabet}).

\subsubsection{Clustering and Mutations}\label{Sect:ClusteringShowcaseRel}
From the filtered probabilities $p_i$, we compute discretised time-derivatives $\mathfrak{s}_i$ of the information according to (\ref{DiscDerivative}), which allows us to cluster the $N$ sequences into clusters. The metrics (\ref{ClusterMetricDisc}) stemming from the cluster probabilities, following different clustering algorithms, are shown in Figure~\ref{Fig:FranceClustersMod} and are compared to the metric calculated from the unclustered, filtered and un-filtered data. Since a higher number of clusters generally leads to a metric that is closer to~the~unclus-

\begin{wrapfigure}{r}{0.55\textwidth}
\vspace{-0.5cm}
\begin{center}
\includegraphics[width=7.5cm]{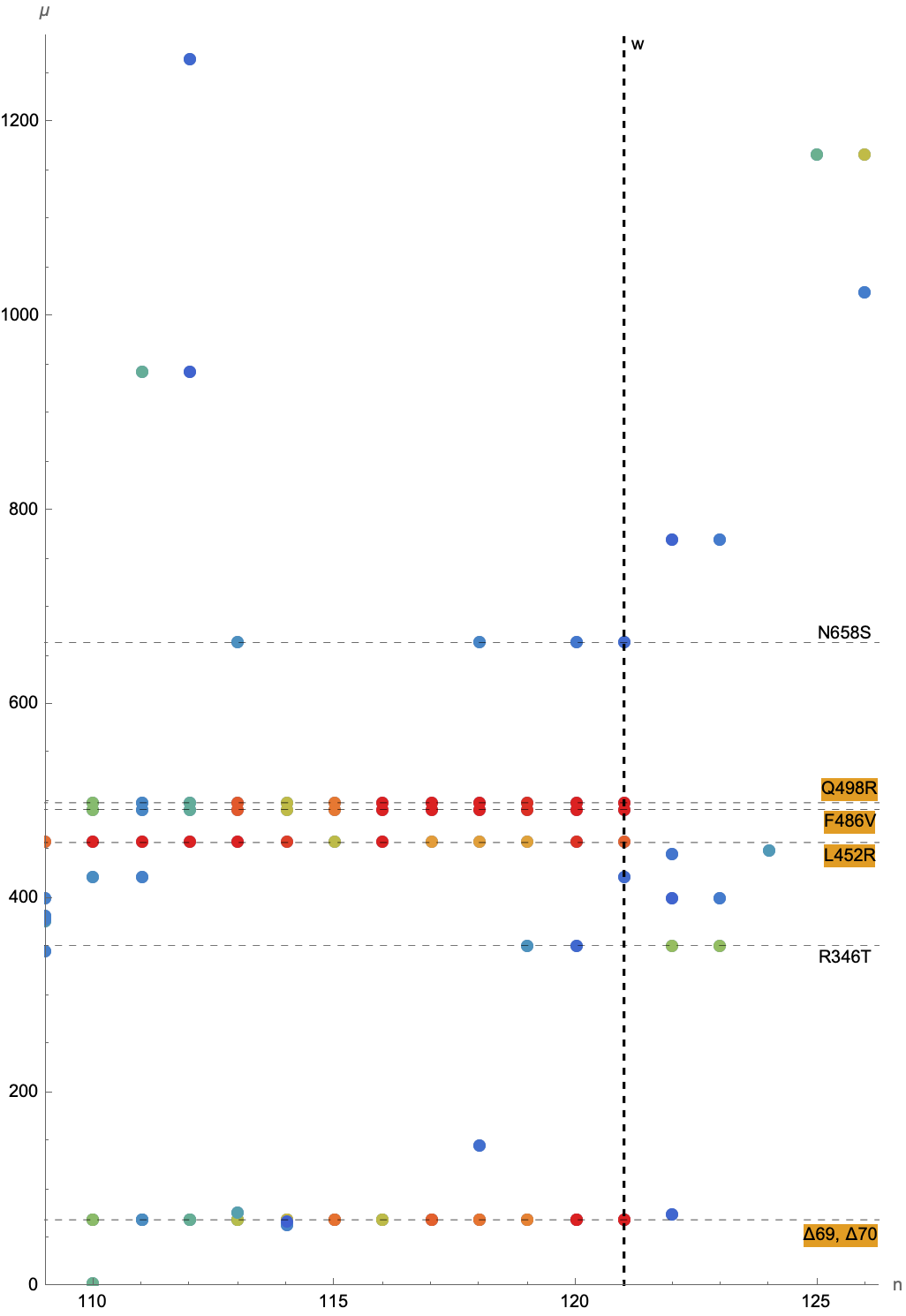}
\end{center}
\vspace{-0.5cm}
\caption{Evolution of the divergences (\ref{AlphaDiv}) as a function of the position $\mu$. Warmer colours represent larger values of $\mathcal{D}$.}
\label{Fig:FranceProbDevelopShowcase}
\end{wrapfigure}

\noindent
tered (filtered) $g_{tt}^{(\text{filter})}$ (the blue line in Figure~\ref{Fig:FranceClustersMod}), algorithms in which a (fixed and sufficiently large) number of clusters is specified in advance, typically lead to a smoother metric. Furthermore, $\ell=6$ or $\ell=7$ leads to a metric that is very close to $g_{tt}^{(\text{filter})}$ and in the following we shall focus on the clustering based on the k-means algorithm with $\ell=6$.

Once the sequences have been grouped into clusters, we systematically compute the divergences (\ref{AlphaDiv}), where for concreteness we choose $\alpha=3/4$. For the weeks 109-121, all $\mathcal{D}(q_\mu[\mathbb{A}_a,n]||q_\mu[\mathbb{A}_b,n])>0.5$ are tabulated in Appendix~\ref{App:TableAlphaDivergences} as functions of their position $\mu$.\footnote{For comparison, Appendix~\ref{App:TableAlphaDivergences} also lists the divergences of the largest cluster from one week to the next, which are generally much smaller than $0.5$. This shows that over long periods of time, the dominant cluster remains stable and is composed of (genetically) very similar sequences.} This information is also graphically represented in Figure~\ref{Fig:FranceProbDevelopShowcase}: for each week $n$, the coloured dots represent the $\mathcal{D}(q_\mu[\mathbb{A}_a,n]||q_\mu[\mathbb{A}_b,n])>1.5$ according to their position $\mu$ in the protein sequence. The colour of the dots indicates the numerical value of $\mathcal{D}$, with warmer colours (red) representing larger values, and thus mutations in the sequences making up the (fast) growing clusters compared to the sequences in the largest (dominant) cluster. The vertical lines mark the positions $\mu\in\{1,\ldots,L=1279\}$ of the important mutations in the spike protein sequence, while the labels identify their position relative to the original Wuhan spike protein (following the notation of GISAID \cite{Gisaid1,Gisaid2,Gisaid3}). More quantitatively, a tally of the divergences over the weeks 109-121 is shown in the left panel of Figure~\ref{Fig:FranceDivergencesSummed}. Concretely, this diagram shows
\begin{align}
\sum_{n=109}^{121}\sum_{b:\,\overline{|\mathbb{A}_b|}<\overline{|\mathbb{A}_a|} }\mathcal{D}(q_\mu[\mathbb{A}_a,n]||q_\mu[\mathbb{A}_b,n])\,,\nonumber
\end{align}
as a function of the amino acid position $\mu\in\{1,\ldots,L\}$ (only those positions with a sum $\geq 8$ are displayed). The blue part of each column is the contribution from the cluster with the smallest $|\mathbb{A}_b|$ each week. This indeed makes it evident that the clusters with the largest change of information compared to the dominant cluster show the largest differences in five positions, which coincide with the following 5 mutations relative to the Wuhan spike protein sequence, as identified for example by the GISAID CovSurver App \cite{Gisaid1,Gisaid2,Gisaid3}
\begin{align}
\begin{tabular}{|l|c|c|c|c|c|}\hline
&&&&&\\[-12pt]
{\bf position} & 69 & 70 & 458 & 492 & 499\\[2pt]\hline
&&&&&\\[-12pt]
{\bf mutation }&\text{H69del}& \text{H70del}&\text{L452R} & \text{F486V} & \text{Q498R}\\[2pt]\hline
\end{tabular}\label{Mutations238}
\end{align}
The positions of these mutations are also highlighted in blue in the tables of $\mathcal{D}(q_\mu[\mathbb{A}_a,n]||q_\mu[\mathbb{A}_b,n])$ in Appendix~\ref{App:TableAlphaDivergences}. Formulated differently, clusters whose information grows faster than the dominant cluster (\emph{i.e.} which have competitive advantages), consistently contain sequences with mutations in the positions of the mutations~(\ref{Mutations238}). The clustering algorithm therefore is capable  of pointing out 'weaknesses' of the dominant cluster.

\begin{figure}[htbp]
\begin{center}
\includegraphics[width=7.5cm]{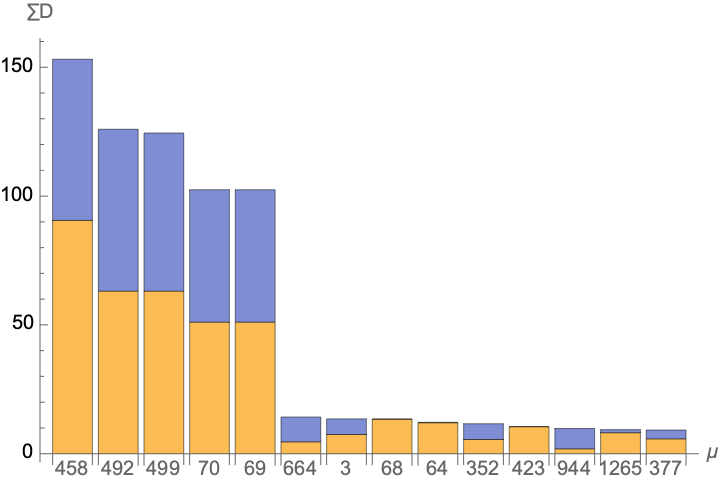}\hspace{1cm}\includegraphics[width=7.5cm]{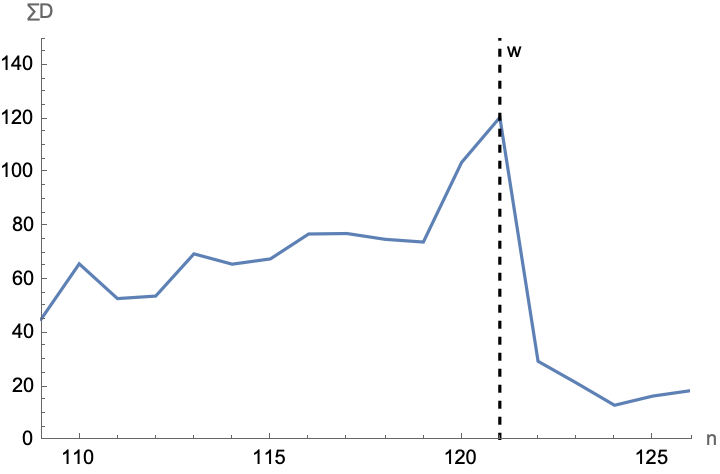}
\end{center}
\caption{Left panel: Tally of the divergences per sequence position. The position labelling follows the uniformised sequence alignment (\emph{i.e.} $\mu\in\{1,\ldots,L=1279$). Right panel: sum (\ref{TotalDivergencesWeek}) over the divergences in all positions of the amino acid sequence per week. The maximum indicated by $w$ in week $125$ marks a change in the nature of the largest (dominant) cluster.}
\label{Fig:FranceDivergencesSummed}
\end{figure}

\noindent
The right panel in Figure~\ref{Fig:FranceDivergencesSummed} shows the sum over the divergences in all positions in a given week
\begin{align}
\sum_{\mu=1}^{L}\sum_{b:\,\overline{|\mathbb{A}_b|}<\overline{|\mathbb{A}_a|} }\mathcal{D}(q_\mu[\mathbb{A}_a,n]||q_\mu[\mathbb{A}_b,n])\,.\label{TotalDivergencesWeek}
\end{align}
Figure~\ref{Fig:FranceDivergencesSummed} shows that starting from week 109, the summed divergences are growing: this suggests, sequences with mutations relative to the dominant cluster (notably in positions 69, 70, 458, 492 and 499) have competitive advantages and are growing. The right panel of Figure~\ref{Fig:FranceDivergencesSummed} shows a distinct maximum in week 121, indicated by the dashed vertical line labeled $w$ (see also Figure~\ref{Fig:FranceProbDevelopShowcase}), followed by a steep drop in the summed divergences. Indeed, between week 121  and 122, the nature of the dominant cluster changes: to understand this change, we show in Figure~\ref{Fig:FranceProbSeqsDominating} the probabilities of a number of important individual sequences, which we have ident-

\begin{wrapfigure}{l}{0.50\textwidth}
\begin{center}
\includegraphics[width=7.5cm]{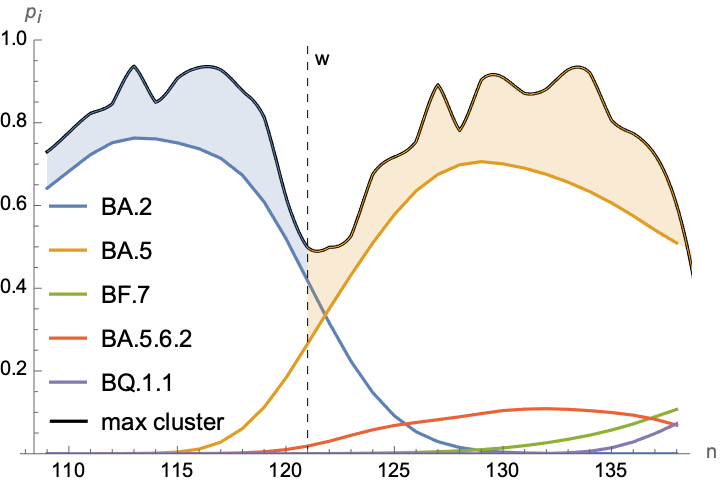}
\end{center}
\caption{Probabilities of selected sequences and combined probability of the largest cluster.}
\label{Fig:FranceProbSeqsDominating}
\end{wrapfigure}

\noindent
ified with the help of the GISAID CovSurver App (coloured lines) along with the (combined) probability of the largest cluster of each week (black line). The plotted sequences have been chosen, since they achieve fairly large probabilities, notably sequences corresponding to the variants BA.2 and BA.5, which become dominant (\emph{i.e.} with probabilities larger than 0.5) in a certain time period. Week 114 is the first week in which the cluster containing sequence BA.5 is the fastest growing one and its overall probability in this week is 0.0034. Week 122 marks the transition, in which the cluster containing sequence BA.5 becomes larger than the cluster containing sequence BA.2 (week 121 immediately before, is indicated by the vertical dashed line and denoted $w$ in Figure~\ref{Fig:FranceProbSeqsDominating}). Finally, BA.5.6.2, BQ.1.1 and BF.7 are sequences that also reach a sizeable probability. 

The Levenshtein distance \cite{Levenshtein1965BinaryCC} (see Appendix~\ref{Sect:GeneticClustering}) between sequence BA.2 and BA.5 is five, which means that their spike-protein sequences differ in five positions. These five are in fact precisely the mutations (\ref{Mutations238}). The positions of these mutations are highlighted in blue in the tables of $\mathcal{D}(q_\mu[\mathbb{A}_a,n]||q_\mu[\mathbb{A}_b,n])$ in Appendix~\ref{App:TableAlphaDivergences}): the dominant clusters $\mathbb{A}_a(n)$ for $n\in\{109,\ldots,121\}$ all contain sequence BA.2, while among the clusters $\mathbb{A}_b$ in these tables, those containing BA.5 are

\begin{wrapfigure}{r}{0.46\textwidth}
\begin{center}
\vspace{-0.5cm}
\includegraphics[width=7.5cm]{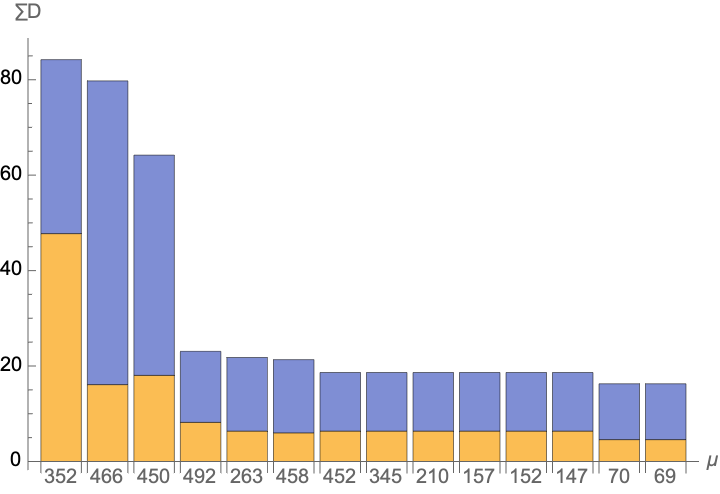}
\end{center}
\vspace{-0.5cm}
\caption{Divergences per amino acid position relative to the cluster containing sequence BA.5.}
\label{Fig:FranceClusterTallyDivSub}
\vspace{-0.5cm}
\end{wrapfigure}

\noindent
marked in orange. As is evident, even clusters that do not contain sequence BA.5 (but have a smaller $\overline{|\mathbb{A}_b|}$ than the dominant cluster), show large deviations in the locations (\ref{Mutations238}) of the spike protein.

It is interesting to continue the analysis past week 121, when the composition of the largest cluster has changed and now notably contains sequence BA.5, while BA.2 is slowly disappearing in the population. Similar to before, we can study the divergences $\mathcal{D}(q_\mu[\mathbb{A}_{\text{BA.5}},n]||q_\mu[\mathbb{A}_b,n])$ for weeks $n\geq 118$, where $\mathbb{A}_{\text{BA.5}}$ is the cluster containing the spike protein sequence of BA.5 (which after week 121 constitutes the largest cluster) and $\mathbb{A}_b$ are chosen such that $\overline{|\mathbb{A}_b|}<\overline{|\mathbb{A}_{\text{BA.5}}|}$. For brevity, here we only present a counting of the divergences per amino acid position in Figure~\ref{Fig:FranceClusterTallyDivSub}. Concretely, we have plotted  the summed divergences
\begin{align}
\sum_{n=124}^{136}\sum_{b:\,\overline{|\mathbb{A}_b|}<\overline{|\mathbb{A}_{38}|} }\mathcal{D}(q_\mu[\mathbb{A}_{38},n]||q_\mu[\mathbb{A}_b,n])\,,\nonumber
\end{align}

\noindent
This suggests that sequences with mutations in certain positions have competitive advantages:
\begin{align}
\begin{tabular}{|l|c|c|c|}\hline
&&&\\[-12pt]
{\bf position} & 352 & 450 & 466 \\[2pt]\hline
&&&\\[-12pt]
{\bf mutation }&\text{R346T}& \text{K444T}&\text{N460K} \\[2pt]\hline\hline
&&&\\[-12pt]
{\bf BA.5.6.2 }& & & $\checkmark$ \\[2pt]\hline
&&&\\[-12pt]
{\bf BF.7 }& $\checkmark$ & &  \\[2pt]\hline
&&&\\[-12pt]
{\bf BQ.1.1 }& $\checkmark$ & $\checkmark$ & $\checkmark$ \\[2pt]\hline
\end{tabular}\nonumber
\end{align}
As indicated, these  positions coincide with the positions of known mutations (with regards to the original Wuhan variant) that in turn appear in certain (classified) variants. Among these three variants, BQ.1.1 (which carries all three mutations) continues to become the largest sequence after week 141, reaching a maximum probability of $0.47$ in week 148. This result is again compatible with the interpretation that the clustering approach outlined above is capable of detecting competitive advantages even when the probabilities of the genomic sequences concerned are still relatively small.



\subsubsection{Correlations Among Mutations}\label{Sect:ShowcaseCorrelations}
After discussing positions of mutations that provide competitive advantages for variants, we shall now consider correlations among mutations. Systematically searching for positions $\mu,\nu\in\{1,\ldots,L\}$ that satisfy (\ref{CondCorrelation}) (while still presenting some amount of mutations), we found two instances of strong correlations:

\noindent
\begin{itemize}
\item $\mu\in\{69,70\}$, which correspond to the position of the mutations H69del and V70del (with respect to the Wuhan variant). Deletions in position 69 and 70 appear in all sequences of the sampling period together: 79 (of the 119) sequences considered show both deletions, while the remaining 40 sequences carry neither of the two. Moreover, throughout the entire time evolution, clusters tend to be composed predominantly of only one type of sequence, \emph{i.e.} those that contain or not contain both deletions. Indeed, the left panel of Figure~\ref{Fig:ProbRebels} shows the fraction of sequences carrying the $\Delta 69-70$ deletion as a function of the size of the cluster: only clusters of relatively small size (\emph{i.e.} $|\mathbb{A}|\lesssim 0.15$ show a perceptible mixture between the two types of sequences).
\item $\mu\in\{24,25,26,27,67,95,143,144,211,411,502,862\}$, which correspond to the position of the following mutations with respect to the Wuhan variant
\begin{align}
\begin{tabular}{|c|c|c|c|c|c|c|c|c|c|c|c|c|}\hline
&&&&&&&&&&&&\\[-12pt]
 24 & 25 & 26 & 27 & 67 & 95 & 143 & 144 & 211 & 212 & 411 & 502 & 862\\[2pt]\hline
&&&&&&&&&&&&\\[-12pt]
\scriptsize{\text{L24del}} & \scriptsize{\text{P25del}}  & \scriptsize{\text{P26del}}& \scriptsize{\text{A27S}} & \scriptsize{\text{A67V}} & \scriptsize{\text{T95I}} & \scriptsize{\text{V143del}} & \scriptsize{\text{Y144del}} & \scriptsize{\text{N211del}} & \scriptsize{\text{L212I}} & \scriptsize{\text{D405N}} & \scriptsize{\text{G496S}} & \scriptsize{\text{N856K}}\\[2pt]\hline
\end{tabular}\nonumber
\end{align}
The 119 sequences considered during the entire time frame (weeks 109-138) can be classed into two groups according to the amino acids in these positions

{\centering
\begin{tabular}{|l|l||c|c|c|c|c|c|c|c|c|c|c|c|c|}\hline
&&&&&&&&&&&&&&\\[-12pt]
type & num. of seq. & 24 & 25 & 26 & 27 & 67 & 95 & 143 & 144 & 211 & 212 & 411 & 502 & 862\\[2pt]\hline
&&&&&&&&&&&&&&\\[-12pt]
I) & 6 & L & P & P & A & V & I & --- & --- & I & D & S & K & K\\[2pt]\hline
&&&&&&&&&&&&&&\\[-12pt]
II) & 113 & --- & --- & --- & S & A & T & V & Y & N & N & G & T & N\\[2pt]\hline
\end{tabular}}

\noindent
The (combined) probabilities of the six sequences labelled I) is shown in Figure~\ref{Fig:ProbRebels}: these sequences are mostly prevalent up to week 123 and are mostly absent at later times. We also remark that we find indications for (weaker) correlation between the positions indicated above, as well as 145 (Y145del), 216-217 (ins214EPE), 452 (G446S) and 553 (T547K): including these mutations, however, the set of sequences is separated into more than two groups, since we also find a number of sequences (with relatively low probabilities) that show mixed mutations.
\end{itemize}

\begin{figure}[htbp]
\begin{center}
\includegraphics[width=7.5cm]{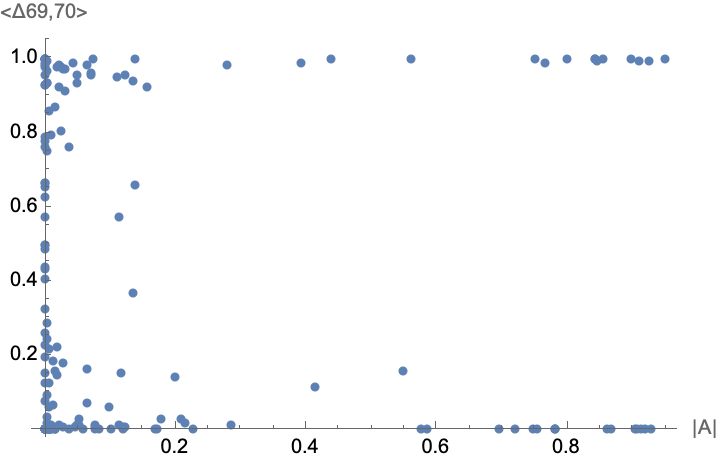}\hspace{1cm}\includegraphics[width=7.5cm]{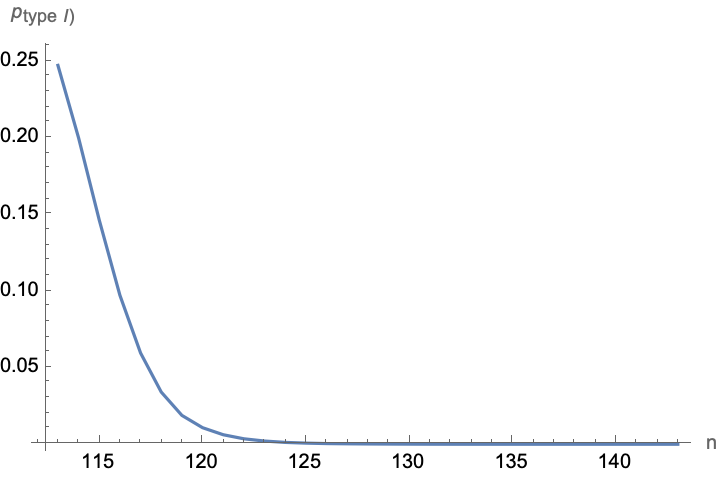}
\end{center}
\caption{Left panel: Probability of $\Delta 69-70$ mutation as function of the cluster size. Right panel: Combined probability of the six sequences labelled I).}
\label{Fig:ProbRebels}
\end{figure}

\subsubsection{Dangerous Variants and Prediction of Evolution}\label{Sect:DangerousVariants}
Following the discussion in Section~\ref{Sect:TheorySpreadDangerous} we shall compare properties of the time-evolution of sequences that achieve high probabilities (notably BA.5) with that of sequences which do not circulate widely in the population. Following the right panel of Figure~\ref{Fig:SchematicAnalyticSolutions}, we have plotted in Figure~\ref{Fig:VshapeSeq} the derivative of the information for each sequence as a function of the probability.

\begin{figure}[htbp]
\begin{center}
\includegraphics[width=7.5cm]{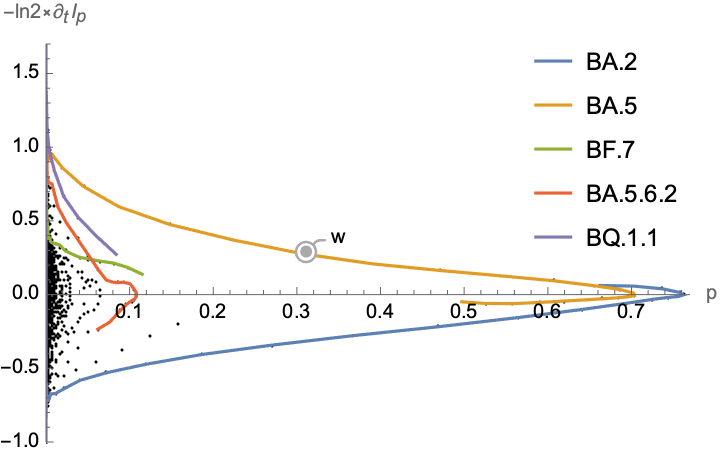}\hspace{1cm}\includegraphics[width=7.5cm]{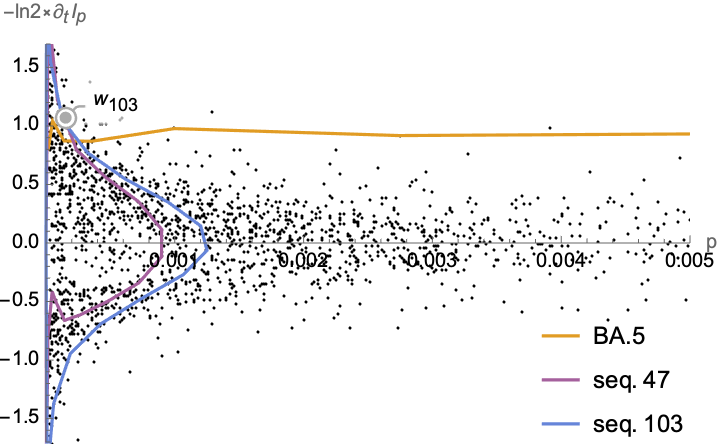}
\end{center}
\caption{Derivative of the information as a function of the probability: each black dot represents the values for one of the 119 sequences in one of the weeks $n\in\{109,\ldots,131\}$. The coloured lines indicate the trajectories of selected sequences: in the left panel sequences that reach a probability of larger than 7.5\%, while the right panel highlights the trajectories of two further unclassified sequences (simply named "seq. 47" and "seq. 103" following an internal labelling) that reach maximal probabilities of around 0.001. $w$ in the left panel denotes week 121 in the trajectory of BA.5, while $w_{103}$ denotes week 126 in the trajectory of seq. 103.}
\label{Fig:VshapeSeq}
\end{figure}

The majority of the data points lie in a region close to the origin (thus representing small probabilities), while sequences that grow to large probabilities follow trajectories that are significantly outside of this region, indeed following the schematic curves shown in the right panel of Figure~\ref{Fig:SchematicAnalyticSolutions}.

Figure~\ref{Fig:PropsCluster} shows the derivative of information per cluster (\emph{i.e.} the quantity $|\mathbb{A}|$ defined in (\ref{DefNormsClusters})) as a function of the average probability of the sequences in the same cluster (\emph{i.e.} $\overline{|\mathbb{A}|}$ defined in (\ref{DefNormsClusters}) divided by the number of sequences in $\mathbb{A}$): more precisely, the left panel shows 
\begin{align}
\delta\overline{|\mathbb{A}_a(n)|}:=|\mathbb{A}_a(n)|-\frac{1}{\ell}\sum_{b=1}^\ell |\mathbb{A}_b(n)|\,,\label{CenterMass}
\end{align}
which is the derivative of information per cluster corrected by the average over all clusters. In this way, the clustering across all weeks become comparable and the $\ell=6$ clusters fall into horizontal bands as indicated in the Figure.\footnote{The definition (\ref{CenterMass}) essentially removes the time-dependent quantity $c(t)$ in (\ref{EnergiesSInR}) from the perspective of the theoretical models.} The right panel highlights the clusters which contain the sequences BA.5 and seq.~103 respectively, and show a similar structure as in Figure~\ref{Fig:VshapeSeq} at the level of individual sequences. Both plots demonstrate that sequences which grow to large probabilities occupy clusters with positive growth for a longer period of time.

\begin{figure}[htbp]
\begin{center}
\includegraphics[width=7.5cm]{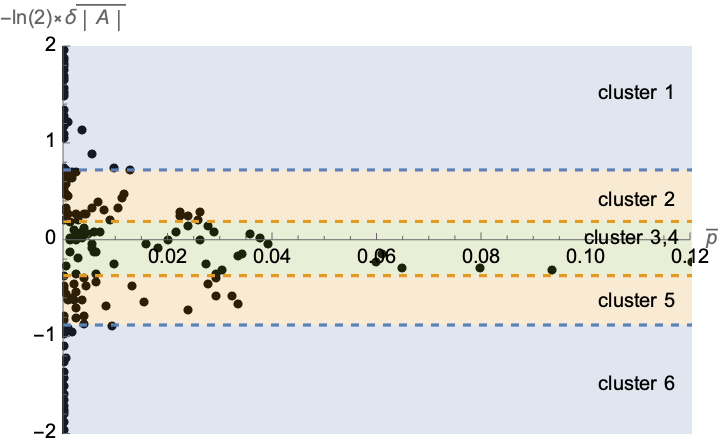}\hspace{1cm}\includegraphics[width=7.5cm]{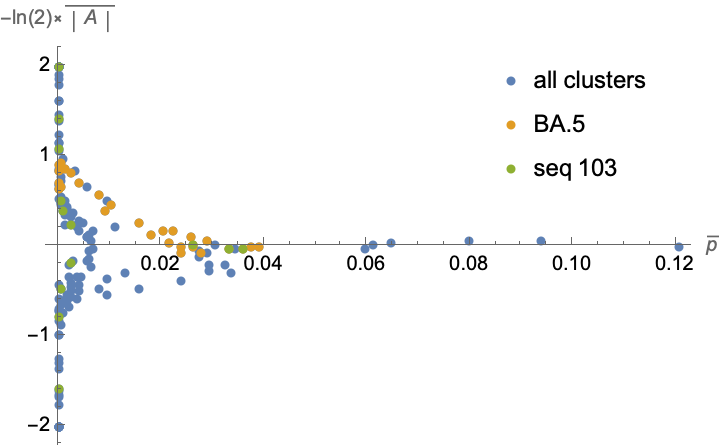}
\end{center}
\caption{Derivative of information per cluster as a function of the average probability for all clusters in the weeks $n\in\{109,\ldots,131\}$. The left panel shows the quantity (\ref{CenterMass}) which allows to compare similar clusters across different weeks, which occupy horizontal bands. The right panel highlights clusters that contain BA.5 and seq. 103 respectively. }
\label{Fig:PropsCluster}
\end{figure}

As explained in (\ref{ProbabilityOccupy}), once the data in a given week are clustered, the probability for \emph{any} sequence to be a member of a given cluster can also be defined purely based on the amino acid sequences. These probabilities show marked differences between sequences that reach sizeable probabilities and those which do not. Figure~\ref{Fig:PropsClusterGenetic} shows the quantity
\begin{align}
\overline{P}^+_i(n)=\frac{\sum_{a\text{ with }\overline{|\mathbb{A}_a|}(n)<0}\mathfrak{h}(i,a,n) \overline{|\mathbb{A}_a|}(n)}{\sum_{a\text{ with }\overline{|\mathbb{A}_a|}(n)<0}1}\,,\label{AverageProbPositive}
\end{align}
\emph{i.e.} the average probability of the sequence $i$ to be in a cluster of shrinking information, for some selected sequences. These plots highlight properties of dangerous sequences that grow to large probabilities, such as BA.5 and BF.7 (in the top row of the plot):

\begin{figure}[htbp]
\begin{center}
\includegraphics[width=7.5cm]{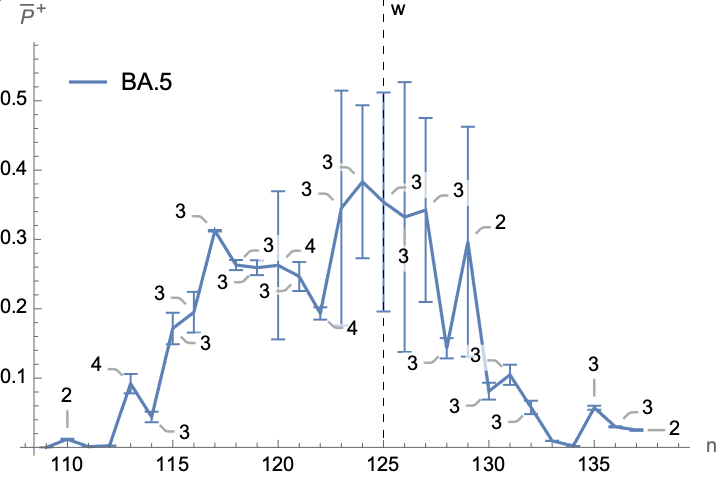}\hspace{1cm}\includegraphics[width=7.5cm]{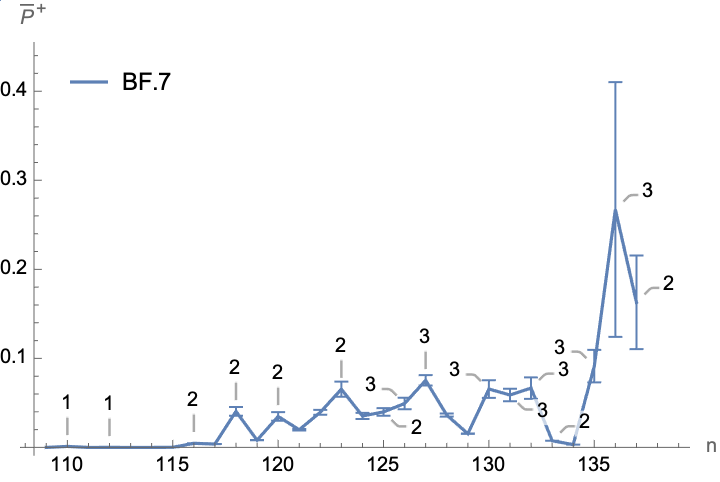}\\
\includegraphics[width=7.5cm]{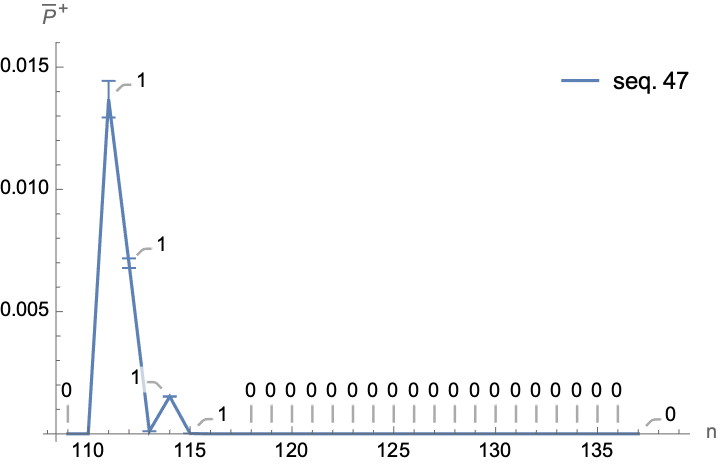}\hspace{1cm}\includegraphics[width=7.5cm]{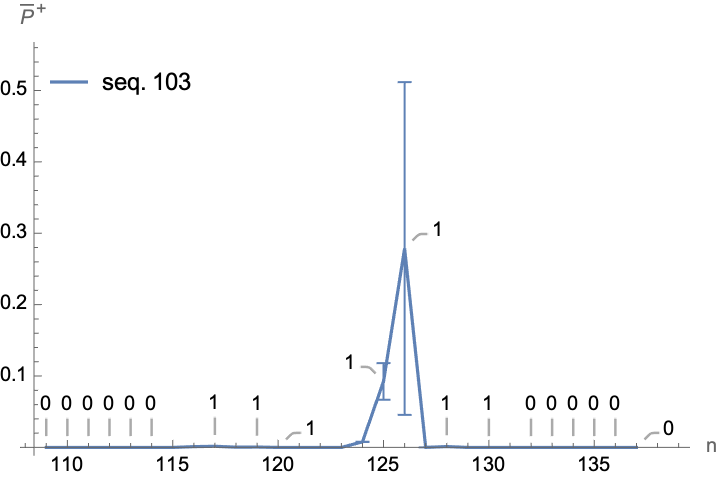}
\end{center}
\caption{Average probability (\ref{AverageProbPositive}) to be a member of a cluster with positive growth for a number of selected sequences. The error bars indicate the variance of the probability (\ref{AverageProbPositive}), while the numbers attached to each point denote the number of clusters for which the probability (\ref{ProbabilityGenetic}) is non-zero. Seq. 47 and 103 are the same sequences appearing in Figure~\ref{Fig:VshapeSeq}.}
\label{Fig:PropsClusterGenetic}
\end{figure}

\begin{itemize}
\item[\emph{(i)}] dangerous sequences have a positive (and sizeable) $\overline{P}^+_i(n)$ for several consecutive weeks
\item[\emph{(ii)}] dangerous sequences have non-vanishing probabilities to be members of more than one cluster (over an extended period of time)
\item[\emph{(iii)}] the variance of $\overline{P}^+_i(n)$ is small (over an extended period of time), indicating that the probabilities to be members of more than one cluster are comparable
\end{itemize}
This suggests that the probabilities $\overline{P}^+_i(n)$ (along with its variance) and the number of clusters a sequence can be a member of are indicators for the growth potential of a sequence. These indicators could help in the future to ascertain if newly appearing variants are dangerous.

\begin{figure}[p]
\begin{center}
\includegraphics[width=5.5cm]{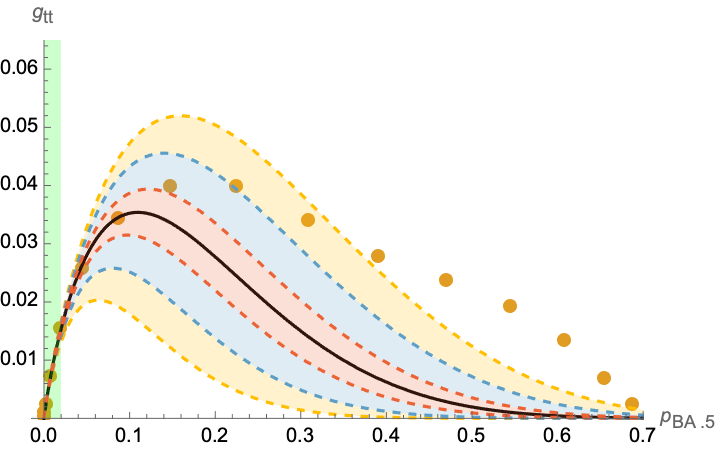}\hspace{0.25cm}\includegraphics[width=5.5cm]{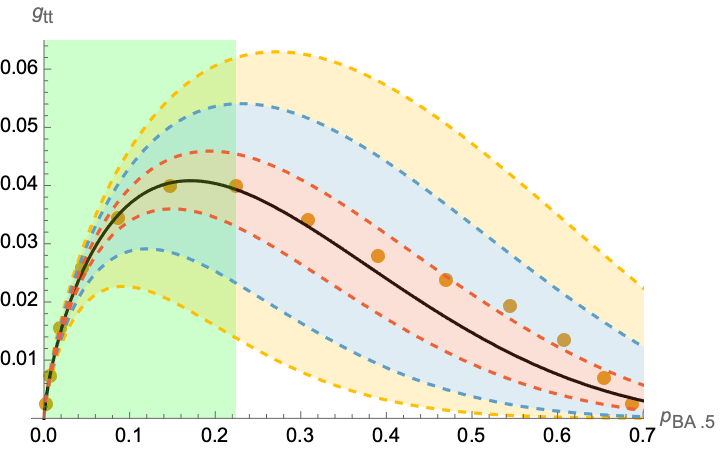}\hspace{0.25cm}\includegraphics[width=5.5cm]{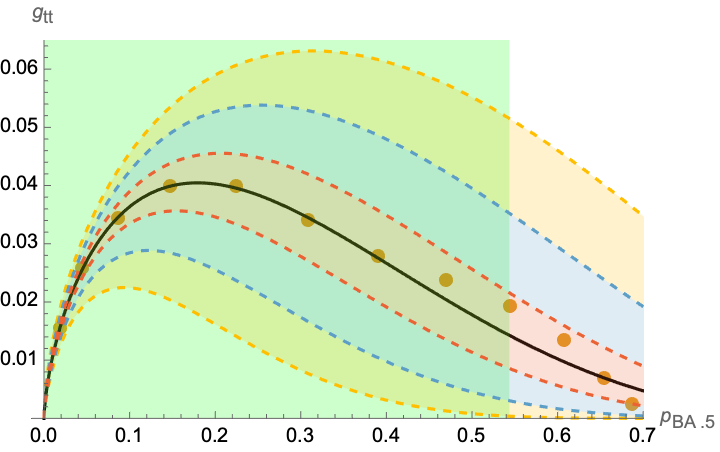}
\end{center}
\caption{Fit of the auxiliary metric (\ref{AuxiliaryMetric}) (solide black line) for the weeks $n_0=119$ (left), $n_0=121$ (middle) and $n_0=123$ (right), compared to the actual data (orange points). The green shaded region represents the data points that have been used for the fit, while the red, blue and yellow shaded regions represent fits based on an assumed error of $\pm10\%$, $\pm 25\%$ and $\pm 40\%$ of the data points, as explained in the text. }
\label{Fig:FrancePredictionMetrics}
\end{figure}

\begin{figure}[p]
\begin{center}
\includegraphics[width=5.5cm]{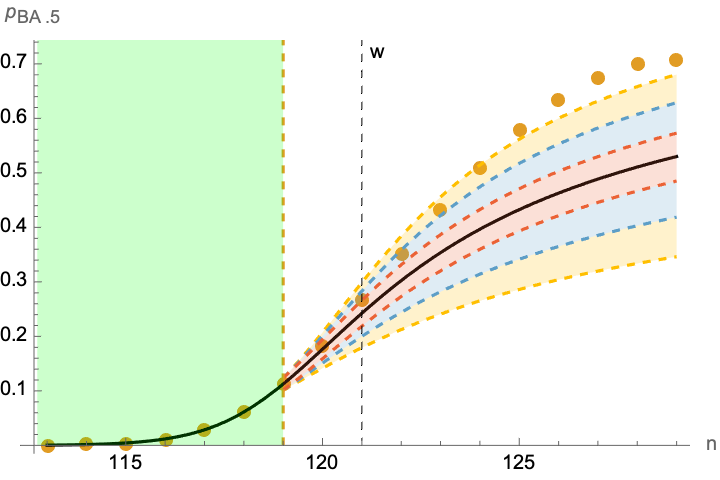}\hspace{0.25cm}\includegraphics[width=5.5cm]{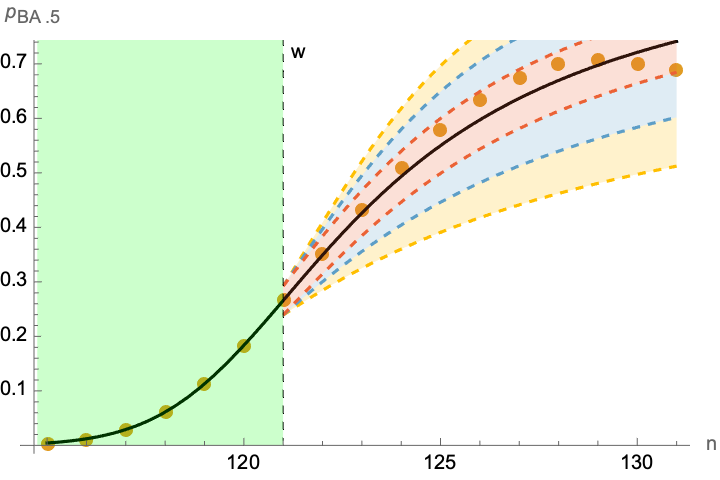}\hspace{0.25cm}\includegraphics[width=5.5cm]{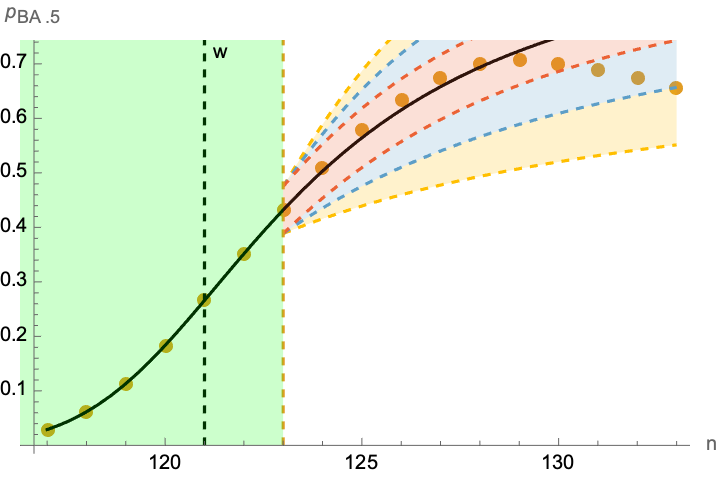}
\end{center}
\caption{Comparison of predicted (black curve) versus actual (orange points) probabilities for the sequence BA.5 for the weeks $n_0=119$ (left), $n_0=121$ (middle) and $n_0=123$ (right) The dashed line labelled $w$ denotes week 121 (\emph{i.e.} the week before the cluster containing BA.5 becomes dominant). The yellow, blue and red bands represent predictions based on the respective fits of the metric in Figure~\ref{Fig:FrancePredictionMetrics} assuming $\pm10\%$, $\pm 25\%$ and $\pm 40\%$ error of the input data. Finally, a $10\%$ error on the initial data in week $n_0$ has been assumed in all three plots.}
\label{Fig:FrancePrediction}
\end{figure}

Once a fast-growing variant has been identified, which is expected to become dominant, we can use the formalism developed in \cite{Filoche:2024xka} (and reviewed in Section~\ref{Sect:TheorySpreadDangerous}) to describe and predict its development. To illustrate the power of this approach, we shall apply (\ref{FlowEquation}) to predict the evolution of BA.5. To this end, we shall assume that the probabilities $\mathfrak{p}_{\text{BA.5}}$ are known up to a certain week $n_0$ and try to predict the evolution $\mathfrak{p}_{\text{BA.5}}(n)$ for $n>n_0$. From these known data, we calculate an auxiliary metric 
\begin{align}
\mathfrak{g}_{tt}(n)=4\left[\left(\sqrt{\mathfrak{p}_{\text{BA.5}}(n+1)}-\sqrt{\mathfrak{p}_{\text{BA.5}}(n)}\right)^2+\left(\sqrt{1-\mathfrak{p}_{\text{BA.5}}(n+1)}-\sqrt{1-\mathfrak{p}_{\text{BA.5}}(n)}\right)^2\right]\,,\label{AuxiliaryMetric}
\end{align}
for all weeks $n<n_0$, which we fit as a function of $\mathfrak{p}_{\text{BA.5}}$ using (\ref{ApproxMetric}) (with $p_1=0$). The result for different values of $n_0$ is shown in Figure~\ref{Fig:FrancePredictionMetrics}: the orange dots represent the data for the entire time period, while the points in the green shaded region are the ones considered known, which have been used for the fit, which is represented by the black solid line. Allowing for an error of $10-40\%$ of the input data has an impact on the fit, as is represented by the coloured bands in Figure~\ref{Fig:FrancePredictionMetrics}.

Using the metric (\ref{AuxiliaryMetric}) as input for (\ref{FlowEquation}) (with initial condition $\mathfrak{p}_{\text{BA.5}}(n_0)$ with a $\pm10\%$ error) leads to predictions of $\mathfrak{p}_{\text{BA.5}}(n)$ for $n>n_0$, which are shown in Figure~\ref{Fig:FrancePrediction}. This plot also shows the uncertainties, due to the errors in fitting the metric in Figure~\ref{Fig:FrancePredictionMetrics}. These Figures demonstrate that the formalism is capable of predicting the time-evolution of the strongly-growing variant BA.5 over several weeks.

We hasten to add that the plots in Figure~\ref{Fig:FrancePrediction} were based on data that were subject to a

\begin{wrapfigure}{r}{0.46\textwidth}
\begin{center}
\vspace{-0.5cm}
\includegraphics[width=7.5cm]{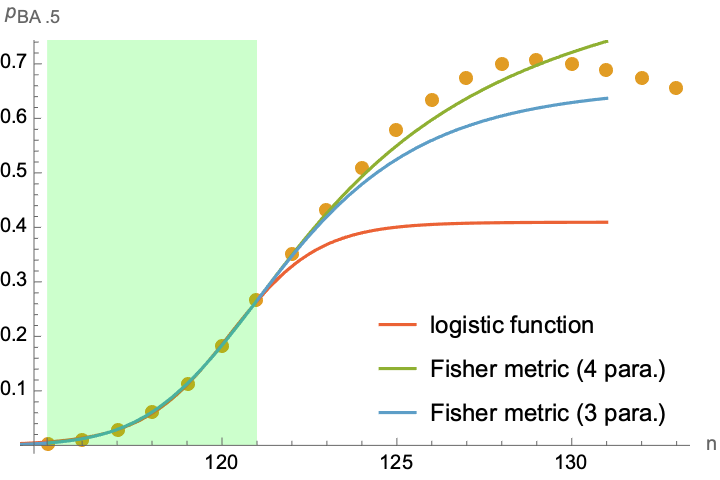}
\end{center}
\vspace{-0.5cm}
\caption{Comparison of predictions to a fit with a logistic function. The dots represent the actual data, while the green and blue curve are predictions based on the Fisher information metric (which is fitted with 4 or 3 parameters respectively) and the red curve is a fit of the data (in the green region) with a logistic function.}
\label{Fig:ComparisonLogistic}
\vspace{-0.5cm}
\end{wrapfigure}

\noindent
Gaussian filtering with $\sigma=4$ weeks, thus smearing the information over a period of time, such that the 'cutoff' up to which information has been used to make the prediction, is not completely sharp. We shall eliminate this problem in Section~\ref{Sect:FrancLongTerm}, when discussing the entire pandemic in France without any filtering. However, at the same time we also point out that the predictions shown in Figure~\ref{Fig:FrancePrediction} capture the data very well and, indeed, better than a simple fitting with a simple logistic function $\mathfrak{p}_{\text{BA.5}(n)}=\frac{A}{1+e^{-\lambda(n-c)}}$ (with $A$, $\lambda$ and $c$ the fit parameters): Figure~\ref{Fig:ComparisonLogistic} compares such a fit to the results obtained using the description in terms of the Fisher metric. All methods only use data points located in the green shaded region. Furthermore, when fitting the Fisher information metric, we use (\ref{ApproxMetric}) with either $(a,b,c,p_2)$ (green curve, with $p_1=0$) or  $(a,b,c)$ (blue curve, with $p_1=0$ and $p_2=0.66$) as free parameters. The latter case is therefore comparable to a direct fit of the data with a logistic function with 3 free parameters. As is evident from the plot, even in this simplified approach, the description in terms of the Fisher information metric represents the data much better than a simple logistic function.


\subsubsection{Comparison to other Forms of Clustering}
Before closing this Section and continuing with a more systematic analysis of the SARS-CoV-2 pandemic in France, we briefly comment on the relation of the clustering proposed in Section~\ref{Sect:SeqMethodology} to other forms of clustering the genomic data, specifically the approach advocated in \cite{MLvariants}. Indeed, in this work, the genomic data of the spike-protein sequences of SARS-CoV-2 (the same data used in Section~\ref{Sect:ClusteringShowcaseRel}) have been clustered with an agglomerative algorithm, based on their Levenshtein distance~\cite{Levenshtein1965BinaryCC}. The details on the algorithm can be found in Appendix~\ref{Sect:GeneticClustering}. Furthermore, based on their dominant sequence (\emph{i.e.} the sequence with the highest probability inside each cluster), clusters of consecutive weeks can be linked to chains, while chains persisting for a sufficiently long time are identified with (emergent) variants. It has been demonstrated in \cite{MLvariants} that this method is a very efficient tool of detecting dangerous (\emph{i.e.} fast growing) variants early on. Compared to the method proposed in Section~\ref{Sect:SeqMethodology}, the clustering in \cite{MLvariants} works very differently, since it uses genomic information for determining the clusters: the approach advocated in the present work is complementary, in the sense that the clustering is based on the growth of (the information of) each sequence, which is used as \emph{input} to detect genomic (dis)similarities among different variants.

Figure~\ref{Fig:OtherClusterVI} provides a comparison between both methods: the left panel compares the Fisher information metric calculated from the pruned data (using a Gaussian filtering with $\sigma=4$), with the metrics obtained from the clustering based on the change of information (orange curve) and clustering based on the Levenshtein distance (green curve) following \cite{MLvariants}. The former method uses the k-means algorithm with a goal of at most 6 clusters, while the latter is also set to terminate with 6 clusters. Under these conditions, in most weeks $g_{tt}^{(\text{KMeans},6)}>g_{tt}^{(\text{Lev},6)}$, suggesting that the clustering based on information is closer to a sufficient statistics.

\begin{figure}[htbp]
\begin{center}
\includegraphics[width=7.5cm]{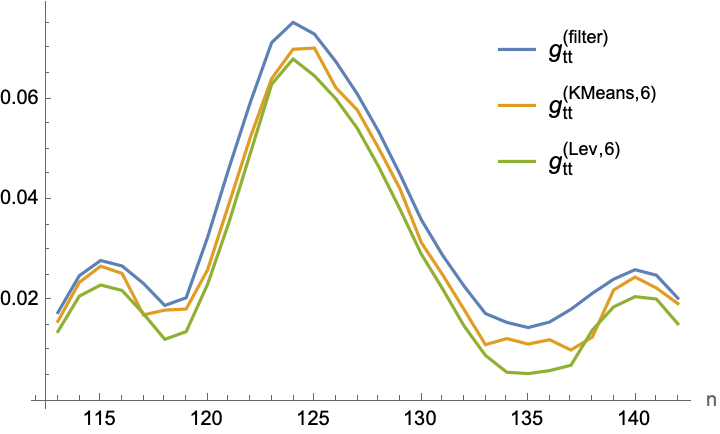}\hspace{1cm}\includegraphics[width=7.5cm]{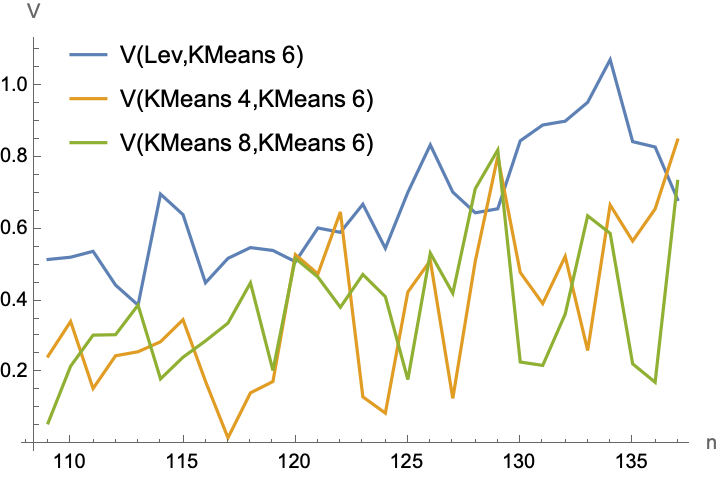}
\end{center}
\caption{Left panel: Comparison of the Fisher information metric from clustering the time-evolution of the information and the Levenshtein distance of the spike-protein of the virus, following a similar approach as in \cite{MLvariants}. Right panel: Variation of information between clustering based on information (KM, 6 clusters) and based on the Levenshtein distance, both methods use 6 clusters. }
\label{Fig:OtherClusterVI}
\end{figure}

The right panel of Figure~\ref{Fig:OtherClusterVI} shows the variation of information (see (\ref{DefVI}) for the definition) for different  clustering methods (as a function of time). The blue curve shows $\mathcal{V}(\mathbb{W}_{\text{Lev,6}},\mathbb{W}_{\text{KMeans},6})$, while (for comparison) the orange curve shows $\mathcal{V}(\mathbb{W}_{\text{KMeans},4},\mathbb{W}_{\text{KMeans},6})$ and the green curve $\mathcal{V}(\mathbb{W}_{\text{KMeans},8},\mathbb{W}_{\text{KMeans},6})$. Here $\mathbb{W}_{\text{Lev},\ell}$ stands for a clustering based on the Levenshtein distance (as explained above) with $\ell$ clusters, while $\mathbb{W}_{\text{KMeans},\ell}$ denotes a clustering according to the differences in the sequence-informations, using the k-means-algorithm with $\ell$ clusters. Figure~\ref{Fig:OtherClusterVI} suggests that the difference between a clustering based on the genomic differences and the sequence-information is larger than the difference between two clusterings of the latter type, but with a different graining (\emph{i.e.} a different number of clusters). However, the fact that $\mathbb{W}_{\text{Lev},6}$ is still fairly close to the Fisher metric obtained without clustering suggests that this method is still correctly capturing important aspects of the time-evolution of the genomic-variants: this is in agreement with the fact that clusterings of the form $\mathbb{W}_{\text{Lev},6}$ have been successfully used for detecting and tracking dangerous variants of SARS-CoV-2 \cite{MLvariants}.

\section{Time Evolution in France}\label{Sect:FrancLongTerm}
In this Section we extend the discussion of the previous Subsection~\ref{Sect:CaseStudyFrance} to the time-period of 27/01/2020--23/10/2023. While we still prune\footnote{This means, we eliminate incomplete sequences and sequences which only which appear only in a very short period of time and/or accumulate a too small
fraction of the sequences.} the data as explained in Section~\ref{Sect:DataTreatment}, we do not apply any Gaussian filtering beforehand, thus the information contained in the sequencing data is no longer smeared in time, 
\subsection{Overview of the Time Evolution}

With the pruned and un-filtered data, we calculate the Fisher information metric (\ref{Fisher1DMetric}), which is plotted in Figure~\ref{Fig:FranceMetricFull} (blue curve). We next apply a k-means algorithm to cluster the time-derivative of the information for each sequence into up to 6 clusters.\footnote{For all $n>50$, the number of clusters is always 6. Only for some weeks $n\leq 50$, in which few sequences are available, the algorithm has grouped the time derivatives of the informations into less than 6 clusters.} The metric associated with the cluster probabilities is shown in Figure~\ref{Fig:FranceMetricFull}. Compared to Figure~\ref{Fig:FranceClustersMod}, using unfiltered data leads to more fluctuations. Nevertheless, the difference $\Delta g_{tt}$ in (\ref{SufficientStatisticCondRel}), is small indicating that this clustering indeed captures the time evolution of the system well.

\begin{figure}[h!]
\begin{center}
\includegraphics[width=7.5cm]{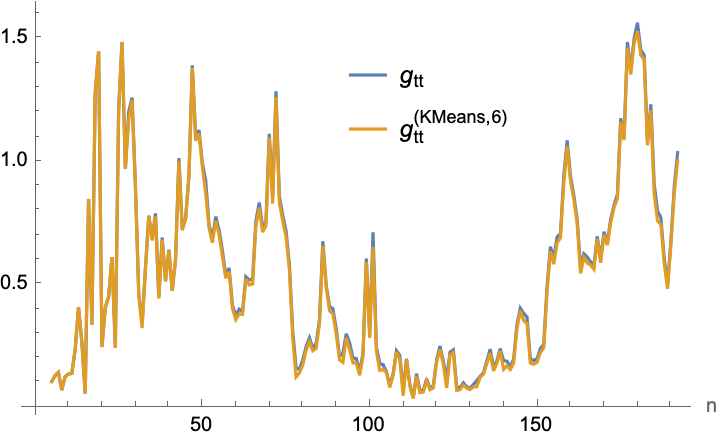}\hspace{1cm}\includegraphics[width=7.5cm]{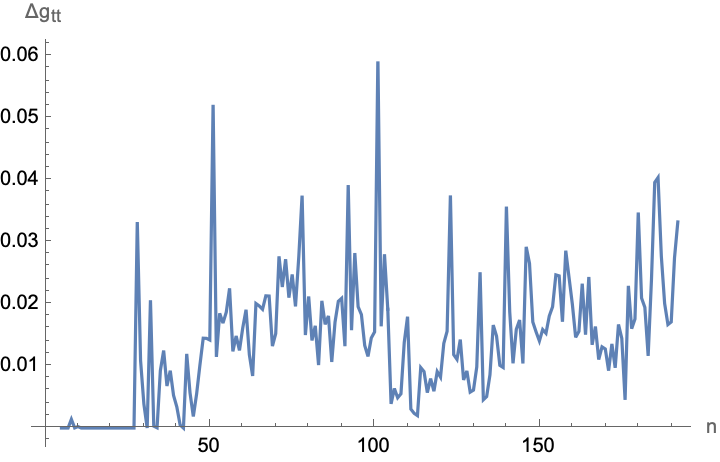}
\end{center}
\caption{Fisher information metric for the evolution of the spike protein of SARS-CoV-2 in France in the period of 27/01/2020--23/10/2023. The blue curve in the left panel represents the metric for the entirety of the pruned data, while the orange curve represents the cluster metric. For better comparison, the right panel shows the difference (\ref{DifferenceMetric}) between these two metrics.
}
\label{Fig:FranceMetricFull}
\end{figure}

In Figure~\ref{Fig:FranceOverviewProbs} we have shown the time evolution of the entire pandemic (more concretely from weeks 10 to 200) by plotting the probabilities of all sequences that in some week reach a probability of more than~$0.1$. The sequences are either denoted through the Pango nomenclature (if possible) \cite{Pango} or through an internal labelling system. All sequences plotted in Figure~\ref{Fig:FranceOverviewProbs} are also characterised in Appendix~\ref{App:SeqMutations} through their mutations relative to the original Wuhan sequence. The latter is designated seq 1 in our internal convention, which in fact does not appear in Figure~\ref{Fig:FranceOverviewProbs}, since by week 10, the dominant variant is already seq 2, which carries the mutation D614G.

\begin{figure}[htbp]
\begin{center}
\includegraphics[width=17cm]{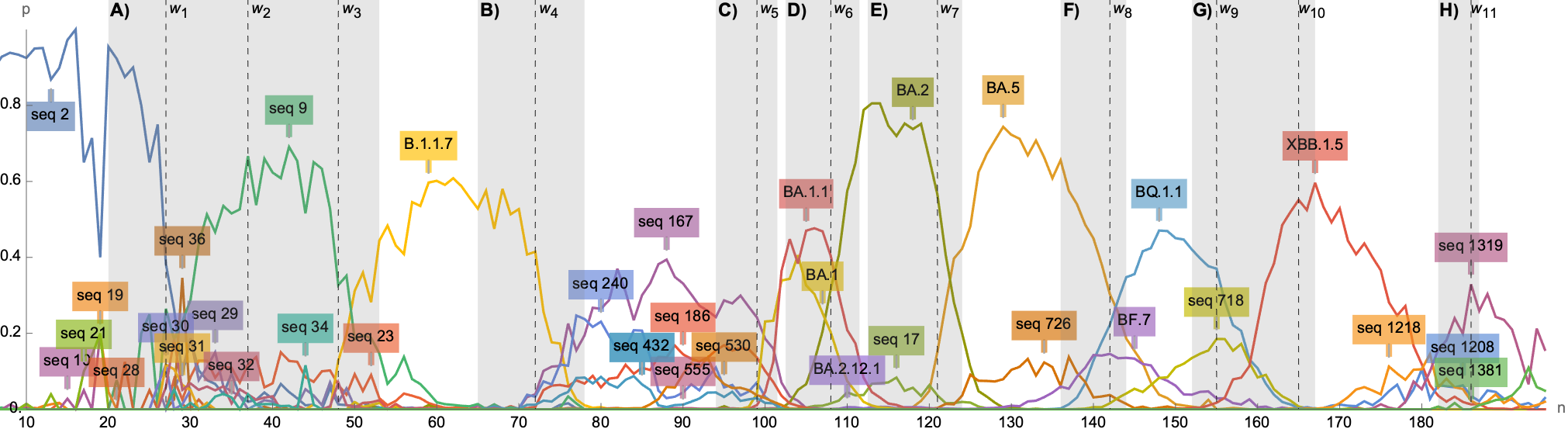}\end{center}
\caption{Probabilities as function of time of all sequences that achieve a maximum larger than~$0.1$. Unless a Pango designation could be obtained, the sequences are labelled according to an internal numbering scheme. All sequences are characterised in Appendix~\ref{App:SeqMutations} through their mutations relative to the original Wuhan variant.}
\label{Fig:FranceOverviewProbs}
\end{figure}

Based on the clustering (with the metric shown in Figure~\ref{Fig:FranceMetricFull}) we can calculate for each week

\begin{wrapfigure}{l}{0.68\textwidth}
\begin{center}
\vspace{-0.5cm}
\includegraphics[width=11.5cm]{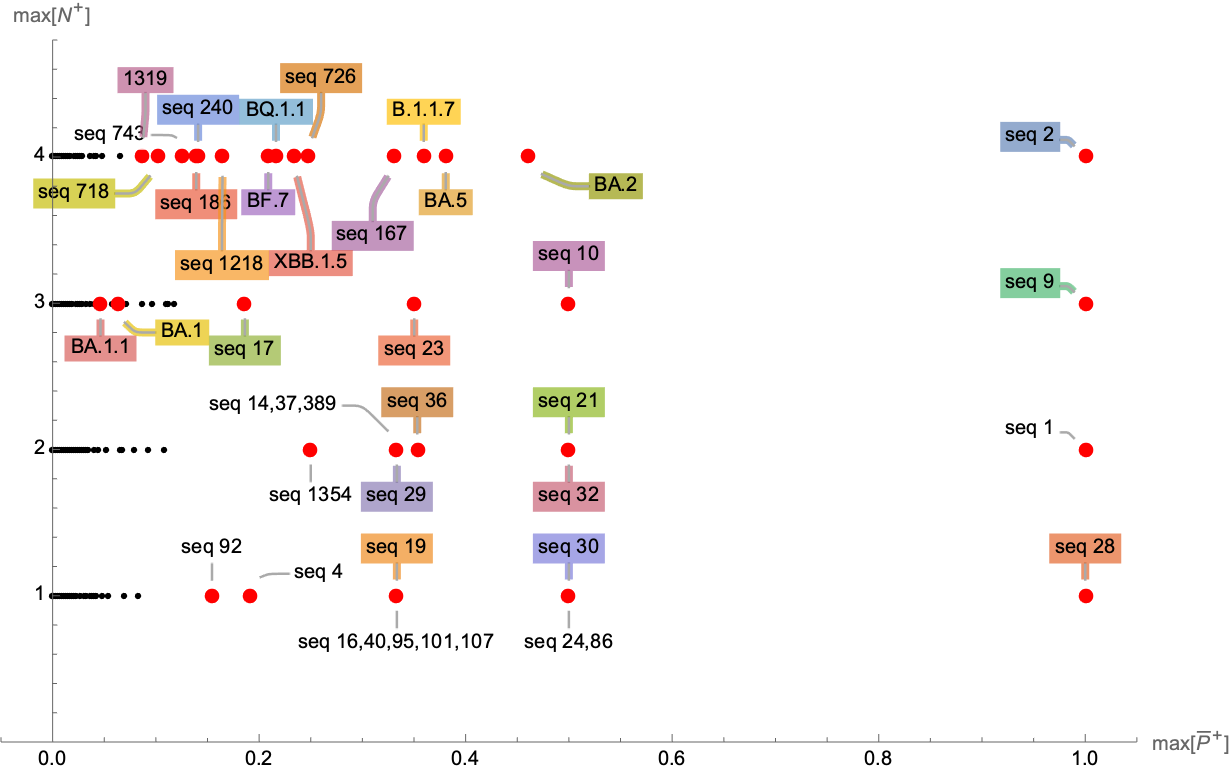}
\end{center}
\vspace{-0.5cm}
\caption{All 1474 sequences as a function of the maximum of $\overline{P}^+_i$ in (\ref{AverageProbPositive}) and $N^+$. The largest values are labelled with the name of the sequence following the colouring scheme of Figure~\ref{Fig:FranceOverviewProbs}.}
\label{Fig:FranceMaxIndicators}
\end{wrapfigure}

\noindent
$n$ the probability $\overline{P}^+_i(n)$ of the sequence $i$ to be part of a cluster with diminishing information as well as the number $N^+(n)$ of such clusters for which $\mathfrak{h}(i,a,n)$ in (\ref{ProbabilityGenetic}) is non-zero (\emph{i.e.} the number of clusters with shrinking information for which sequence $i$ has non-vanishing probability to be part of). For certain sequences we shall discuss these quantities as a function of $n$ in more detail below. To provide an overview, the maxima of these values are plotted in Figure~\ref{Fig:FranceMaxIndicators}, with the largest values labelled with the respective sequences, following the same convention as in Figure~\ref{Fig:FranceOverviewProbs}. This plot shows that, sequences, which reach a large probability at some point in time, also reach the large maximal values of $\overline{P}^+_i(n)$ and $N^+(n)$. As already argued in Section~\ref{Sect:DangerousVariants}, these quantities are therefore useful indicators to judge the capability of a sequence to spread throughout the population.

\subsection{Specific Time Periods}
In the following we shall discuss certain time-periods in more detail, which are indicated by the regions shaded in grey in Figure~\ref{Fig:FranceOverviewProbs}: they correspond to the appearance of important variants, which become (nearly) dominant. A summary of our major conclusions drawn from this analysis as well as more general remarks can be found in Section~\ref{Sect:Summary} at the end.

\subsubsection{Time Period A): Appearance of B.1.1.7 (Alpha)}
We first focus on the time frame of weeks 20-53 (18/05/2020 -- 10/01/2021), labelled as time

\begin{wrapfigure}{r}{0.55\textwidth}
\vspace{-0.5cm}
\begin{center}
\includegraphics[width=9cm]{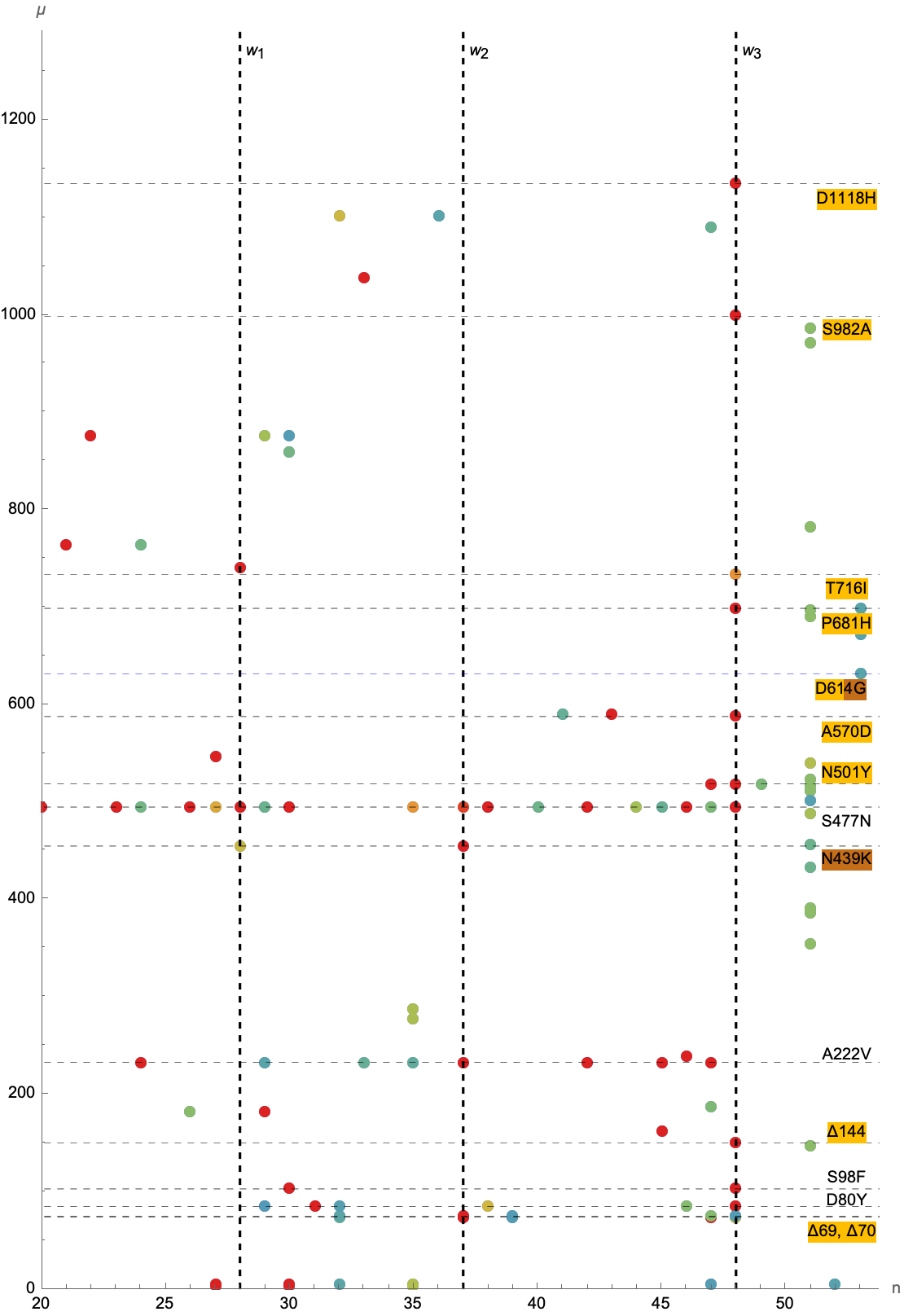}
\end{center}
\vspace{-0.5cm}
\caption{Evolution of the divergences (\ref{AlphaDiv}) for all protein sequence positions $\mu$ during the time period {\bf A)}. $w_{1,2,3}$ indicate maxima of (\ref{TotalDivergencesWeek}) as in Figure~\ref{Fig:AlphaSummedDivergences}.}
\label{Fig:FullTimeAlphaFrance}
\vspace{-0.5cm}
\end{wrapfigure}

\noindent
period {\bf{A)}} in Figure~\ref{Fig:FranceOverviewProbs}. Figure~\ref{Fig:FullTimeAlphaFrance} shows the divergences (\ref{AlphaDiv}) during this time period, where warmer colours represent larger values of $\mathcal{D}$. For better orientation, the weeks $w_{1,2,3}$ represent certain maxima of the summed divergences (\ref{TotalDivergencesWeek}), which are plotted in Figure~\ref{Fig:AlphaSummedDivergences}. In particular, the less pronounced maximum $w_1$ coincides with the appearance of seq 36, while $w_3$ marks the week in which B.1.1.7 becomes the dominant sequence in France (such that the dominant cluster changes in nature). Furthermore, the horizontal dashed lines in Figure~\ref{Fig:FullTimeAlphaFrance} indicate the positions where very large divergences occur and correlate them with mutations (which are labelled relative to the original Wuhan sequence): mutations shaded in yellow are carried by B.1.1.7, while those shaded in brown are carried by seq 36. 

Both Figure~\ref{Fig:FullTimeAlphaFrance} and \ref{Fig:AlphaSummedDivergences} indicate that the largest divergences occur related to the appearance of B.1.1.7 (\emph{i.e.} the Alpha variant of SARS-CoV-2): indeed, while there are maxima of the summed divergences (\emph{e.g.} $w_{1,2}$ in Figure~\ref{Fig:AlphaSummedDivergences}) and while some of the mutations carried by B.1.1.7 are present before $w_3$ (and thus in sequences before the appearance of B.1.1.7) in Figure~\ref{Fig:FullTimeAlphaFrance}, many of them appear only at $w_3$ (indicated through red dots). Furthermore, immediately after the point in time, when B.1.1.7 has become part of the dominant cluster (\emph{i.e.} the largest cluster in terms of summed probabilities) in France, divergences in these positions are very~small or zero. This

\begin{wrapfigure}{l}{0.5\textwidth}
\vspace{-0.5cm}
\begin{center}
\includegraphics[width=7.5cm]{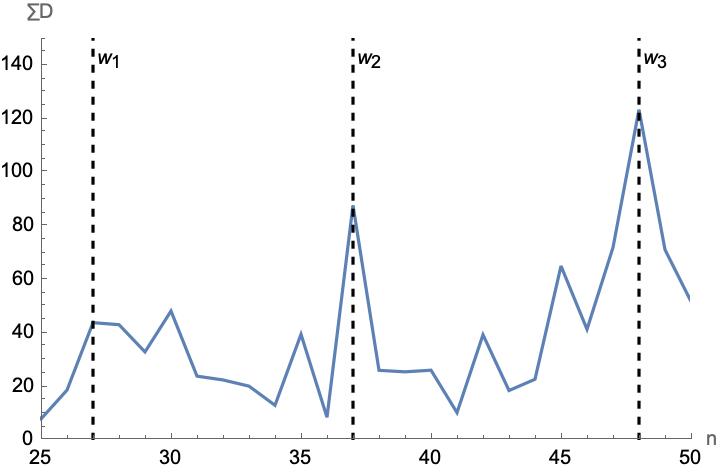}
\end{center}
\vspace{-0.5cm}
\caption{Evolution of the summed divergences  (\ref{TotalDivergencesWeek}) during the time period {\bf A)}. The weeks $w_{1,2,3}$ are the same as in Figure~\ref{Fig:FullTimeAlphaFrance} and mark certain maxima for better orientation.}
\label{Fig:AlphaSummedDivergences}
\end{wrapfigure}

\noindent
indicates that the nature of the dominant cluster has indeed changed. These findings reflect the epidemiological development during this period: the first case of B.1.1.7 was detected in September 2020 (week 34 in our counting) in Kent (England) and the first sample in France was detected in 26/12/2020 (week 48 in our counting) \cite{FranceAlpha}. The first sequence of B.1.1.7 in the GISAID data analysed in our study occurs also in week 48, after which the variant becomes dominant very quickly, as can be seen from Figure~\ref{Fig:FullTimeAlphaFrance}. The fact that B.1.1.7 (very quickly) has become the dominant sequence is also reflected by its probability to be part of a cluster of shrinking information: in the rightmost panel of Figure~\ref{Fig:IndicatorsA}, we have plotted the average probability as defined in (\ref{AverageProbPositive}) (with the variance indicate by the error bars), while the numbers associated with each data point indicate the number $N^+$ of clusters with diminishing information for which the sequence has non-vanishing probability to be a member. This plot indicates that already in week $w_3$ this average probability is around 0.15 and there are 2 growing clusters (\emph{i.e.} also one additional cluster besides the one into which the clustering algorithm has assigned B.1.1.7 based on the derivative of its information) for which B.1.1.7 has non-zero probability to be a member of. The remaining two panels in Figure~\ref{Fig:IndicatorsA} show the same probabilities for other important sequences in this time-period, namely seq 36 (leftmost) and 23, 29 (middle) panel. All of these sequences are also plotted in Figure~\ref{Fig:FranceMaxIndicators} and are thus among those which achieve maximal probabilities of larger than 0.1.

\begin{figure}[htbp]
\begin{center}
\includegraphics[width=5.5cm]{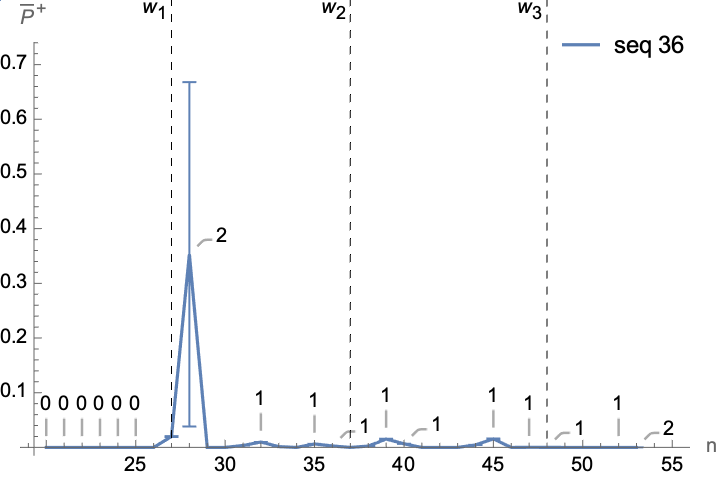}\hspace{0.25cm}\includegraphics[width=5.5cm]{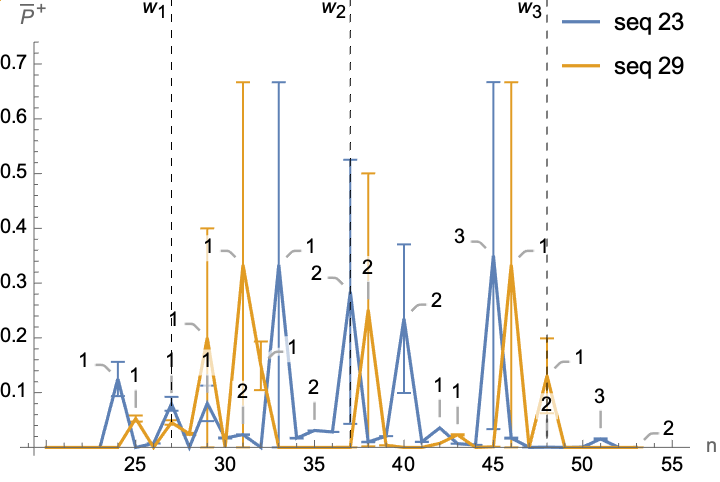}\hspace{0.25cm}\includegraphics[width=5.5cm]{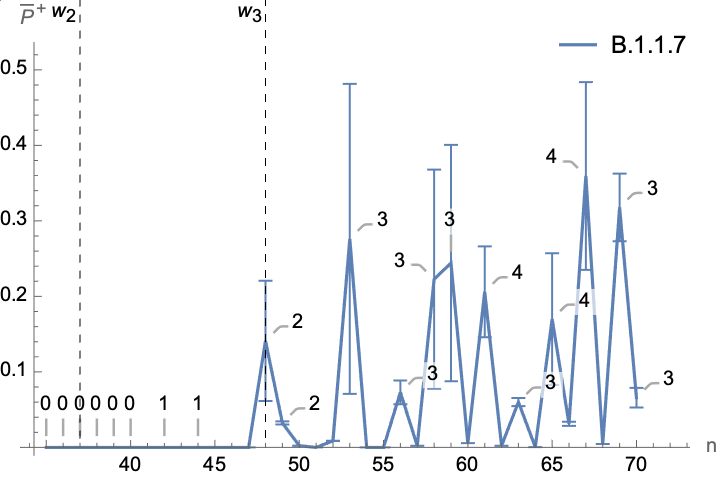}
\end{center}
\caption{Average probabilities (\ref{AverageProbPositive}) for certain sequences to be part of a cluster with shrinking information as a function of $n$, as well as the number of such clusters for which the same sequence has a non-vanishing probability to be a part of. The left-most panel shows seq 36, the middle panel seq 23 and 29 and the rightmost panel B.1.1.7.}
\label{Fig:IndicatorsA}
\end{figure}

Since B.1.1.7 is clearly the sequence which becomes dominant during time period {\bf A)} (and stays dominant for a certain period of time afterwards), we can use (\ref{FlowEquation}) to predict the time

\begin{wrapfigure}{r}{0.5\textwidth}
\vspace{-0.5cm}
\begin{center}
\includegraphics[width=7.5cm]{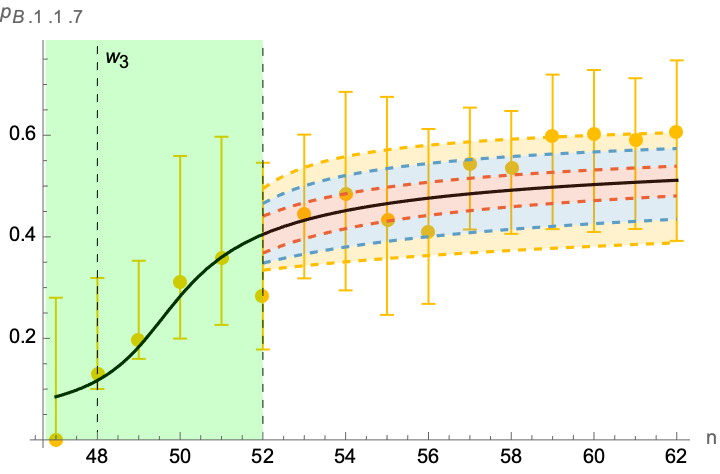}
\end{center}
\vspace{-0.5cm}
\caption{Prediction of the probability $\mathfrak{p}_{\text{B.1.1.7}}$: the data points represent the actual values of the probability of B.1.1.7 along with uncertainties based on the number of discarded incomplete sequences, while the black line shows a solution of (\ref{FlowEquation}) using the data in the green shaded region as input. The coloured bands show deviations of the prediction taking into account uncertainties of the input data.}
\label{Fig:PredictionTimeAlphaFrance}
\end{wrapfigure}

\noindent
evolution of its probability $p_{\text{B.1.1.7}}$ of the variant B.1.1.7. Due to the high uncertainty of the data points because of the small number of available (complete) sequences per week during period {\bf{A)}} (see Figure~\ref{Fig:FranceTotalOverview}), reliable predictions are only possible after a certain number of weeks after the first appearance of the variant. Furthermore, since B.1.1.7 rises to large probabilities very quickly, this implies that a prediction\footnote{Predictions with higher uncertainties are possible already at earlier weeks.} is only available once the variant has already reached a probability of around 0.4, which is shown in Figure~\ref{Fig:PredictionTimeAlphaFrance}: as input, we have used all data points in the green shaded region (\emph{i.e.} up to week 52), from which we have calculated a fit for an auxiliary metric $\mathfrak{g}_{tt}$ in the same manner as explained in (\ref{AuxiliaryMetric}).We remark that the numerical fitting procedure yields better results when using a function of the form $\mathfrak{g}_{tt}(p)=a\, p^b\, (0.66-p)^c$, with $a,b,c$ free parameters (\emph{i.e.} with $p_2=0.66$ fixed in (\ref{ApproxMetric})). We shall use this function throughout in the following. The solution of (\ref{FlowEquation}) based on this metric $\mathfrak{g}_{tt}$ is shown as the black line in Figure~\ref{Fig:PredictionTimeAlphaFrance} (with initial values imposed at $n=51$). As in Section~\ref{Sect:DangerousVariants}, the coloured bands show deviations of the solution, assuming $\pm10\%$, $\pm 25\%$ and $\pm 40\%$ error of the initial input data. Finally, for comparison, we have added uncertainties (represented as error bars) for the (measured) probabilities $\mathfrak{p}_{\text{B.1.1.7}}$ (represented by the yellow dots in Figure~\ref{Fig:PredictionTimeAlphaFrance}) based on the relative number of discarded incomplete sequences in each week $n$: the upper end of this error assumes that all incomplete sequences that have been removed from the dataset (see left panel of Figure~\ref{Fig:FranceTotalOverview}) have been of the type B.1.1.7, while the lower end assumes that none of the removed sequences has been of this variant. Taking these uncertainties into account, the prediction in Figure~\ref{Fig:PredictionTimeAlphaFrance} represents the actual data very well. This result therefore presents a further validation of the universal model proposed in \cite{Filoche:2024xka} for a real epidemiological system.

\subsubsection{Time Period B): Appearance of B.1.617 (Delta)}
We next consider time period {\bf{B)}} in Figure~\ref{Fig:FranceOverviewProbs}, namely the weeks 65-78 (19/04/2021--19/07/2021). Figure~\ref{Fig:FullTimeDeltaFrance} shows the divergences (\ref{AlphaDiv}) during this time period, where warmer colours represent larger values of $\mathcal{D}$. For better orientation, the week $w_{4}=72$ represents the maximum of the summed divergences (\ref{TotalDivergencesWeek}), which is plotted in Figure~\ref{Fig:DeltaSummedDivergences}. This week (and the following week 73) coincide with the appearance of a number of variants that are part of the B.1.617 lineage

\newpage
\begin{wrapfigure}{r}{0.55\textwidth}
\vspace{-0.5cm}
\begin{center}
\includegraphics[width=9cm]{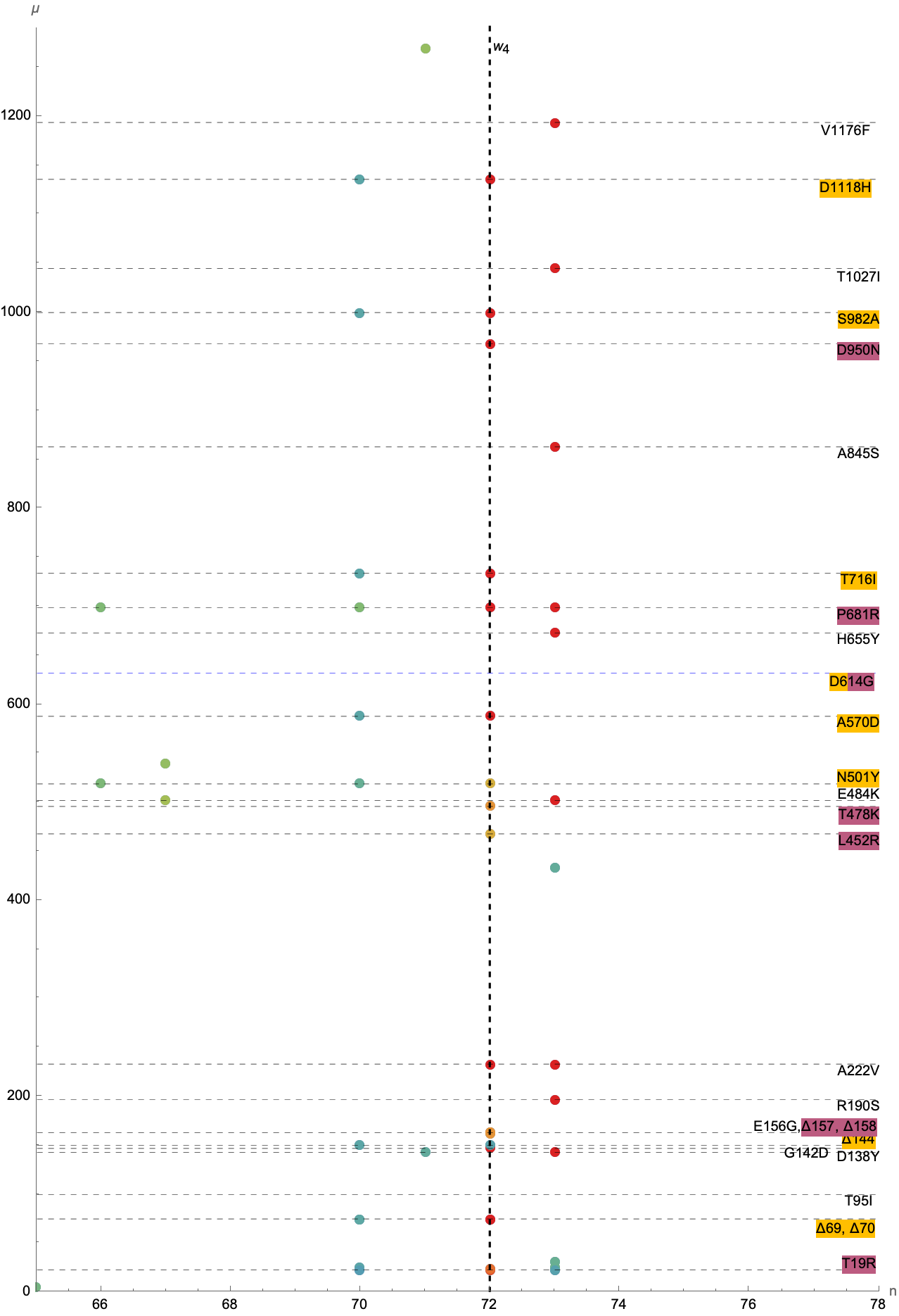}
\end{center}
\vspace{-0.5cm}
\caption{Evolution of the divergences (\ref{AlphaDiv}) for sequence positions $\mu$ during the time period {\bf B)}.}
\label{Fig:FullTimeDeltaFrance}
\vspace{-0.1cm}
\end{wrapfigure}

\noindent
(also called Delta),  notably seq 186, seq 240, seq 167, seq 432, seq 555 and seq 530, which are also plotted in Figure~\ref{Fig:FranceOverviewProbs}. The horizontal dashed lines in Figure~\ref{Fig:FullTimeDeltaFrance} indicate the positions on the spike protein sequence where very large divergences occur and correlate them with mutations (which are labelled relative to the original Wuhan sequence): mutations shaded in yellow are carried by B.1.1.7, while those shaded in purple are carried by B.1.617.

As in the case of Alpha in the previous Subsubsection, Figures~\ref{Fig:FullTimeDeltaFrance} and \ref{Fig:DeltaSummedDivergences} suggest that the largest divergences occur related to the appearance of B.1.617: indeed, the summed divergences in Figure~\ref{Fig:DeltaSummedDivergences} show a very pronounced maximum in week $w_4$ and Figure~\ref{Fig:FullTimeDeltaFrance} shows strong divergences at the positions of the mutations carried by B.1.617. After week $w_4$, the sequences in the B.1.617 lineage have become dominant (\emph{i.e.} they are part of the largest cluster) and further divergences of the type (\ref{AlphaDiv}) in these positions are very small or in fact zero: this indicates that~indeed the nature of the largest cluster has changed (since Figure~\ref{Fig:FullTimeDeltaFrance} only shows divergences between clusters with shrinking information and the dominant cluster, as explained in Section~\ref{Sect:SeqMethodology}). The fact that sequences of the B.1.617 lin-

\begin{wrapfigure}{l}{0.5\textwidth}
\vspace{-0.5cm}
\begin{center}
\includegraphics[width=7.5cm]{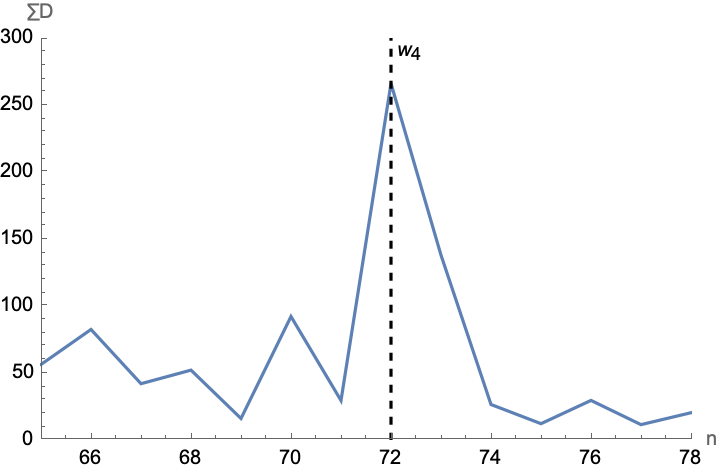}
\end{center}
\vspace{-0.5cm}
\caption{Evolution of the summed divergences (\ref{TotalDivergencesWeek}) during the time period {\bf B)}. The week $w_{4}$ is the same as in Figure~\ref{Fig:FullTimeDeltaFrance}.}
\label{Fig:DeltaSummedDivergences}
\vspace{-0.5cm}
\end{wrapfigure}

\noindent
eage become dominant during the time period {\bf{B)}} is also reflected by their probabilities to be part of a cluster of diminishing information. This is shown by three examples in Figure~\ref{Fig:IndicatorsB}, namely seq 186, seq 240 and seq 167, which show the largest average probability as defined in (\ref{AverageProbPositive}) (with the variance indicate by the error bars). As in previous cases Figure~\ref{Fig:IndicatorsB} also shows the number $N^+$ of clusters with shrinking information for which the sequence has non-vanishing probability to be a member, for each data point: already before $w_4$, this number is $>1$ for all sequences in Figure~\ref{Fig:IndicatorsB}, \emph{i.e.} there is at least one additional cluster besides the one into which the clustering algorithm has assigned them to based on the derivative of its information, for which they have non-zero probability to be a member of.

\begin{figure}[htbp]
\begin{center}
\includegraphics[width=5.5cm]{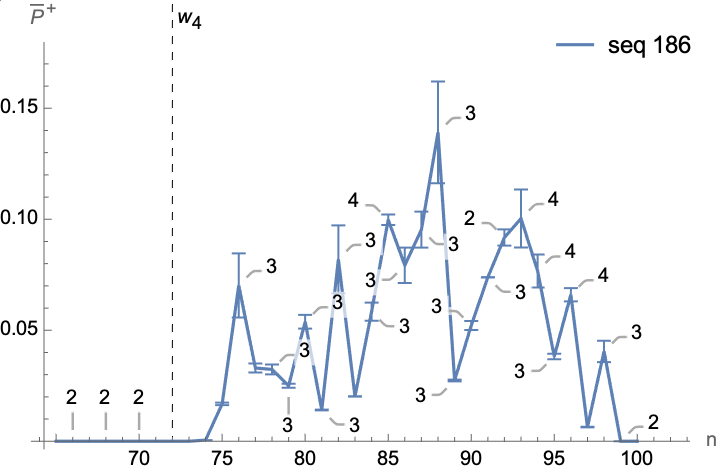}\hspace{0.25cm}\includegraphics[width=5.5cm]{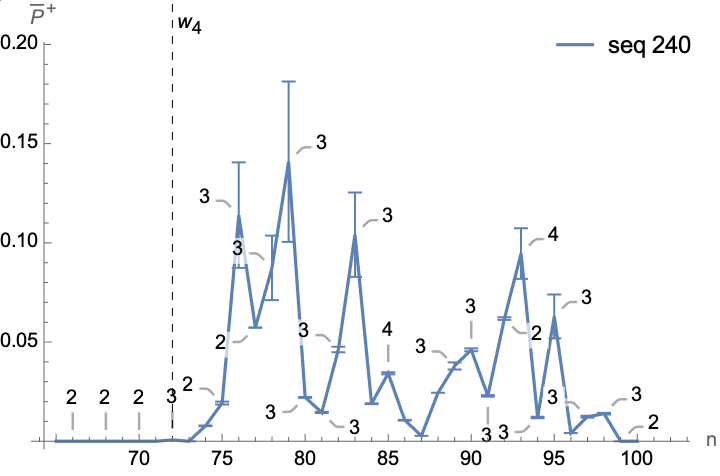}\hspace{0.25cm}\includegraphics[width=5.5cm]{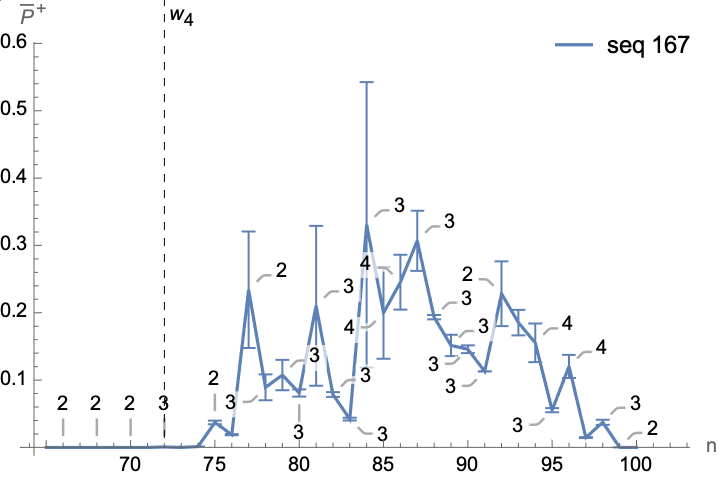}
\end{center}
\caption{Average probabilities (\ref{AverageProbPositive}) for certain sequences to be part of a cluster with diminishing information as a function of $n$, as well as the number of such clusters for which the same sequence has a non-vanishing probability to be a part of. The sequences seq 186 (left), seq 240 (middle) and seq 167 (right) are members of the B.1.617 (Delta) lineage.}
\label{Fig:IndicatorsB}
\end{figure}

As for the case of Alpha in the previous Subsubsection, we can predict the evolution of

\begin{wrapfigure}{r}{0.5\textwidth}
\vspace{-0.5cm}
\begin{center}
\includegraphics[width=7.5cm]{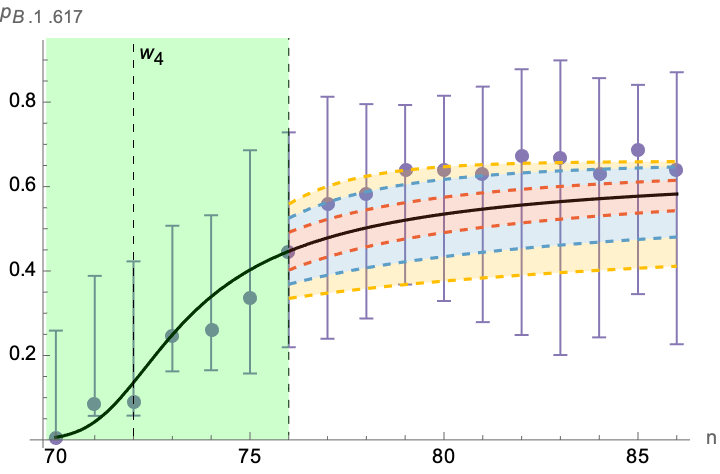}
\end{center}
\vspace{-0.5cm}
\caption{Prediction of the probability $\mathfrak{p}_{\text{B.1.617}}$: the data points represent the actual values of the probability of B.1.617 along with uncertainties based on the number of discarded incomplete sequences, while the black line shows a solution of (\ref{FlowEquation}) using the data in the green shaded region as input. The coloured bands show deviations of the prediction taking into account uncertainties of the input data.}
\label{Fig:PredictionTimeDeltaFrance}
\end{wrapfigure}

\noindent
 the probabilities for certain sequences. Since we have multiple sequences that are part of the B.1.617 lineage and that achieve high probabilities, we consider the combined probability
\begin{align}
\mathfrak{p}_{\text{B.1.617}}=\,&\mathfrak{p}_{186}+\mathfrak{p}_{240}+\mathfrak{p}_{167}\nonumber\\
&+\mathfrak{p}_{432}+\mathfrak{p}_{555}+\mathfrak{p}_{530}\,.
\end{align}
Similar to the case of the time period {\bf{A)}} (see Figure~\ref{Fig:PredictionTimeAlphaFrance}), the number of (complete) sequences per week is still rather small, at least in the beginning of the time interval, which is relevant for our prediction (see Figure~\ref{Fig:FranceTotalOverview}). As for the prediction of $\mathfrak{p}_{1.1.7}$, due to the large uncertainties inherent in the data, accurate predictions are therefore possible only at relatively late points in time.\footnote{Moving the point at which the prediction is made back in time, increases further the uncertainty.} An example for such a prediction is shown in Figure~\ref{Fig:PredictionTimeDeltaFrance}: as input, we have used all data points in the green shaded region (\emph{i.e.} up to week 76), from which we have calculated a fit for an auxiliary metric $\mathfrak{g}_{tt}$ in the same manner as explained in (\ref{AuxiliaryMetric}). The solution of the flow equation (\ref{FlowEquation}) based on this $\mathfrak{g}_{tt}$ (with initial values imposed at $n=76$) is shown as the black line in Figure~\ref{Fig:PredictionTimeDeltaFrance}. As in Section~\ref{Sect:DangerousVariants}, the coloured bands show deviations of the solution, assuming $\pm10\%$, $\pm 25\%$ and $\pm 40\%$ error of the initial input data. These need to be contrasted with the uncertainties (represented as error bars) for the (measured) probabilities $\mathfrak{p}_{\text{B.1.617}}$ (represented by the blue dots in Figure~\ref{Fig:PredictionTimeDeltaFrance}) based on the relative number of discarded incomplete sequences in each week $n$. Taking all uncertainties into account, the prediction in Figure~\ref{Fig:PredictionTimeDeltaFrance} represents the actual data very well. We recall that $\mathfrak{p}_{\text{B.1.617}}$ is a combined probability, which therefore shows that the universal model proposed in \cite{Filoche:2024xka} is also applicable in this case.

\subsubsection{Time Period C): Appearance of BA.1 (Omicron)}
The time period {\bf{C)}} in Figure~\ref{Fig:FranceOverviewProbs} comprises the weeks 94-102 (08/11/2021--03/01/2022). The

\begin{wrapfigure}{r}{0.55\textwidth}
\vspace{-0.5cm}
\begin{center}
\includegraphics[width=9cm]{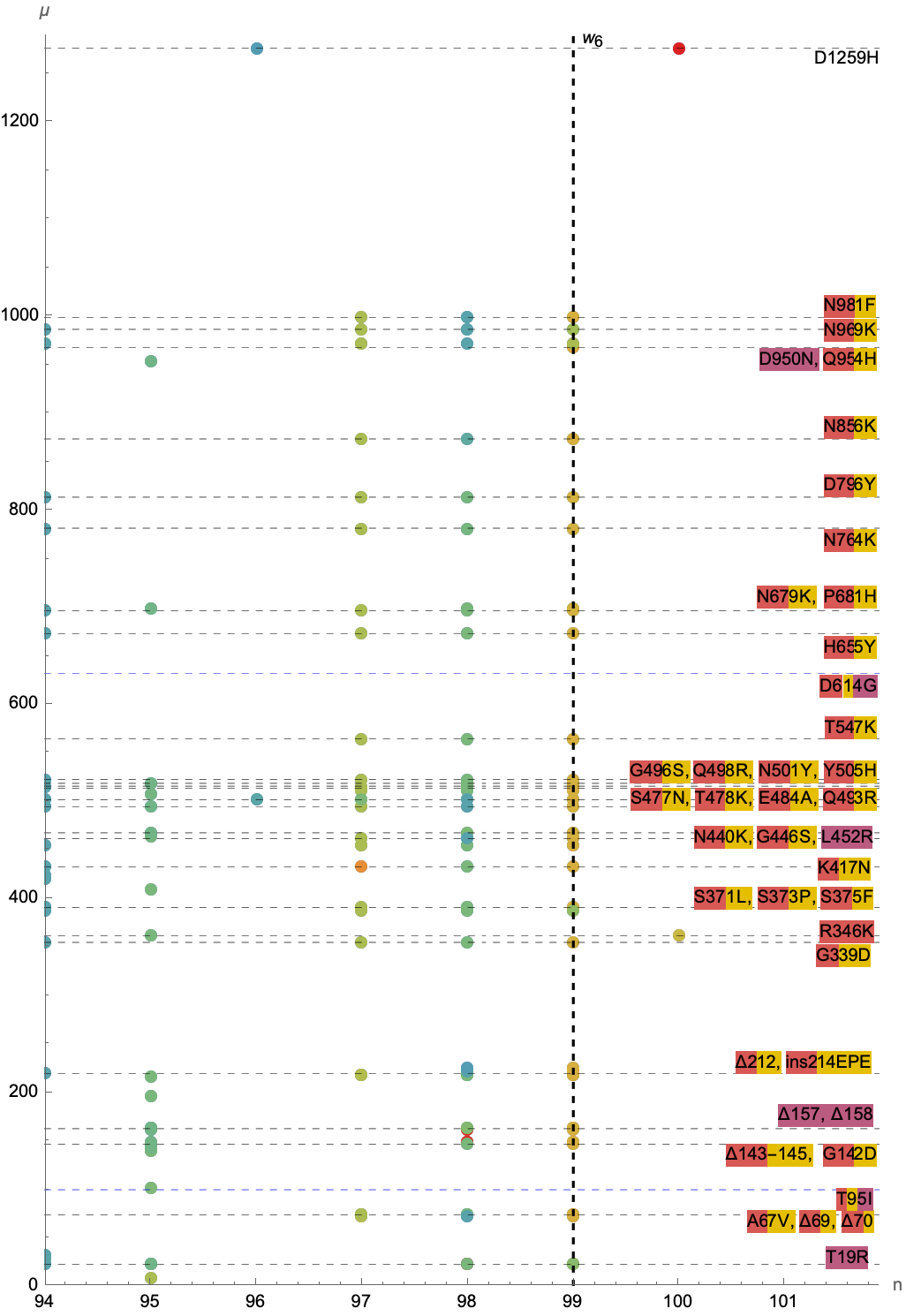}
\end{center}
\vspace{-0.5cm}
\caption{Evolution of the divergences (\ref{AlphaDiv}) for all protein sequence positions $\mu$ during the time period {\bf C)}. $w_{5}$ indicates a maximum of (\ref{TotalDivergencesWeek}) as shown in Figure~\ref{Fig:BA1SummedDivergences}.}
\label{Fig:FullTimeBA1France}
\end{wrapfigure}

\noindent
divergences as a function of the position on the spike protein sequence are shown in Figure~\ref{Fig:FullTimeBA1France}, where as before warmer colours represent larger values of $\mathcal{D}$. For better orientation, the week $w_{5}=99$ represents the maximum of the summed divergences (\ref{TotalDivergencesWeek}), which is plotted in Figure~\ref{Fig:BA1SummedDivergences}. This week marks the last week before the sequences of BA.1 and BA.1.1 (the first variants of Omicron) become part of the largest cluster and thus dominant. Indeed, after this week, divergences of the type (\ref{AlphaDiv}) are very small, indicating that the dominant sequences have indeed changed. The horizontal dashed lines in Figure~\ref{Fig:FullTimeBA1France} indicate the positions on the spike protein sequence where very large divergences occur and correlate them with mutations (which are labelled relative to the original Wuhan sequence): mutations shaded in purple are carried by B.1.617, while mutations shaded in yellow are carried by BA.1 and mutations shaded in pink are carried by BA.1.1. Notice that most of the new mutations are carried by both Omicron variants (and are thus shaded in both colours), except for R346K (which is only carried by BA.1.1). This mutation (and its impact on the infectivity of the virus) have been discussed for example in \cite{Cao,Mohandas,Cele}. We also remark that the maximum of the summed divergences in week $w_5$ in Figure~\ref{Fig:BA1SummedDivergences} is quite pronounced, indicating a quite abrupt change with many quite distinct modifications on the spike protein sequence. This indeed agrees with the epidemiological evolution 

\newpage
\begin{wrapfigure}{l}{0.5\textwidth}
\vspace{-0.5cm}
\begin{center}
\includegraphics[width=7.5cm]{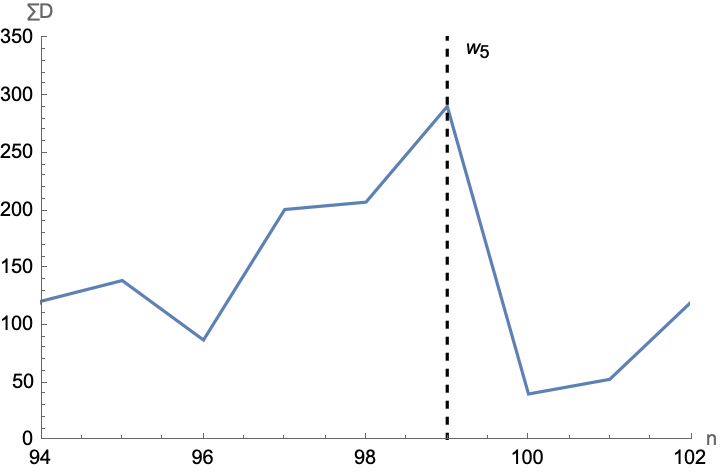}
\end{center}
\vspace{-0.5cm}
\caption{Evolution of the summed divergences (\ref{TotalDivergencesWeek}) during the time period {\bf C)}. The week $w_{5}$ is the same as in Figure~\ref{Fig:FullTimeBA1France}.}
\label{Fig:BA1SummedDivergences}
\end{wrapfigure}

\noindent
of the first Omicron variant: the first confirmed samples were collected on 08/11/2021 in South Africa and on 09/11/2021 in Botswana (week 94) \cite{OmicronFirst} and spread very quickly to other countries. The first samples in France were reported on 02/12/2021 (week 97) \cite{FranceOmicron}. As is evident from Figure~\ref{Fig:FranceOverviewProbs}, in France, Omicron spread very quickly and became the dominant variant in week 100. The quick rise of the probability of BA.1 and BA.1.1 are also reflected by their probabilities to be part of a growing cluster. Figure~\ref{Fig:IndicatorsC} shows the average probabilities (\ref{AverageProbPositive})  (with the variance indicate by the error bars) for both variants, along with the number $N^+$ of clusters with shrinking information, for which they have non-vanishing probability to be a

\begin{figure}[t]
\begin{center}
\includegraphics[width=7.5cm]{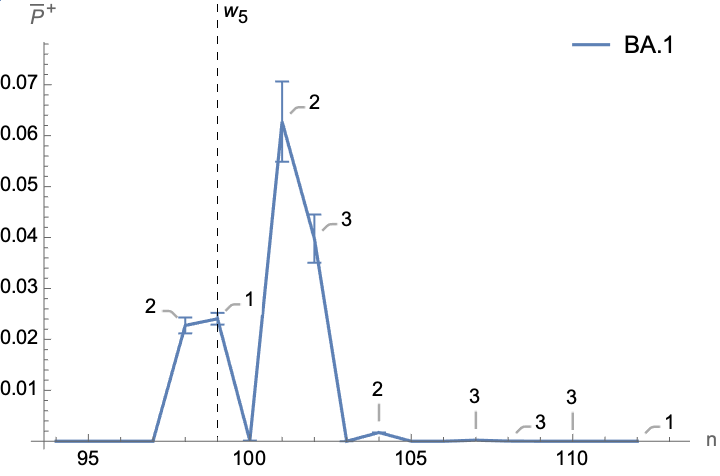}\hspace{1cm}\includegraphics[width=7.5cm]{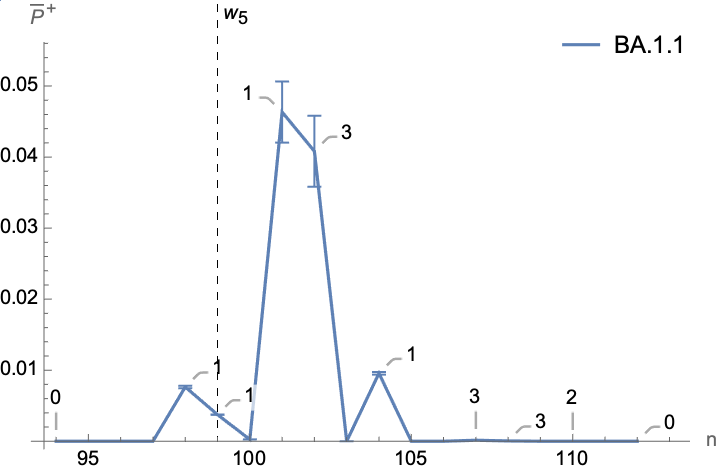}
\end{center}
\caption{Average probabilities (\ref{AverageProbPositive}) for BA.1 (left) and BA.1.1 (right) to be part of a cluster with diminishing information as a function of $n$, as well as the number of such clusters for which the same sequence has a non-vanishing probability to be a part of.}
\label{Fig:IndicatorsC}
\end{figure}

\begin{wrapfigure}{r}{0.5\textwidth}
\vspace{-0.5cm}
\begin{center}
\includegraphics[width=7.5cm]{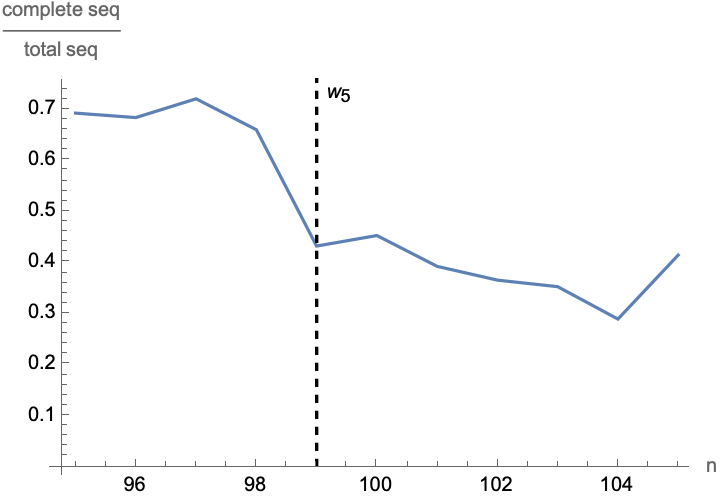}
\end{center}
\vspace{-0.5cm}
\caption{Fraction of complete sequences for the period {\bf{C)}}.}
\label{Fig:CompleteSequences}
\vspace{-0.5cm}
\end{wrapfigure}

\noindent
member of. While the latter numbers are generally larger than 1 (ranging even up to three), compared to similar previous plots (see Figure~\ref{Fig:IndicatorsA} for sequences during period {\bf{A)}} and Figure~\ref{Fig:IndicatorsB} for sequences during period {\bf{B)}}) it is striking that the average probabilities are much smaller. This is also evident from the maxima of $\overline{P}^+_{\text{BA.1}}$ and $\overline{P}^+_{\text{BA.1.1}}$ as well as $N^{+}_{\text{BA.1}}$ and $N^{+}_{\text{BA.1.1}}$, which are compared directly with other sequences in Figure~\ref{Fig:FranceMaxIndicators}.

A possible explanation for this difference is shown in Figure~\ref{Fig:CompleteSequences}, which plots the fraction of complete sequences relative to the total number of sequences extracted from GISAID during the period {\bf{C)}}. While in the beginning of this time period, around 70\% of the available sequences could be used for our analysis, this value dropped sharply with the appearance of BA.1 and BA.1.1 to around 40\% after $w_5$. In other words, during the period in which Omicron established itself as the dominant variant, almost 2/3 of the available sequences had to be discarded. This sharp drop around week $w_5$ suggests that in this period the appearance of incomplete sequence is no longer a random stochastic phenomenon, but biased against the Omicron variants. This bias could interfer with the calculation of the probabilities and in turn with the clustering, thus potentially explaining, why the probabilities shown in Figure~\ref{Fig:IndicatorsC} are much smaller than in previous cases. In order to address this issue, a probability analysis (similar to the one discussed in \cite{MLvariants}) could be applied that 'repairs' incomplete sequences with the most likely combinations of amino acids by comparing with the remaining sequences of this week. Since this causes further ambiguities and uncertainties and since in this paper we are only interested in demonstrating the viability of our information theoretical approach to real epidemiological data, we refrain from attempting this computation here and leave a more com-

\begin{wrapfigure}{l}{0.5\textwidth}
\vspace{-0.5cm}
\begin{center}
\includegraphics[width=7.5cm]{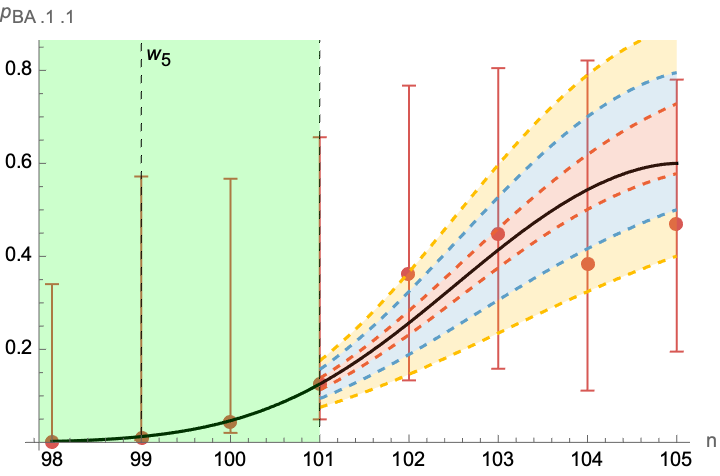}
\end{center}
\vspace{-0.5cm}
\caption{Prediction of the probability $\mathfrak{p}_{\text{BA.1.1}}$: the data points represent the actual values of the probability of BA.1.1 along with uncertainties based on the number of discarded incomplete sequences, while the black line shows a solution of (\ref{FlowEquation}) using the data in the green shaded region as input. The coloured bands show deviations of the prediction taking into account uncertainties of the input data.}
\label{Fig:PredictionTimeBA11France}
\end{wrapfigure}

\noindent
prehensive study for future work.

This same problem also makes an accurate prediction of the evolution of the probabilities of the sequences BA.1 and BA.1.1 difficult and feasible only at relatively late times. In Figure~\ref{Fig:PredictionTimeBA11France}, we show as an example a prediction for the probability $\mathfrak{p}_{\text{BA.1.1}}$ for the variant BA.1.1: as input, we have used all data points in the green shaded region (\emph{i.e.} up to week 101), from which we have calculated a fitted auxiliary metric $\mathfrak{g}_{tt}$, as described in (\ref{AuxiliaryMetric}). Substituting this metric into the flow equation (\ref{FlowEquation}) (with initial values imposed at $n=101$) we find as solution the black line in Figure~\ref{Fig:PredictionTimeBA11France}. As before, the coloured bands show corrections to the solution, assuming $\pm10\%$, $\pm 25\%$ and $\pm 40\%$ error of the initial input data. These need to be contrasted with the uncertainties (represented as error bars) for the (measured) probabilities $\mathfrak{p}_{\text{BA.1.1}}$ (represented by the red dots in Figure~\ref{Fig:PredictionTimeBA11France}) based on the relative number of discarded incomplete sequences in each week $n$. Taking all uncertainties into account (notably the relatively large number of discarded sequences in this time period), the prediction in Figure~\ref{Fig:PredictionTimeBA11France} represents the actual data very well.

\newpage
\subsubsection{Time Period D): Appearance of BA.2 }
The time period {\bf{D)}} follows immediately after {\bf{C)}} and comprises the weeks 102-112 (03/01/2022

\begin{wrapfigure}{r}{0.55\textwidth}
\vspace{-0.5cm}
\begin{center}
\includegraphics[width=9cm]{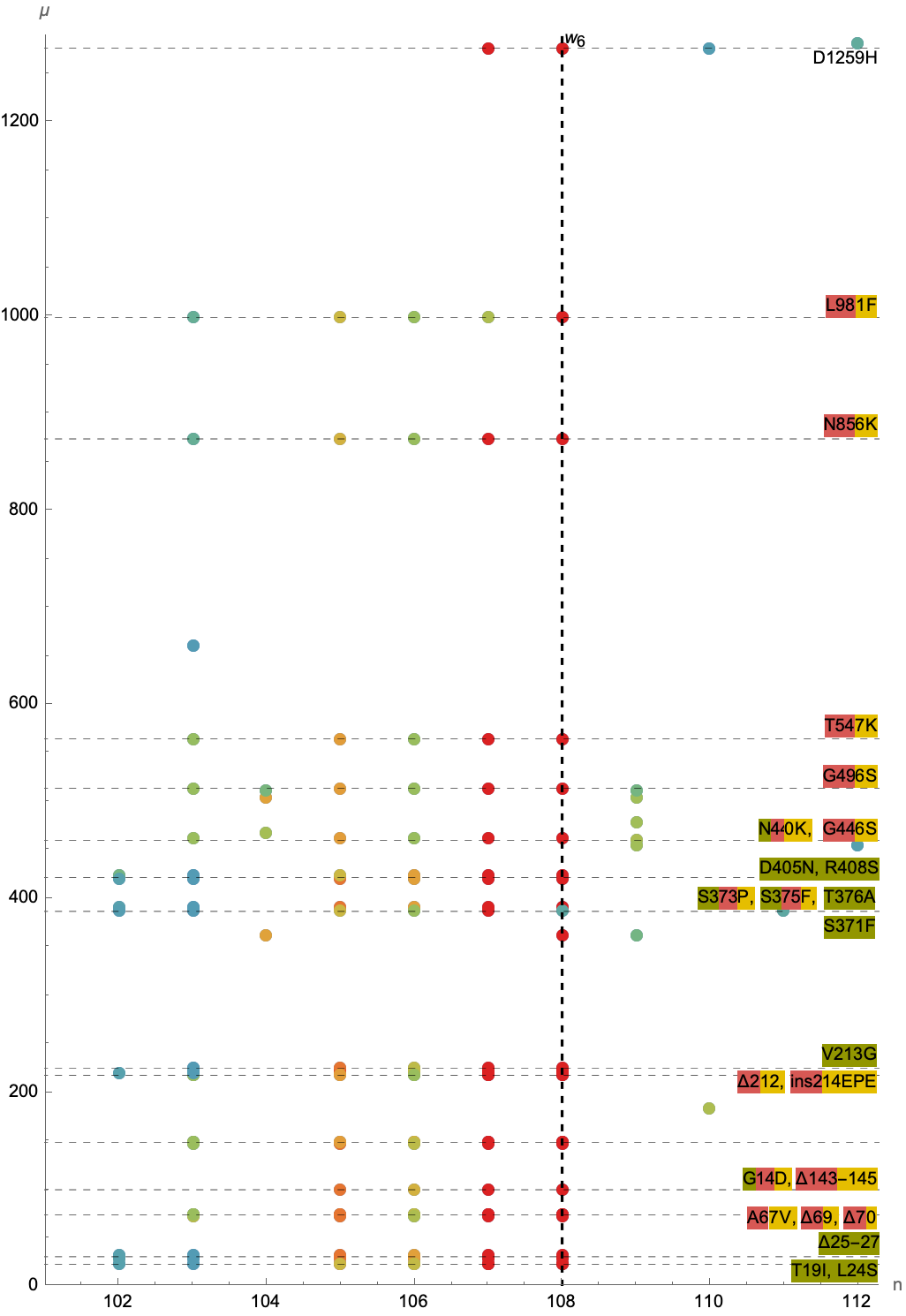}
\end{center}
\vspace{-0.5cm}
\caption{Evolution of the divergences (\ref{AlphaDiv}) for all protein sequence positions $\mu$ during the time period {\bf D)}. $w_{6}$ indicates a maximum of (\ref{TotalDivergencesWeek}) as shown in Figure~\ref{Fig:BA2SummedDivergences}.}
\label{Fig:FullTimeBA2France}
\end{wrapfigure}

\noindent
--14/03/2022). Figure~\ref{Fig:FullTimeBA2France} shows the divergences (\ref{AlphaDiv}) during this time period, where for better orientation, the week $w_{6}=108$ indicates the maximum of the summed divergences (\ref{TotalDivergencesWeek}), which is plotted in Figure~\ref{Fig:BA2SummedDivergences}. In the week following immediately after (\emph{i.e.} $n=109$), the probabilities of the variant BA.2 have become the largest among all sequences and the variant has become part of the largest cluster. The horizontal dashed lines in Figure~\ref{Fig:FullTimeBA2France} indicate the positions on the spike protein sequence where very large divergences (represented by red points) occur and correlate them with mutations (which are labelled relative to the original Wuhan sequence): mutations shaded in yellow and pink are carried by BA.1 and BA.1.1 respectively, while mutations shaded in green are carried by BA.2. Figure~\ref{Fig:FullTimeBA2France} indicates that, in contrast to the appearance of Alpha (see Figure~\ref{Fig:FullTimeAlphaFrance}) or Delta (see Figure~\ref{Fig:FullTimeDeltaFrance}), the divergences in these protein sequence positions are large already several weeks prior to $w_6$. This can be explained by the presence of numerous other sequences that carry some of these mutations (but not all of those of BA.2). This agrees with the observation \cite{ReviewDevelop} that Omicron variants tend to accumulate advantageous mutations over a certain period of time. We shall encounter a similar behaviour also for other variants of Omicron in subsequent Subsubsections. After week $w_6$, the divergences drop very sharply (see Figure~\ref{Fig:BA2SummedDivergences}) indicating that the nature of the dominant cluster has changed and the dominant sequence is now indeed BA.2. The quick rise of BA.2 to very large probabilities (we find a maximum of $\mathfrak{p}_{\text{BA.2}}(113)=0.8055$ in week 113) is also reflected by the probability of BA.2 to be part of a growing cluster. The left panel of Figure~\ref{Fig:IndicatorsD} shows the average probabilities (\ref{AverageProbPositive})  (with the variance indicate by the error bars) per week $n$, along with the number $N^+$ of clusters with shrinking information, for which BA.2 has non-vanishing probability to be a~mem-

\newpage

\begin{wrapfigure}{l}{0.5\textwidth}
\vspace{-0.5cm}
\begin{center}
\includegraphics[width=7.5cm]{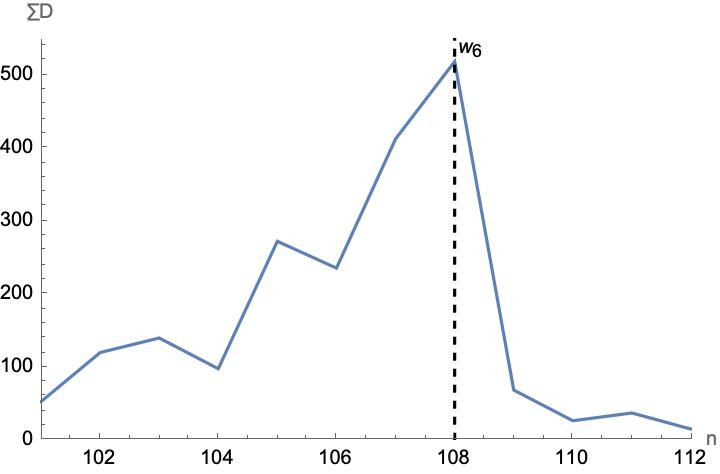}
\end{center}
\vspace{-0.5cm}
\caption{Evolution of the summed divergences (\ref{TotalDivergencesWeek}) during the time period {\bf D)}. The week $w_{6}$ is the same as in Figure~\ref{Fig:FullTimeBA2France}.}
\label{Fig:BA2SummedDivergences}
\end{wrapfigure}

\noindent
ber of. 

In the right panel of Figure~\ref{Fig:IndicatorsD} we have plotted a prediction for the probability  $\mathfrak{p}_{\text{BA.2}}$ for the variant BA.2: as input, we have used all data points in the green shaded region (\emph{i.e.} up to week 106). Notice that this point is before $w_6$, \emph{i.e.} before BA.2 has become the dominant sequence. Due to the availability of a larger number of (complete) sequences per week (see Figure~\ref{Fig:FranceTotalOverview}) throughout the entire time period, an accurate prediction is already possible at a rather early point in time. Indeed, from the input data, we have calculated a fitted auxiliary metric $\mathfrak{g}_{tt}$, as described in (\ref{AuxiliaryMetric}).\footnote{Due to the higher number of data, we could fit the metric with a 4-parameter function of the form $g_{tt}(p)\sim a\,p^b\,(p_2-p)^c$, with $a,b,c,p_2$ free parameters.} Substituting this metric into the flow equation (\ref{FlowEquation}) (with initial values imposed at $n=106$) we find as solution the black line in the right panel of Figure~\ref{Fig:IndicatorsD}. As before, the coloured bands show corrections to the solution, assuming $\pm10\%$, $\pm 25\%$ and $\pm 40\%$ error of the initial input data. These need to be contrasted with the uncertainties (represented as error bars) for the (measured) probabilities $\mathfrak{p}_{\text{BA.2}}$ (represented by the green dots in Figure~\ref{Fig:IndicatorsD}) based on the relative number of discarded incomplete sequences in each week $n$. Even without the uncertainties, the prediction in Figure~\ref{Fig:IndicatorsD} represents the actual data very well.

\begin{figure}[htbp]
\begin{center}
\includegraphics[width=7.5cm]{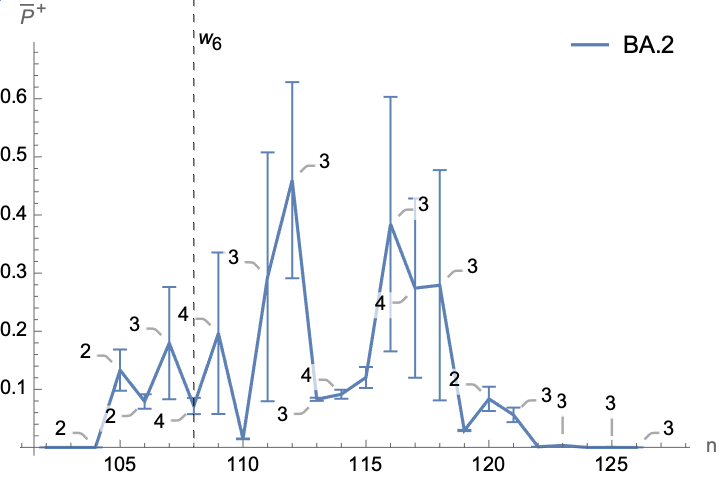}\hspace{1cm}\includegraphics[width=7.5cm]{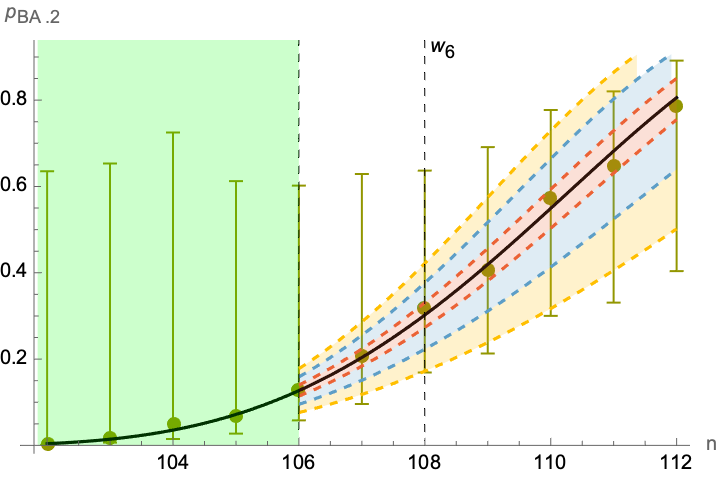}
\end{center}
\caption{Left panel: Average probabilities (\ref{AverageProbPositive}) for BA.2 to be part of a cluster with diminishing information as a function of $n$, as well as the number of such clusters for which the same sequence has a non-vanishing probability to be a part of. Right panel: Prediction of the probability $\mathfrak{p}_{\text{BA.2}}$: the data points represent the actual values of the probability of BA.2 along with uncertainties based on the number of discarded incomplete sequences, while the black line shows a solution of (\ref{FlowEquation}) using the data in the green shaded region as input. The coloured bands show deviations of the prediction taking into account uncertainties of the input data.}
\label{Fig:IndicatorsD}
\end{figure}

\subsubsection{Time Period E): Appearance of BA.5}

The time period {\bf{E)}} comprises the weeks 112-124 (14/03/2022--16/05/2022) and coincides~most-

\begin{wrapfigure}{r}{0.55\textwidth}
\vspace{-0.6cm}
\begin{center}
\includegraphics[width=9cm]{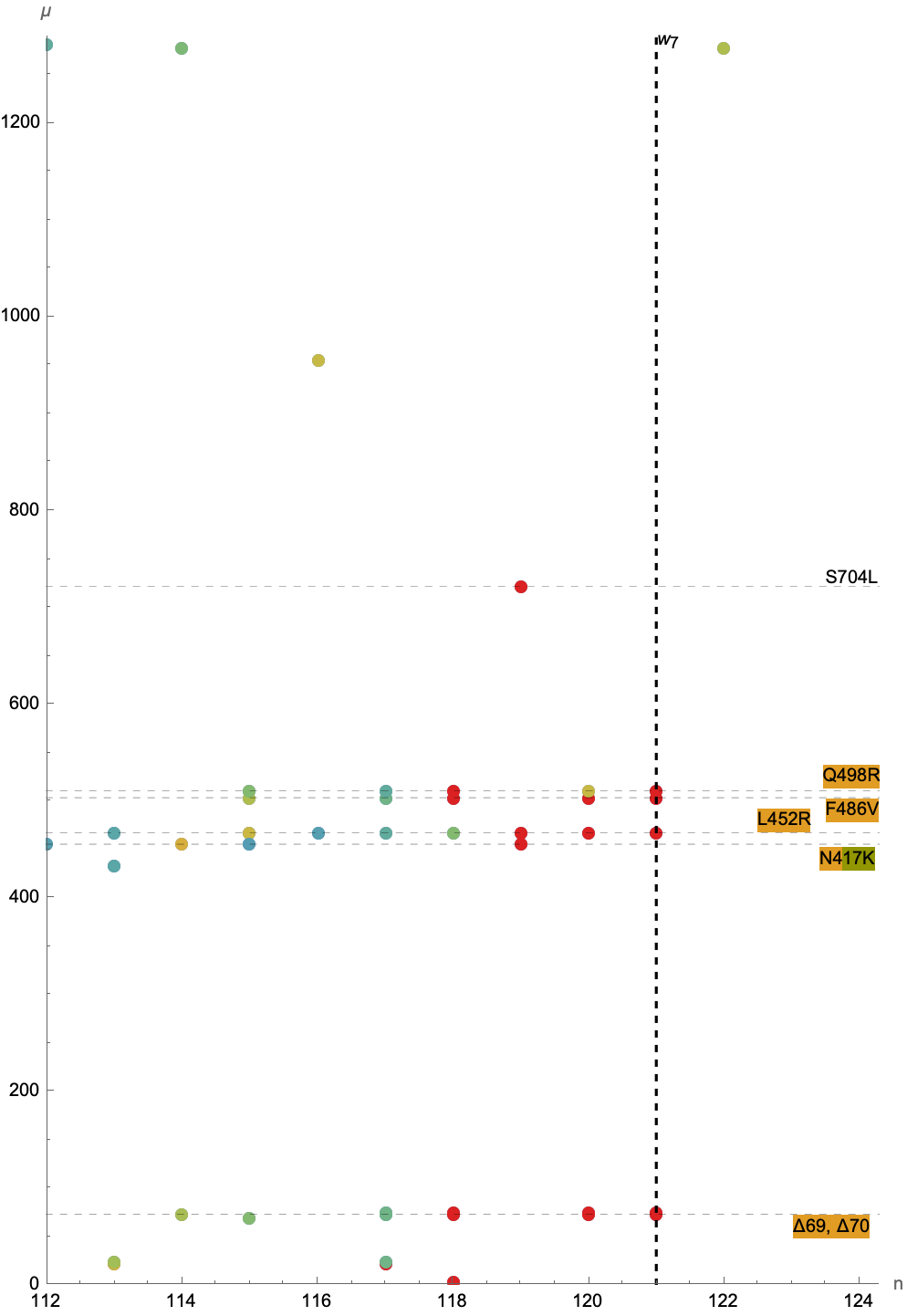}
\end{center}
\vspace{-0.5cm}
\caption{Evolution of the divergences (\ref{AlphaDiv}) for all protein sequence positions $\mu$ during the time period {\bf E)}. $w_{7}$ indicates a maximum of (\ref{TotalDivergencesWeek}) as shown in Figure~\ref{Fig:BA5SummedDivergences}.}
\label{Fig:FullTimeBA5France}
\end{wrapfigure}

\noindent
ly with the time period discussed in Section~\ref{Sect:CaseStudyFrance}. Nevertheless, we shall discuss the time period {\bf{E)}} here in detail, since the data used in Section~\ref{Sect:CaseStudyFrance} were subjected to a (strong) Gaussian filtering, which we shall not apply here. We thereby demonstrate that our main conclusions and results do not depend on this simplification and smoothening of the data.

The divergences as a function of the position on the spike protein sequence are shown in Figure~\ref{Fig:FullTimeBA5France}, where as before warmer colours represent larger values of $\mathcal{D}$. For better orientation, the week $w_{7}=121$ represents the maximum of the summed divergences (\ref{TotalDivergencesWeek}), which is plotted in Figure~\ref{Fig:BA5SummedDivergences}. Moreover, $w_7$ indicates the week before the variant BA.5 has reached the largest probability of all sequences and thus has become dominant. The horizontal dashed lines in Figure~\ref{Fig:FullTimeBA5France} point out the positions along the spike protein sequence where very large divergences occur and correlate them with mutations (which are labelled relative to the original Wuhan sequence): mutations shaded in green are carried by BA.2 while mutations shaded in orange are carried by BA.5 (mutations carried by both are shown in both colours). The differences between these two variants are indeed the same mutations (\ref{Mutations238}) represented by very strong divergences in Figure~\ref{Fig:FullTimeBA5France}.  As is visible from Figure~\ref{Fig:BA5SummedDivergences}, the summed divergences are already quite high before week $w_7$, which is in fact compatible with the findings in Section~\ref{Sect:CaseStudyFrance}: many other sequences have carried some of the mutations (\ref{Mutations238}) already prior to week $w_7$ and the first appearance of BA.5 in week 116.

The variant BA.5 achieves a maximum probability of $p_{\text{BA.5}}(129)=0.7436$ in week 129. This potential to grow very strongly is also reflected by the probability of BA.5 to be part of a growing cluster. Indeed, the left panel of Figure~\ref{Fig:IndicatorsE} gives the average probabilities (\ref{AverageProbPositive})  (with the variance indicate by the error bars) per week $n$, along with the number $N^+$ of clusters with shrinking information, for which BA.5 has non-vanishing probability to be a member of. As in the cases before, the right panel of Figure~\ref{Fig:IndicatorsE} shows a prediction of the growth of BA.5 by plotting a prediction for the probability $\mathfrak{p}_{\text{BA.5}}$ for the variant BA.5: as input, we have used all

\begin{wrapfigure}{l}{0.5\textwidth}
\vspace{-0.5cm}
\begin{center}
\includegraphics[width=7.5cm]{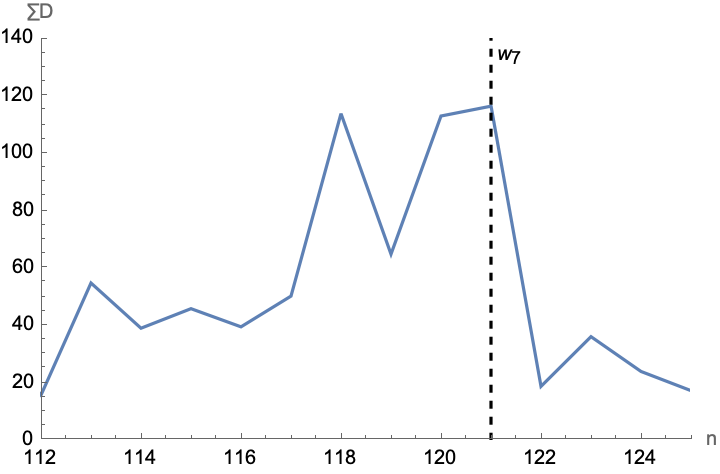}
\end{center}
\vspace{-0.5cm}
\caption{Evolution of the summed divergences (\ref{TotalDivergencesWeek}) during the time period {\bf E)}. The week $w_{7}$ is the same as in Figure~\ref{Fig:FullTimeBA5France}.}
\label{Fig:BA5SummedDivergences}
\end{wrapfigure}

\noindent
data points in the green shaded region (\emph{i.e.} up to week 122): due to fluctuations of the input probabilities, this is the first week, which allows for an accurate prediction of the probabilities. This is to be contrasted with the plots shown in Figure~\ref{Fig:FrancePrediction}: due to the Gaussian filtering (and thus a much smoother curve for the probabilities), predictions are possible at earlier times, \emph{i.e.} when the variant BA.5 still had only a rather small probability. Based on the input data in Figure~\ref{Fig:IndicatorsE}, we have calculated a fitted auxiliary metric $\mathfrak{g}_{tt}$, as described in (\ref{AuxiliaryMetric}).\footnote{Due to the higher number of data, we could fit the metric with a 4-parameter function of the form $g_{tt}(p)\sim a\,p^b\,(p_2-p)^c$, with $a,b,c,p_2$ free parameters.} Substituting this metric into the flow equation (\ref{FlowEquation}) (with initial values imposed at $n=122$) we find as solution the black line in the right panel of Figure~\ref{Fig:IndicatorsE}. As before, the coloured bands show corrections to the solution, assuming $\pm10\%$, $\pm 25\%$ and $\pm 40\%$ error of the initial input data. These need to be contrasted with the uncertainties (represented as error bars) for the (measured) probabilities $p_{\text{BA.5}}$ (represented by the yellow dots in Figure~\ref{Fig:IndicatorsE}) based on the relative number of discarded incomplete sequences in each week $n$.

\begin{figure}[htbp]
\begin{center}
\includegraphics[width=7.5cm]{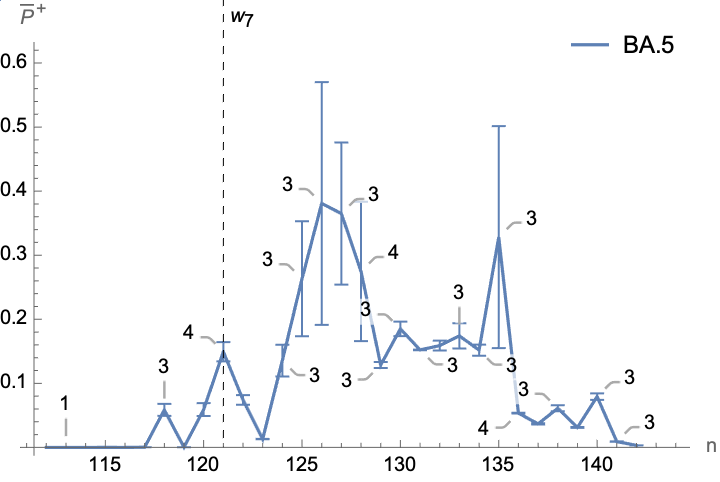}\hspace{1cm}\includegraphics[width=7.5cm]{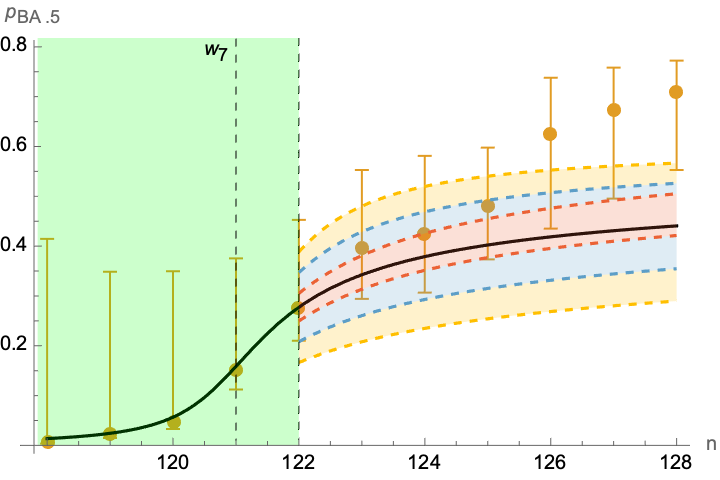}
\end{center}
\caption{Left panel: Average probabilities (\ref{AverageProbPositive}) for BA.5 to be part of a cluster with diminishing information as a function of $n$, as well as the number of such clusters for which the same sequence has a non-vanishing probability to be a part of. Right panel: Prediction of the probability $\mathfrak{p}_{\text{BA.5}}$: the data points represent the actual values of the probability of BA.5 along with uncertainties based on the number of discarded incomplete sequences, while the black line shows a solution of (\ref{FlowEquation}) using the data in the green shaded region as input. The coloured bands show deviations of the prediction taking into account uncertainties of the input data.}
\label{Fig:IndicatorsE}
\end{figure}

\subsubsection{Time Period F): Appearance of BQ.1.1}

The time period {\bf{F)}} comprises the weeks 136-144 (29/08/2022--24/10/2022), with the domi-

\begin{wrapfigure}{r}{0.55\textwidth}
\vspace{-0.6cm}
\begin{center}
\includegraphics[width=9cm]{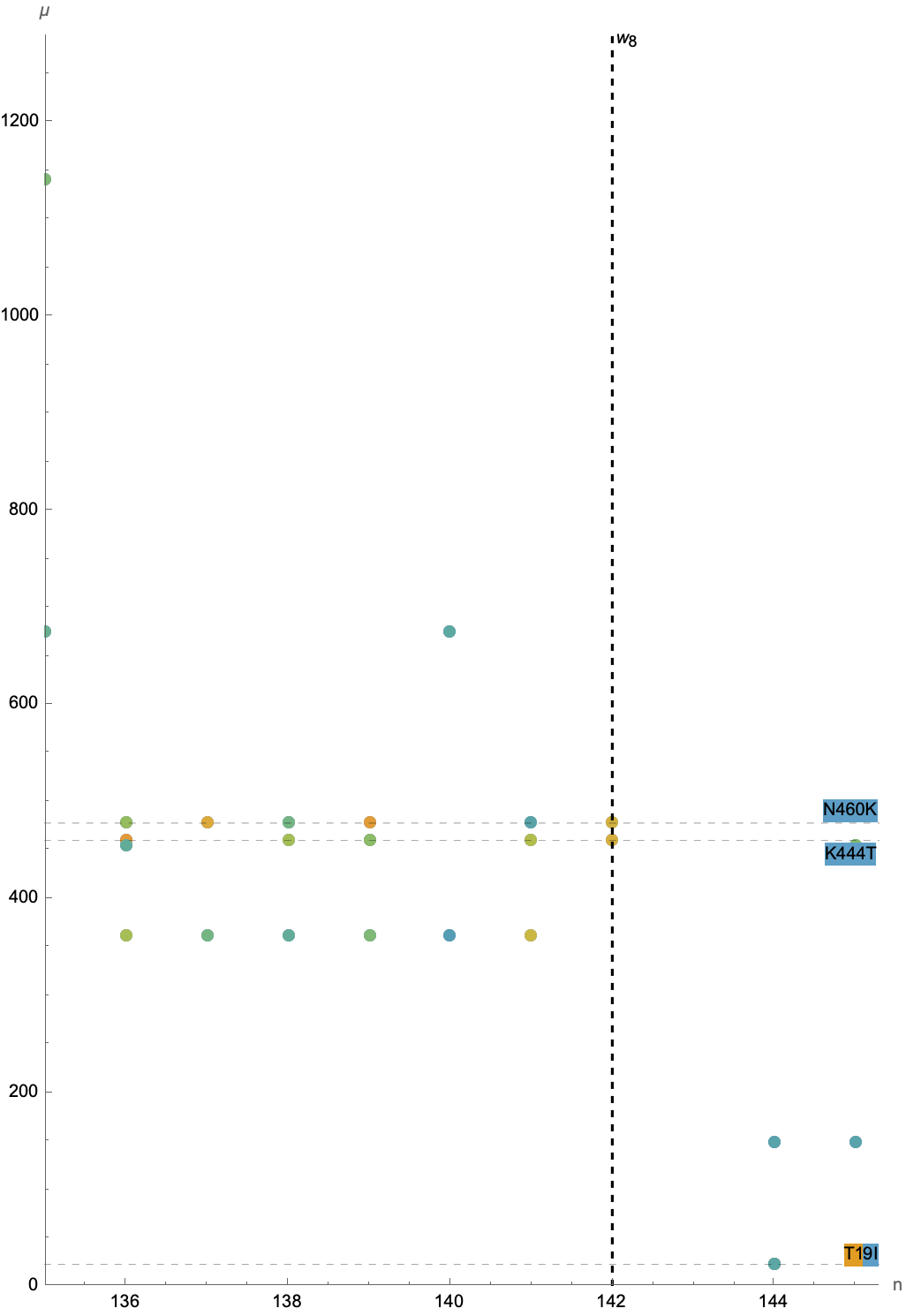}
\end{center}
\vspace{-0.5cm}
\caption{Evolution of the divergences (\ref{AlphaDiv}) for all protein sequence positions $\mu$ during the time period {\bf F)}. $w_{8}$ indicates a local maximum of (\ref{TotalDivergencesWeek}) as shown in Figure~\ref{Fig:BQ11SummedDivergences}.}
\label{Fig:FullTimeBQ11France}
\end{wrapfigure}

\noindent
nant sequence given by BQ.1.1, which, however, never reaches a probability over 0.5. The divergences as a function of the position on the spike protein sequence are shown in Figure~\ref{Fig:FullTimeBQ11France}, where, as in all cases above, warmer colours represent larger values of $\mathcal{D}$. For better orientation, the week $w_{8}=142$ represents a local maximum of the summed divergences (\ref{TotalDivergencesWeek}), which is plotted in Figure~\ref{Fig:BQ11SummedDivergences}. The latter, however, is not particularly pronounced, which correlates with the fact that the largest sequence BQ.1.1 never reaches complete dominance (\emph{i.e.} a probability larger than 0.5). The week $w_8$ marks the last week before the sequences of BQ.1.1 becomes part of the largest cluster. Indeed, after this week, divergences of the type (\ref{AlphaDiv}) are very small, indicating that the dominant sequences have indeed changed. The horizontal dashed lines in Figure~\ref{Fig:FullTimeBQ11France} indicate the positions on the spike protein sequence where very large divergences occur and correlate them with mutations (which are labelled relative to the original Wuhan sequence): mutations shaded in blue are carried by BQ.1.1, while mutations shaded in yellow are carried by BA.5 (with mutations carried by both shaded in yellow and blue). As is evident from Figure~\ref{Fig:FullTimeBQ11France}, the divergences in these positions are already high in the weeks before $w_8$ and are also carried by some other sequences. The variant BQ.1.1 reaches a maximal probability of $p_{\text{BQ.1.1}}(148)=0.4706$ in week 148. Its potential for growth is also reflected by the probability to be part of a growing cluster. Indeed, the left panel of Figure~\ref{Fig:IndicatorsF} gives the average probabilities (\ref{AverageProbPositive})  (with the variance indicate by the error bars) per week $n$, along with the number $N^+$ of clusters with shrinking information, for which BQ.1.1 has non-vanishing pro-

\newpage
\begin{wrapfigure}{l}{0.5\textwidth}
\vspace{-0.5cm}
\begin{center}
\includegraphics[width=7.5cm]{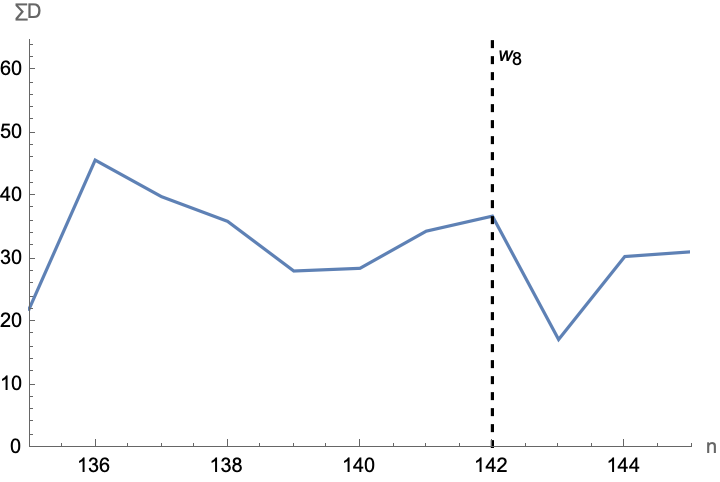}
\end{center}
\vspace{-0.5cm}
\caption{Evolution of the summed divergences (\ref{TotalDivergencesWeek}) during the time period {\bf F)}. The week $w_{8}$ is the same as in Figure~\ref{Fig:FullTimeBQ11France}.}
\label{Fig:BQ11SummedDivergences}
\end{wrapfigure}

\noindent
bability to be a member of. As in the cases before, the right panel of Figure~\ref{Fig:IndicatorsF} shows a prediction of the growth of BQ.1.1 by plotting a prediction for the probability $p_{\text{BQ.1.1}}$ for the variant BQ.1.1. As input, we have used all data points in the green shaded region (\emph{i.e.} up to week 139): notice that this value is 3 weeks before $w_8$ in which BQ.1.1 has become part of the dominant cluster. Based on the input data in Figure~\ref{Fig:IndicatorsE}, we have calculated a fitted auxiliary metric $\mathfrak{g}_{tt}$, using a 4-parameter function of the form $\mathfrak{g}_{tt}(p)\sim a\,p^b\,(p_2-p)^c$, with $a,b,c,p_2$ free parameters. Substituting this metric into the flow equation (\ref{FlowEquation}) (with initial values imposed at $n=139$) we find as solution the black line in the right panel of Figure~\ref{Fig:IndicatorsF}. As before, the coloured bands show corrections to the solution, assuming $\pm10\%$, $\pm 25\%$ and $\pm 40\%$ error of the initial input data. These need to be contrasted with the uncertainties (represented as error bars) for the (measured) probabilities $\mathfrak{p}_{\text{BQ.1.1}}$ (represented by the yellow dots in Figure~\ref{Fig:IndicatorsF}) based on the relative number of discarded incomplete sequences in each week $n$.

\begin{figure}[htbp]
\begin{center}
\includegraphics[width=7.5cm]{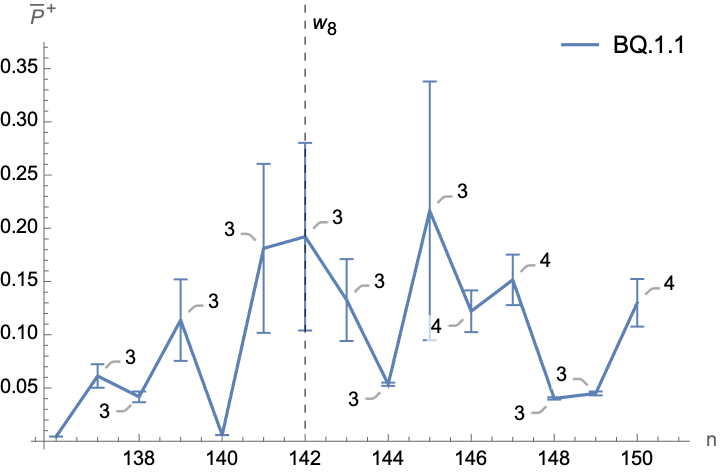}\hspace{1cm}\includegraphics[width=7.5cm]{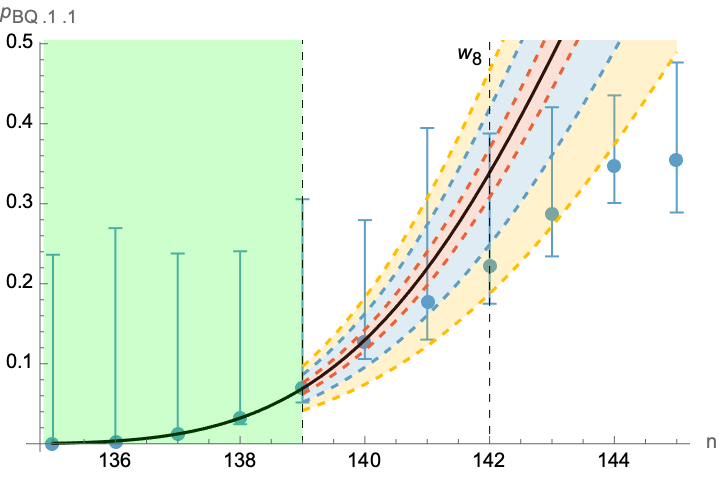}
\end{center}
\caption{Left panel: Average probabilities (\ref{AverageProbPositive}) for BQ.1.1 to be part of a cluster with diminishing information as a function of $n$, as well as the number of such clusters for which the same sequence has a non-vanishing probability to be a part of. Right panel: Prediction of the probability $\mathfrak{p}_{\text{BQ.1.1}}$: the data points represent the actual values of the probability of BQ.1.1 along with uncertainties based on the number of discarded incomplete sequences, while the black line shows a solution of (\ref{FlowEquation}) using the data in the green shaded region as input. The coloured bands show deviations of the prediction taking into account uncertainties of the input data.}
\label{Fig:IndicatorsF}
\end{figure}

\newpage
\subsubsection{Time Period G): Appearance of XBB.1.5}
The time period {\bf{G)}} comprises the weeks 152--167 (19/12/2022--03/04/2023). The diver-

\begin{wrapfigure}{r}{0.55\textwidth}
\vspace{-1.1cm}
\begin{center}
\includegraphics[width=9cm]{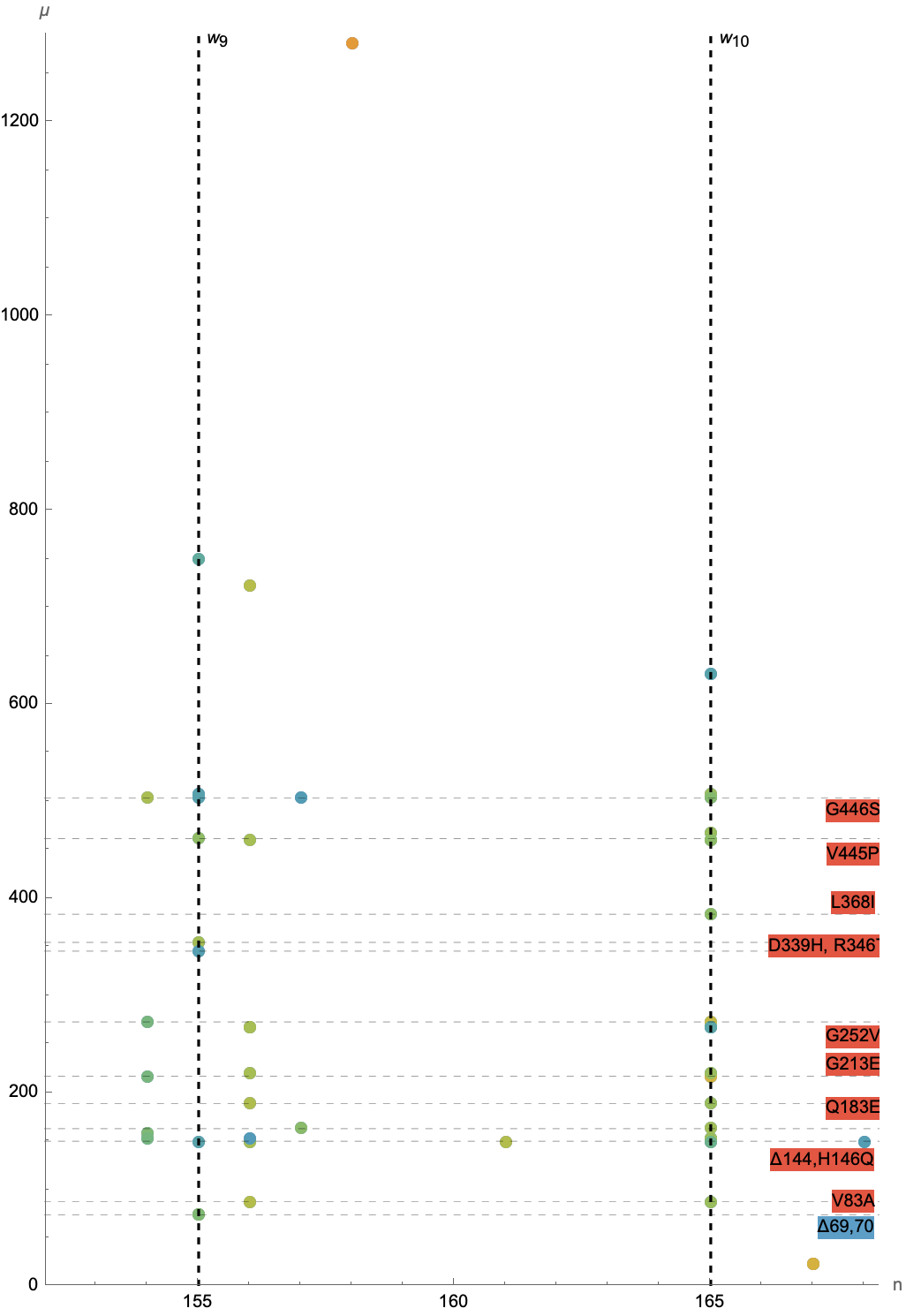}
\end{center}
\vspace{-0.7cm}
\caption{Evolution of the divergences (\ref{AlphaDiv}) for all positions $\mu$ during the time period {\bf G)}.}
\label{Fig:FullTimeXBB15France}
\vspace{-0.5cm}
\end{wrapfigure}

\noindent
gences as a function of the position on the spike protein sequence are shown in Figure~\ref{Fig:FullTimeXBB15France}. Notice that all data points are marked by cool colours, indicating that the divergences are comparatively small. Indeed, the total summed divergences are shown in Figure~\ref{Fig:XBB15SummedDivergences}, where for better orientation the weeks $w_{9}=155$ and $w_{10}=165$, which correspond to local maxima of the summed divergences, are marked by vertical dashed lines. $w_{10}$ furthermore is the week prior to the variant XBB.1.5 becoming part of the largest cluster, after which the divergences indeed drop, reflecting the change in nature of the dominant cluster. The horizontal dashed lines in Figure~\ref{Fig:FullTimeXBB15France} indicate the positions on the spike protein sequence where the larger divergences occur and correlate them with mutations (which are labelled relative to the original Wuhan sequence): mutations shaded in red are carried by XBB.1.5, while mutations shaded in blue are carried by BQ.1.1. As is evident from Figure~\ref{Fig:FullTimeBQ11France}, some of these divergences are already high in the weeks before $w_{10}$ and are notably carried by seq 718. The variant XBB.1.5 reaches a maximal probability of $p_{\text{XBB.1.5}}(167)=0.5964$ in week 167. As in previous cases, this potential for growth is also reflected by the probability to be part of a growing cluster. Indeed, the right panel of Figure~\ref{Fig:IndicatorsG} shows the average probabilities (\ref{AverageProbPositive})  (with the 

\begin{wrapfigure}{l}{0.5\textwidth}
\vspace{-0.5cm}
\begin{center}
\includegraphics[width=7.5cm]{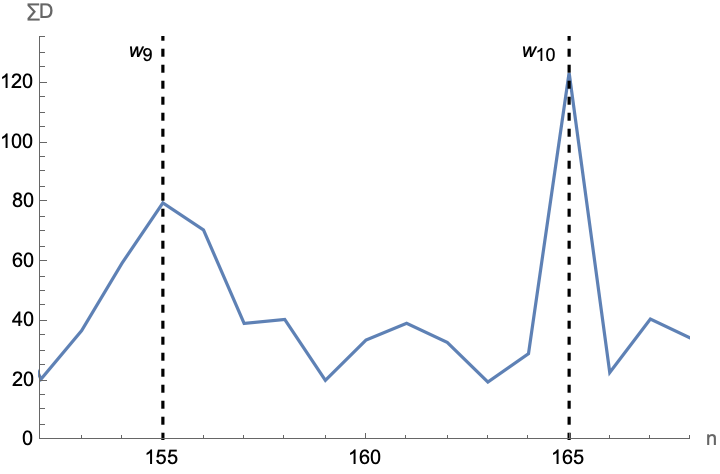}
\end{center}
\vspace{-0.6cm}
\caption{Evolution of the summed divergences (\ref{TotalDivergencesWeek}) during the time period {\bf G)}.}
\label{Fig:XBB15SummedDivergences}
\vspace{-1.3cm}
\end{wrapfigure}

\noindent
variance indicate by the error bars) per week $n$, along with the number $N^+$ of clusters with shrinking information, for which XBB.1.5 has non-vanishing probability to be a member of. For comparison, the left panel shows the same data for seq 718, which equally reaches sizeable probabilities (as is also evident from Figure~\ref{Fig:FranceOverviewProbs}). However, the average $\overline{P}^+_{718}$ remains small compared to $\overline{P}^+_{\text{XBB.1.5}}$, in the same manner as the maximum of $p_{718}$ is smaller than the maximum of $p_{\text{XBB.1.5}}$. The latter sequence clearly becomes dominant during the time period {\bf{G)}}. We can therefore use (\ref{FlowEquation}) to predict the time evolution of the probability $\mathfrak{p}_{\text{XBB.1.5}}$. Figure~\ref{Fig:PredictionTimeXBB15France} shows such a prediction of the growth of XBB.1.5, starting from the week 156: as input, we have used all data points in the green shaded region (\emph{i.e.} up to week 156). This week

\begin{figure}[htbp]
\begin{center}
\includegraphics[width=7.5cm]{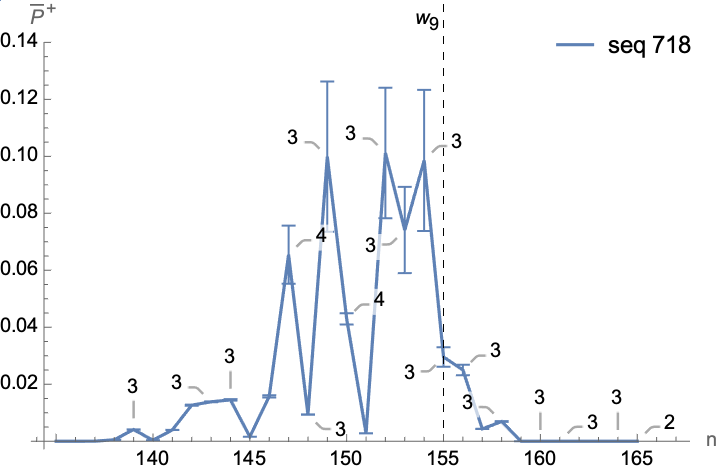}\hspace{1cm}\includegraphics[width=7.5cm]{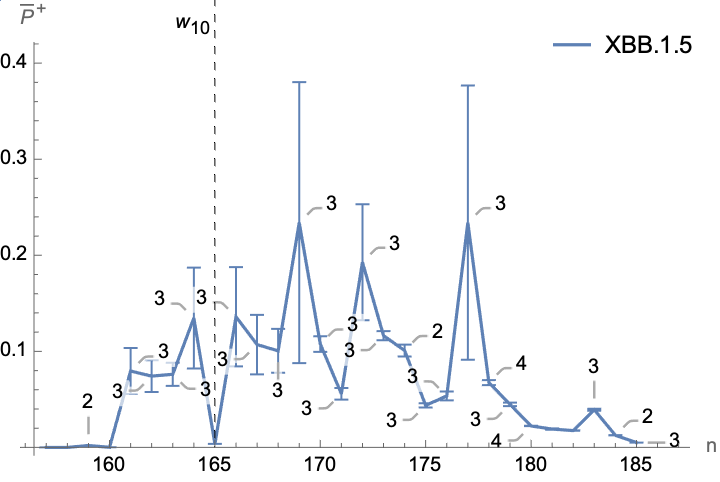}
\end{center}
\caption{Average probabilities (\ref{AverageProbPositive}) for seq 718 (left) and XBB.1.5 (right) to be part of a cluster with diminishing information as a function of $n$, as well as the number of such clusters for which the same sequence has a non-vanishing probability to be a part of.}
\label{Fig:IndicatorsG}
\end{figure}

\begin{wrapfigure}{r}{0.5\textwidth}
\vspace{-0.5cm}
\begin{center}
\includegraphics[width=7.5cm]{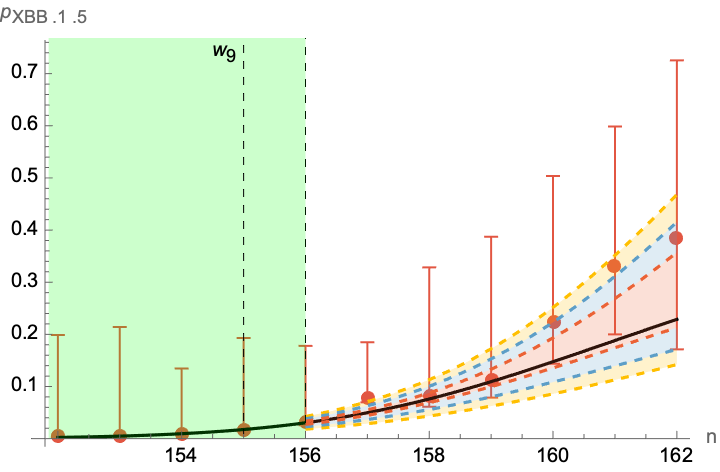}
\end{center}
\vspace{-0.5cm}
\caption{Prediction of the probability $\mathfrak{p}_{\text{XBB.1.5}}$: the data points represent the actual values of the probability of XBB.1.5 along with uncertainties based on the number of discarded incomplete sequences, while the black line shows a solution of (\ref{FlowEquation}) using the data in the green shaded region as input. The coloured bands show deviations of the prediction taking into account uncertainties of the input data.}
\label{Fig:PredictionTimeXBB15France}
\end{wrapfigure}

\noindent
is the week directly after $w_{10}$, \emph{i.e.} the first week in which XBB.1.5 is part of the largest cluster. Notice, however, also that $p_{\text{XBB.1.5}}(156)=0.0303$ such that the the probability of this variant is still quite small. Based on the input data in Figure~\ref{Fig:PredictionTimeXBB15France}, we have calculated a fitted auxiliary metric $\mathfrak{g}_{tt}$, using a 4-parameter function of the form $\mathfrak{g}_{tt}(p)\sim a\,p^b\,(p_2-p)^c$, with $a,b,c,p_2$ free parameters. Substituting this metric into the flow equation (\ref{FlowEquation}) (with initial values imposed at $n=156$) we find as solution the black line in Figure~\ref{Fig:PredictionTimeXBB15France}. As before, the coloured bands show corrections to the solution, assuming $\pm10\%$, $\pm 25\%$ and $\pm 40\%$ error of the initial input data. These need to be contrasted with the uncertainties (represented as error bars) for the (measured) probabilities $p_{\text{XBB.1.5}}$ (represented by the red dots in Figure~\ref{Fig:PredictionTimeXBB15France}) based on the relative number of discarded incomplete sequences in each week $n$. Taking into account the different uncertainties, the prediction describes the actual data quite well.

\subsubsection{Time Period H): Appearance of seq 1319}
The time period {\bf{H)}} comprises the weeks 182--187 (17/07/2023--21/08/2023). The diver-

\begin{wrapfigure}{r}{0.55\textwidth}
\vspace{-0.6cm}
\begin{center}
\includegraphics[width=9cm]{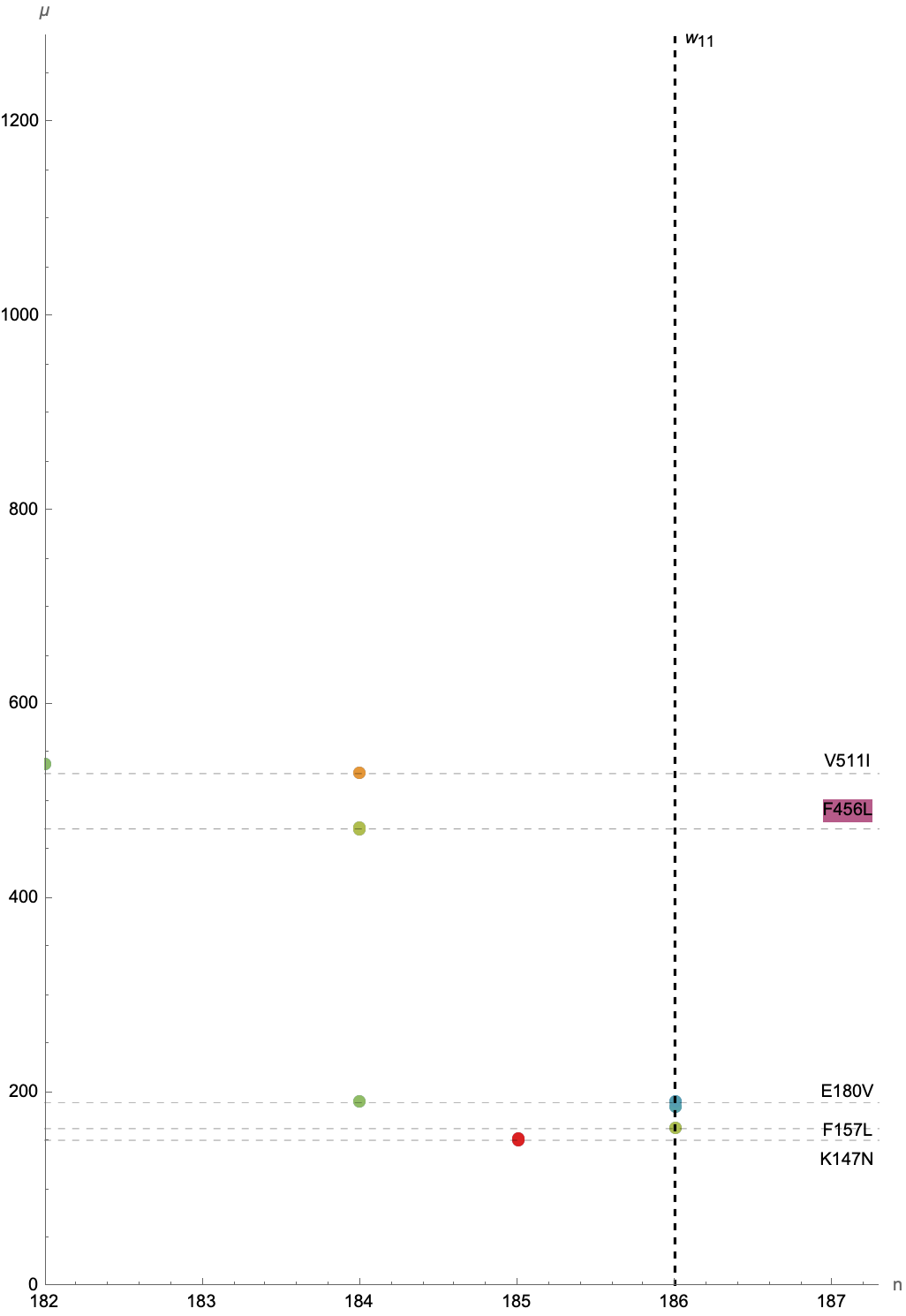}
\end{center}
\vspace{-0.5cm}
\caption{Evolution of the divergences (\ref{AlphaDiv}) for all protein sequence positions $\mu$ during the time period {\bf H)}. $w_{11}$ indicates a local maximum of (\ref{TotalDivergencesWeek}) as shown in Figure~\ref{Fig:1319SummedDivergences}.}
\label{Fig:FullTime1319France}
\vspace{-1cm}
\end{wrapfigure}

\noindent
gences as a function of the position on the spike protein sequence are shown in Figure~\ref{Fig:FullTime1319France}. There are relatively few data points, which in addition also mark small divergences: the total summed divergences are shown in Figure~\ref{Fig:1319SummedDivergences}, where for better orientation the week $w_{11}=186$, which correspond to a local maximum , are marked by vertical dashed lines. Furthermore, $w_{11}$ is the week prior to seq 1319 becoming part of the largest cluster. The horizontal dashed lines in Figure~\ref{Fig:FullTime1319France} indicate the positions on the spike protein sequence where the largest divergences occur and correlate them with mutations (which are labelled relative to the original Wuhan sequence): most mutations are in fact carried by sequences only reaching a few percent, thus explaining the low divergences in Figures~\ref{Fig:FullTime1319France} and \ref{Fig:1319SummedDivergences}. Only the mutation F456L is carried by seq 1319, which is becoming dominant during the time period {\bf{H)}}. This sequence is very close the variant EG.5 (also called Eris), but in addition also carries the mutations Q52H and F456L on the spike protein. This sequence reaches a maximal probability $p_{1319}(186)=0.3291$ in week 186 and thus never becomes fully dominant. Nevertheless, this potential for growth is also reflected by the probability to be part of a growing cluster. Indeed, the left panel of Figure~\ref{Fig:IndicatorsG} shows the average probabilities (\ref{AverageProbPositive})  (with the variance indicate by the 

\begin{wrapfigure}{l}{0.5\textwidth}
\vspace{-0.5cm}
\begin{center}
\includegraphics[width=7.5cm]{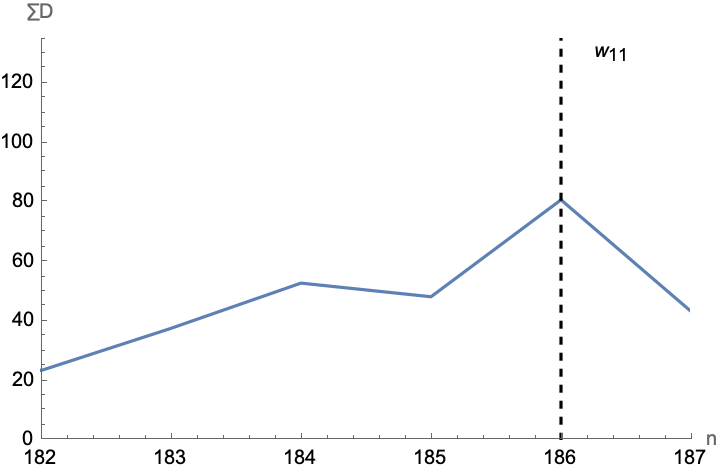}
\end{center}
\vspace{-0.5cm}
\caption{Evolution of the summed divergences (\ref{TotalDivergencesWeek}) during the time period {\bf H)}. }
\label{Fig:1319SummedDivergences}
\vspace{-0.5cm}
\end{wrapfigure}

\noindent
error bars) per week $n$, along with the number $N^+$ of clusters with shrinking information, for which seq 1319 has non-vanishing probability to be a member of. As in the cases before, the right panel of Figure~\ref{Fig:IndicatorsF} shows a prediction of the growth of 1319 by plotting a prediction for the probability $\mathfrak{p}_{1319}$ for the variant seq 1319: as input, we have used all data points in the green shaded region (\emph{i.e.} up to week 184): notice that this value is 2 weeks before $w_{11}$. Based on the input data in Figure~\ref{Fig:IndicatorsH}, we have calculated a fitted auxiliary metric $\mathfrak{g}_{tt}$, using a 4-parameter function of the form $\mathfrak{g}_{tt}(p)\sim a\,p^b\,(p_2-p)^c$, with $a,b,c,p_2$ free parameters. Substituting this metric into the flow equation (\ref{FlowEquation}) (with initial values imposed at $n=184$) we find as solution the black line in the right panel of Figure~\ref{Fig:IndicatorsH}. The coloured bands show corrections to the solution, assuming $\pm10\%$, $\pm 25\%$ and $\pm 40\%$ error of the initial input data. These need to be contrasted with the uncertainties (represented as error bars) for the (measured) probabilities $\mathfrak{p}_{1319}$ (represented by the yellow dots in Figure~\ref{Fig:IndicatorsH}) based on the relative number of discarded incomplete sequences in each week $n$.

\begin{figure}[htbp]
\begin{center}
\includegraphics[width=7.5cm]{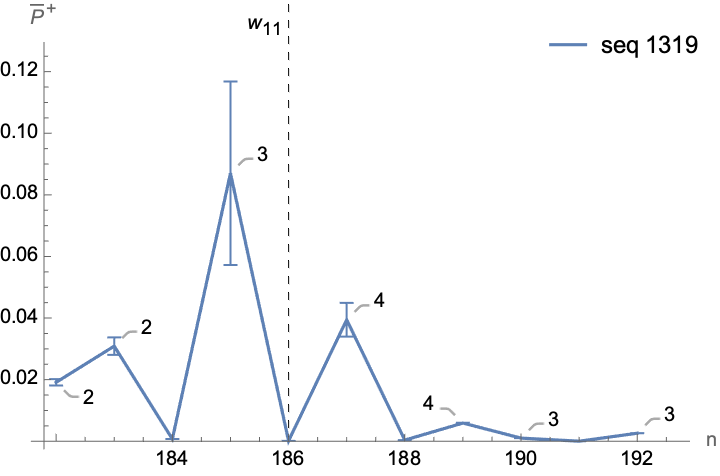}\hspace{1cm}\includegraphics[width=7.5cm]{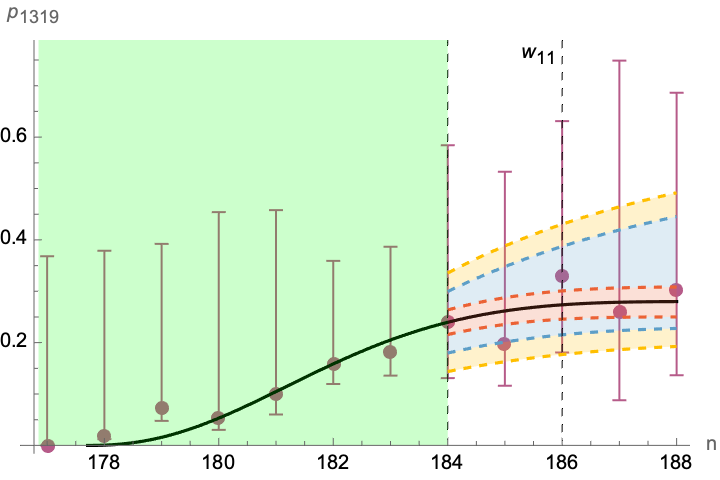}
\end{center}
\caption{Left panel: Average probabilities (\ref{AverageProbPositive}) for seq 1319 to be part of a cluster with diminishing information as a function of $n$, as well as the number of such clusters for which the same sequence has a non-vanishing probability to be a part of. Right panel: Prediction of the probability $\mathfrak{p}_{1319}$: the data points represent the actual values of the probability of seq 1319 along with uncertainties based on the number of discarded incomplete sequences, while the black line shows a solution of (\ref{FlowEquation}) using the data in the green shaded region as input. The coloured bands show deviations of the prediction taking into account uncertainties of the input data.}
\label{Fig:IndicatorsH}
\end{figure}

\subsubsection{Summary and General Remarks}\label{Sect:Summary}
In the previous Subsubsections we have discussed 8 time periods, during each of which a particularly dangerous variant (in terms of the reached probabilities) have appeared and risen to dominance (in the sense of becoming the largest variant in circulation). Our analysis using information theoretical methods and tools, has highlighted a number of common features, but has also revealed certain differences between the 8 cases, which reflect virological and epidemiological properties of each variant.

The analysis of the divergences (\ref{AlphaDiv}) between clusters with a negative derivative of the information and the largest (\emph{i.e.} dominant) cluster, highlights systematically positions in the spike protein sequence of large differences. These correctly capture the mutations of the new dominant variant, compared to the previous ones (see the red points in Figures~\ref{Fig:FullTimeAlphaFrance}, \ref{Fig:FullTimeDeltaFrance}, \ref{Fig:FullTimeBA1France}, \ref{Fig:FullTimeBA2France}, \ref{Fig:FullTimeBA5France}, \ref{Fig:FullTimeBQ11France}, \ref{Fig:FullTimeXBB15France} and \ref{Fig:FullTime1319France}). How far in advance (\emph{i.e.} before the new variant becomes dominant) these mutations are highlighted, depends on the variant under consideration: in the cases of B.1.1.7 (Alpha) and B.1.167 (Delta), (most of) the relevant sequence positions exhibit large divergences only very briefly before the change of the dominant cluster (see Figures~\ref{Fig:FullTimeAlphaFrance} and \ref{Fig:FullTimeDeltaFrance}). This is compatible with the suggested epidemiological history of these variants, \emph{e.g.} having developed in a chronically infected (potentially immunocompromised) patient (Alpha \cite{10.1093/ve/veac080}) outside of France and then spread throughout the population. In contrast, many of the Omicron variants, \emph{e.g.} BA.2 (see Figure~\ref{Fig:FullTimeBA2France}) or BA.5 (see Figure~\ref{Fig:FullTimeBA5France}), exhibit large divergences in the relevant sequence positions already weeks prior to becoming dominant, due to mutations carried by other variants. This is compatible with the suggestion \cite{ReviewDevelop} that Omicron variants tend to accumulate advantageous mutations over a longer period of time. These observations show that the information theoretic analysis of the data can help to systematically study the evolution and relevance of mutations over long periods of time.

Studying the summed divergences (\ref{TotalDivergencesWeek}) (see Figures~\ref{Fig:AlphaSummedDivergences}, \ref{Fig:DeltaSummedDivergences}, \ref{Fig:BA1SummedDivergences}, \ref{Fig:BA2SummedDivergences}, \ref{Fig:BA5SummedDivergences}, \ref{Fig:BQ11SummedDivergences}, \ref{Fig:XBB15SummedDivergences} and \ref{Fig:1319SummedDivergences}) shows generally extrema when the structure of the clusters changes. The size and sharpness of these maxima varies from case to case. Pronounced and large extrema indicate a strong change between the largest cluster and growing clusters and thus generally represent the appearance of a new dominant variant. However, also the appearance of variants that do not become dominant (but still reach sizeable probabilities) can cause local maxima (see \emph{e.g.} Figure~\ref{Fig:AlphaSummedDivergences}). Nevertheless, the quantity (\ref{TotalDivergencesWeek}) is useful to detect shifts in the landscape of sequences and thus adaptations of the virus to its environment, in particular when considered in combination with other quantities. 

Another such quantity which we have shown to be capable of estimating the potential of a sequence to reach large probabilities, is the average probability (\ref{AverageProbPositive}) to be part of a growing cluster. As is evident from Figures~\ref{Fig:IndicatorsA}, \ref{Fig:IndicatorsB}, \ref{Fig:IndicatorsC}, \ref{Fig:IndicatorsD}, \ref{Fig:IndicatorsE}, \ref{Fig:IndicatorsF}, \ref{Fig:IndicatorsG} and \ref{Fig:IndicatorsH}, for sequences that reach high probabilities, not only the average probability (\ref{AverageProbPositive}) is high (\emph{i.e.} generally $>0.2$), but also the number of growing clusters for which a dangerous variant has non-trivial probability to be part of is in general $\geq 2$ for a longer period of time. The only exceptions to this observation seem to the the variants BA.1 and BA.1.1 (see Figure~\ref{Fig:IndicatorsG}), for which we have, however, identified reasons related to the quality of the available data (see further down below).

Once a variant with large growth potential has been identified, we have provided a mechanism to predict its growth, based on the universal model proposed in \cite{Filoche:2024xka} and briefly reviewed in Section~\ref{Sect:TheorySpreadDangerous}. Our examples (see Figures~\ref{Fig:PredictionTimeAlphaFrance}, \ref{Fig:IndicatorsB}, \ref{Fig:IndicatorsC}, \ref{Fig:IndicatorsD}, \ref{Fig:IndicatorsE}, \ref{Fig:IndicatorsF}, \ref{Fig:PredictionTimeXBB15France} and \ref{Fig:IndicatorsG}) show that this model not only describes the data well, but is also capable of predicting the future evolution of (dominant) variants, in particular when taking into account the various systematic errors inherent in the data. A key quantity in this context is the point in time (either relative to the first appearance of a new variant within the data or with respect to the point in time when the variant reaches its maximal probability) at which an accurate prediction is possible. For our examples we have chosen timings that provide good estimates of the future evolution. They show that around 4-7 weeks of data are required (after the first appearance of the variant within the dataset), by which time the variant has reached a probability between 0.05 to 0.4, while predictions based on less data usually lead to much larger uncertainties. The determining factor, for this timing, however, is the the quality of the available data, \emph{i.e.} notably the ratio of complete sequences among all available data and notably whether or not they correctly represent the epidemiological and virological situation in the population. In all plots, we have indicated the impact of uncertainties and fluctuations of the input data on the predictions.

\subsection{Correlation of Mutations}
After having discussed the evolution of point mutations of the spike protein sequence in various different time periods, we shall now discuss correlations among such mutations, generalising the discussion of Section~\ref{Sect:ShowcaseCorrelations} to the entire pandemic. Analysing the entire time period for positions $\mu,\nu\in\{1,\ldots,L\}$ that satisfy the condition (\ref{CondCorrelation}), but still show a certain amount of mutations, we find a number of strong correlations among (multiple) positions: 
\begin{enumerate}
\item $\mu\in\{74,75\}$, which correspond to the position of the mutations H69del and V70del (with respect to the Wuhan variant). As in Section~\ref{Sect:ShowcaseCorrelations}, we find a very strong correlation between the deletions in these two adjacent protein positions almost during the entire pandemic: out of the 1474 sequences considered in the analysis, 568 have gaps -- and 895 have the amino acids HV in the positions 74 and 75. Only 7 sequences have other amino acids  and only 4 have the combination F--. The combined probability of these 11 sequences reaches a maximum of 0.0421 in week 61. We also remark that the different positions compared to before (\emph{i.e.} $\mu\in\{74,75\}$ here and $\mu\in\{69,70\}$ in Section~\ref{Sect:ShowcaseCorrelations}) are due to a different alignment of all sequences. This in turn is due to the fact that we are considering a much larger set of sequences here than before.
\item $\mu\in\{29, 30, 31, 32,  392, 421\}$, which correspond to the positions of the mutations L24del, P25del, P26del, A27S, T376A and D405N. The 1474 sequences only have 11 possible combinations of amino acids in these positions, most of which fall into the following two groups
\begin{center}
\begin{tabular}{|l||c|c|c|c|c|c|}\hline
&&&&&&\\[-12pt]
 num. of seq. & 29 & 30 & 31 & 32 & 392 & 421 \\[2pt]\hline\hline
&&&&&&\\[-12pt]
767 & S & --- & --- & --- & A & N \\[2pt]\hline
&&&&&&\\[-12pt]
 679 & L & P & P & A & T & D \\[2pt]\hline
\end{tabular}
\end{center}


\noindent
Only 22 remaining sequences (all of which reaching only small probabilities) have slightly different amino acid combinations, such that the set of sequences is essentially divided into two by these combinations. 
This shows a quite strong correlation among the corresponding sequence positions. We also find a certain correlation of the positions $\mu\in\{29, 30, 31, 32,  392, 421\}$ with $\mu=220$, the latter corresponding to the location of the mutation V213G: In this case, the set of sequences falls into three groups, carrying the combinations $\{\text{S},-,-,-,\text{A},\text{N},\text{G}\}$, $\{\text{S},-,-,-,\text{A},\text{N},\text{E}\}$, $\{\text{L},\text{P},\text{P},\text{A},\text{T},\text{D},\text{V}\}$ of amino acids. The combined probabilities of sequences carrying these combinations is plotted in Figure~\ref{Fig:CombinedCorrelation220} and show a clear temporal evolution.

\item $\mu\in\{88,152,189,384\}$, which correspond to the positions of the mutations V83I, H146del, Q183E and L368I. The 1474 sequences only have 11 possible combinations of amino acids in these positions, who are mainly distributed as follows:

\begin{center}
\begin{tabular}{|l||c|c|c|c|}\hline
&&&&\\[-12pt]
 num. of seq. & 88 & 152 & 189 & 384 \\[2pt]\hline
&&&&\\[-12pt]
1085 & V & H & Q & L \\[2pt]\hline
&&&&\\[-12pt]
372 & A & --- & E & I \\[2pt]\hline
\end{tabular}
\end{center}

\noindent 


\noindent
and only 17 sequences have different combinations of amino acids. The set of sequences is therefore divided into two groups,
thus showing quite strong correlation among these sequence positions. The combined probabilities of sequences carrying these combinations is plotted in Figure~\ref{Fig:CombinedCorrelation220} and again show a clear temporal evolution.

\item $\mu\in\{148, 149, 150, 218, 514, 565, 874, 999\}$, which correspond to the positions of the mutations V143del, Y144del, Y145del, N211del, L212I, G496S, T547K, N856K and L981F. The 1474 sequences have 24 possible combinations of amino acids in these positions, however, they are mainly distributed as follows:

\begin{center}
\begin{tabular}{|l||c|c|c|c|c|c|c|c|}\hline
&&&&&&&&\\[-12pt]
 num. of seq. & 148 & 149 & 150 & 218 & 514 & 565 & 874 & 999 \\[2pt]\hline
&&&&&&&&\\[-12pt]
732 & V & Y & Y & N & G & T & N & L \\[2pt]\hline
&&&&&&&&\\[-12pt]
344 & V & Y & Q & N & G & T & N & L \\[2pt]\hline
&&&&&&&&\\[-12pt]
314 & V & Y & --- & I & S & K & K & F \\[2pt]\hline
&&&&&&&&\\[-12pt]
24 & --- & --- & --- & I & S & K & K & F \\[2pt]\hline
\end{tabular}
\end{center}
 The remaining 60 sequences only reach small probabilities during the entire time period. The combined probabilities of sequences carrying the above combinations of amino acids is plotted in Figure~\ref{Fig:CombinedCorrelation220} and again shows a clear temporal evolution.

\item $\mu\in\{162,163,164,519,699,968\}$, which correspond to the positions of the mutations E156G, F157del, R158del,  N501Y, P681R, and D950N. The 1474 sequences have 18 possible combinations of amino acids in these positions, however, they are mainly distributed as follows:

\begin{center}
\begin{tabular}{|l||c|c|c|c|c|c|}\hline
&&&&&&\\[-12pt]
 num. of seq. & 162 & 163 & 164 & 519 & 699 & 968  \\[2pt]\hline
&&&&&&\\[-12pt]
933 & E & F & R & Y & H & D  \\[2pt]\hline
&&&&&&\\[-12pt]
211 & G & --- & --- & N & R & N  \\[2pt]\hline
&&&&&&\\[-12pt]
189 & E & F & R & N & P & D  \\[2pt]\hline
&&&&&&\\[-12pt]
75 & E & L & R & Y & H & D \\[2pt]\hline
\end{tabular}
\end{center}
 The remaining 66 sequences only reach small probabilities during the entire time period. The combined probabilities of sequences carrying the above combinations of amino acids is plotted in Figure~\ref{Fig:CombinedCorrelation220}.

\item $\mu\in\{355,389,391,516,523,697,782,972,987\}$, which correspond to the positions of the mutations G339D, S373P, Q498R, Y505H, N679K, N764K, Q954H and N969K. The 1474 sequences have 12 possible combinations of amino acids in these positions, however, they are mainly distributed as follows:

\begin{center}
\begin{tabular}{|l||c|c|c|c|c|c|c|c|c|}\hline
&&&&&&&&&\\[-12pt]
 num. of seq. & 355 & 389 & 391 & 516 & 523 & 697 & 782 & 972 & 987  \\[2pt]\hline
&&&&&&&&&\\[-12pt]
675 & G & S & S & Q & Y & N & N & Q & N  \\[2pt]\hline
&&&&&&&&&\\[-12pt]
434 & H & P & F & R & H & K & K & H & K  \\[2pt]\hline
&&&&&&&&&\\[-12pt]
353 & D & P & F & R & H & K & K & H & K   \\[2pt]\hline
\end{tabular}
\end{center}
 The remaining 12 sequences only reach small probabilities during the entire time period. The combined probabilities of sequences carrying the above combinations of amino acids is plotted in Figure~\ref{Fig:CombinedCorrelation220}.

\item $\mu\in\{588,1000,1136\}$, which correspond to the positions of the mutations A570D, S982A and D1118H. The 1474 sequences have 5 possible combinations of amino acids in these positions, however, they are mainly distributed as follows:

\begin{center}
\begin{tabular}{|l||c|c|c|}\hline
&&&\\[-12pt]
 num. of seq. & 588 & 1000 & 1136   \\[2pt]\hline
&&&\\[-12pt]
1238 & A & S & D \\[2pt]\hline
&&&\\[-12pt]
229 & D & A & H \\[2pt]\hline
\end{tabular}
\end{center}
 The remaining 7 sequences only reach small probabilities during the entire time period. The combined probabilities of sequences carrying the above combinations of amino acids is plotted in Figure~\ref{Fig:CombinedCorrelation220}.

\end{enumerate}

\begin{figure}[hp]
\begin{center}
\includegraphics[width=7.5cm]{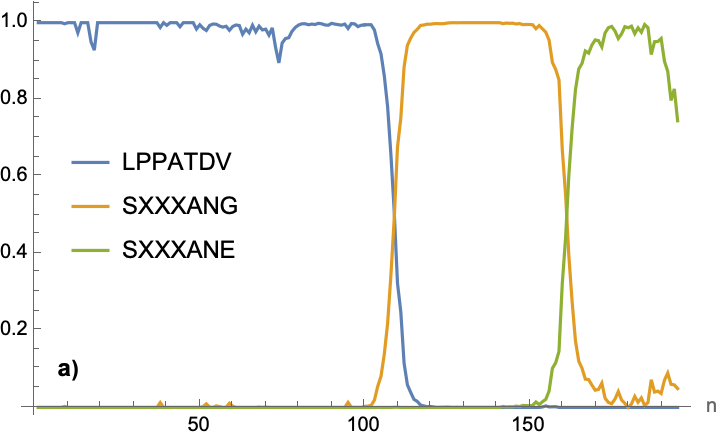}\hspace{1cm}\includegraphics[width=7.5cm]{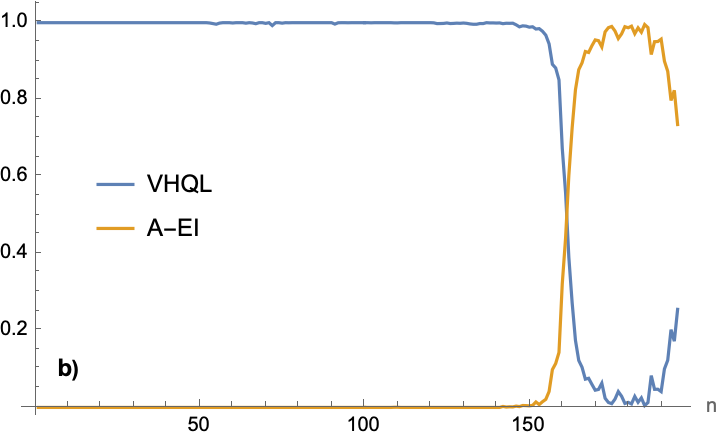}
\includegraphics[width=7.5cm]{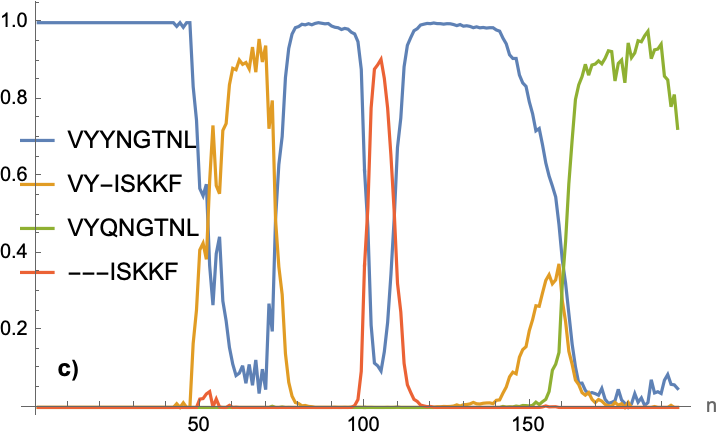}\hspace{1cm}\includegraphics[width=7.5cm]{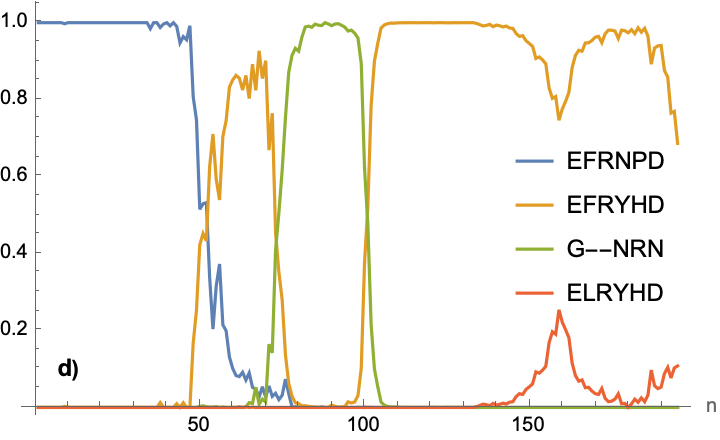}
\includegraphics[width=7.5cm]{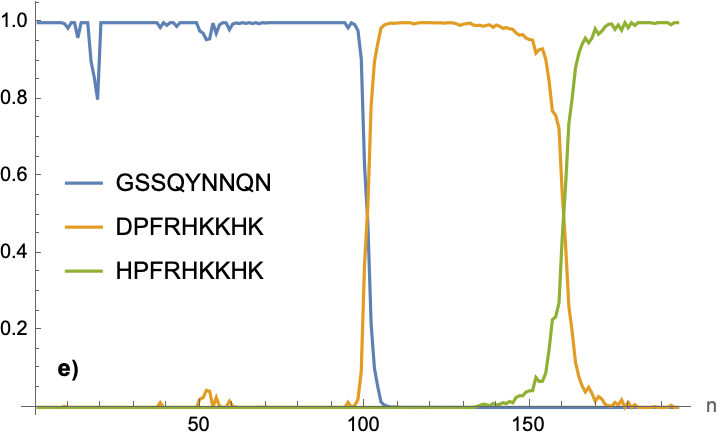}\hspace{1cm}\includegraphics[width=7.5cm]{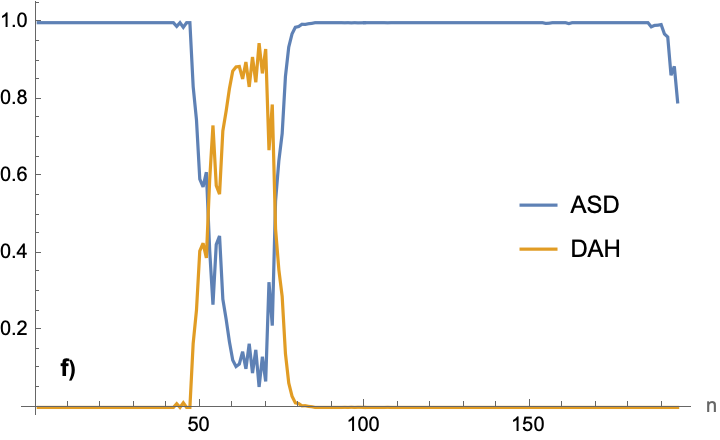}
\end{center}
\caption{{\bf{Panel a)}}: Combined probabilities of sequences carrying the amino acid combinations $\{\text{S},-,-,-,\text{A},\text{N},\text{G}\}$, $\{\text{S},-,-,-,\text{A},\text{N},\text{E}\}$, $\{\text{L},\text{P},\text{P},\text{A},\text{T},\text{D},\text{V}\}$ in positions $\{29, 30, 31, 32,  392, 421,220\}$. {\bf Panel  b)}: Combined probabilities of sequences carrying the amino acid combinations  $\{\text{V,H,Q,L}\}$ and $\{\text{A},-,\text{EI}\}$ in positions $\{88,152,189,384\}$. {\bf{Panel c)}}: Combined probabilities of sequences carrying the amino acid combinations $\{\text{V,Y,Y,N,G,T,N,L}\}$, $\{\text{V,Y},-,\text{I,S,K,F,F}\}$,  $\{\text{V,Y,Q,N,G,T,N,L}\}$ and  $\{-,-,-,\text{I,S,K,K,F}\}$ in positions $\{29, 30, 31, 32,  392, 421\}$. {\bf{Panel d)}}: Combined probabilities of sequences carrying the amino acid combinations $\{\text{E,F,R,Y,H,D}\}$, $\{\text{G,},-,-,\text{N,R,N}\}$,  $\{\text{E,F,R,N,P,D}\}$ and  $\{\text{E,L,R,Y,H,D}\}$ in positions $\{162,163,164,519,699,968\}$. {\bf{Panel e)}}: Combined probabilities of sequences carrying the amino acid combinations $\{\text{G,S,S,Q,Y,N,N,Q,N}\}$, $\{\text{H,P,F,R,H,K,K,H,K}\}$,  and  $\{\text{D,P,F,R,H,K,K,H,K}\}$ in positions $\{355,389,391,516,523,697,782,972,987\}$. {\bf{Panel f)}}: Combined probabilities of sequences carrying the amino acid combinations $\{\text{A,S,D}\}$, and  $\{\text{D,A,H}\}$ in positions $\{588,1000,1136\}$.}
\label{Fig:CombinedCorrelation220}
\end{figure}

\noindent
These examples highlight the capability of our information theoretic approach to detect and track correlations among point mutations. We expect this to be useful in the future in order to understand the interplay between changes in the protein structure which in turn may be useful to better understand their functionality, notably to capture their three-dimensional structure. 


\section{Conclusions}\label{Sect:Conclusions}
In this paper we have analysed the clustering of epidemiological and virological data from an information theoretic perspective. Concretely, we have considered scenarios in which $n$ different variants of a given pathogen (in concrete examples SARS-CoV-2) circulate in a population. We have interpreted the relative abundance of each variant as as a time-dependent probability distribution $p_i(t)$ (for $i=1,\ldots,n$). In order to condense the information contained in these probabilities, we have considered clusterings as surjective maps $f:\,\mathbb{V}=\{1,\ldots,n\}\to \{\mathbb{A}_1,\ldots,\mathbb{A}_\ell\}$ with $\mathbb{A}_a$ a set of disjoint subsets of $\mathbb{V}$ (with $\cup_{a=1}^\ell \mathbb{A}_a=\mathbb{V}$). Based on the initial probabilities $p_i(t)$, we can also associate a probability distribution to the clusters, as defined in (\ref{ClusterProbabilities}). In order for the clusters to encode the same information about the time evolution of the variants of the pathogen, we impose that $f$ is (approximately) a sufficient statistic, \emph{i.e.} leaves the Fisher information metric (at least approximately) invariant (see (\ref{ClusterMetric})). We have analysed this condition for simple compartmental models with multiple variants in Section~\ref{Sect:CompartmentalComputations}, and have demonstrated that it can be realised by grouping together variants according to a simple quantity that characterises how they couple to the full dynamical system (see (\ref{ClusteringCondition}) or (\ref{ClusterInfoVaccines}) in the case of simple examples). Although these so-called couplings do not only depend on the probability distribution (and thus would require to solve the complete dynamics of the entire system), we have shown that the same grouping can be achieved by clustering the variants according to the time-derivative of their information $\mathfrak{I}_{p_i}$ as defined in (\ref{InfoDef}) and which is entirely defined through the probability distribution. This opens the possibility to probe specific interactions of each variant with its environment through a calculationally and conceptually simple clustering of epidemiological data.

We have validated this idea by studying the the evolution  of SARS-CoV-2 in France in the period of 21/01/2020--23/10/2023. For simplicity we have represented variants of this virus through amino acid sequences of its spike protein. We have downloaded the relevant sequencing data from GISAID and pruned them in order to define probability distributions for each week $n$ and calculated a (discretised) derivative of the information for each sequence (see (\ref{DiscDerivative})). Furthermore, we have characterised each cluster as a set of probability distributions for the amino acids at each position of the protein sequence (see (\ref{PropDistributionsCluster})). This allows us to define differences between two clusters at a given protein sequence position in the form of divergences (\ref{AlphaDiv}) as well as an average probability for a variant to be part of a specific cluster (\ref{ProbabilityGenetic}). Using these tools, we have in Section~\ref{Sect:FrancLongTerm} systematically analysed 8 different time periods in which new and dangerous variants of SARS-CoV-2 appear. Our concise summary can be found in Section~\ref{Sect:Summary} and comprises three major points
\begin{itemize}
\item[\emph{(i)}] our algorithm highlights and tracks the locations of strong changes in the amino acid sequences, which coincide with mutations that provide competitive advantages for new variants.

\item[\emph{(ii)}] The average probability of a variant to be part of a growing cluster is an important indicator to gauge the potential of a new variant to reach large probabilities and become dominant. Once such a variant has been identified, we can use the universal description proposed in \cite{Filoche:2024xka} to predict its further development: in the cases studied, such predictions work better than simple fits with logistic functions and can thus complement other effective prediction models, \emph{e.g.}~\cite{DellaMorte:2020wlc,DellaMorte:2020qry,Cacciapaglia:2020mjf,cacciapaglia2020second,cacciapaglia2020mining,cacciapaglia2020evidence,cacciapaglia2020multiwave,cacciapaglia2020better,cacciapaglia2020us,MeRG}.

\item[\emph{(iii)}] By comparing dissimilarities between the probability distributions that characterise the clustering, we can deduce correlations among the point mutations.
\end{itemize}
In this work we have used pruned data of SARS-CoV-2 as a proof of concept for our information theoretic treatment of clustering. We have demonstrated the high potential inherent in this approach, which can be generalised and extended in various different manners: First of all, even continuing to study SARS-CoV-2, the algorithm can be turned into a full diagnostic tool to track, monitor and predict (at least at short term) the evolution of the virus. To this end, the full genome can be studied to include all mutations and the pruning of the data can be removed. Furthermore, our approach can straight-forwardly be extended to other pathogens, provided sufficient data of quality exist. 

Furthermore, in the practical applications in our work, we have only considered probability distributions based on the number of (newly) infected individuals each week. With more refined data, other probability distributions can be studied in the same manner. Notably, the associated clusterings can provide valuable insights into various aspects of the interaction between the pathogen and individuals in the population. For example, a clustering of the probabilities of patients recovering or dying from the pathogen (based on clinical data) could provide information about the severity of new variants. The combined study of the clusterings of several different probability distributions covering the same time period can therefore be used to better understand aspects of the infectivity and severity of pathogens, notably as functions of external factors, such as vaccinations or non-pharmaceutical interventions. Our approach would provide a direct link to the microbiological structure of different variants thus promising further insights in virology and bacteriology.

Finally, we also envision that our approach can be useful for the analysis of the viral load in waste-water analyses. Indeed, the probability for a given variant to be part of a growing cluster (\ref{ProbabilityGenetic}), also works for partial sequences (with a certain ambiguity). This could help to estimate the danger (in an epidemiological sense) of samples extracted from sewage analyses and could therefore further improve the effectiveness of this monitoring tehcnique.

\section*{Acknowledgements}
We thank Francesco Sannino for collaboration at the early stages of this work and many valuable discussions on related topics. SH is furthermore grateful to Francesco Cirotto for invaluable help with extracting data from GISAID. We also thank Bruno Buonomo, Mattia Carrino, Andrea Cimarelli, Marika D'Avanzo and Marta Nunes for useful discussions. We are grateful for support and hospitality from the Quantum Theory Center ($\hbar$QTC) at the University of Southern Denmark where part of this work was performed.
\appendix

\section{Clustering Algorithms}\label{App:OneDimClustering}
For completeness, we briefly describe in this appendix two clustering algorithms that are used throughout the main body of this text. These are the k-means algorithm for clustering real numbers and an algorithm advocated in \cite{MLvariants} (based on the Ward-distance and the Levenshtein distance \cite{Levenshtein1965BinaryCC}) to cluster spike-protein sequences of SARS-CoV-2.
\subsection{k-means Algorithm}
When clustering sets of real numbers (\emph{e.g.} the time derivatives of the informations associated with different sequences), we mostly relay on the \emph{k-means} algorithm, first described in \cite{MR90073,MR214227}, with an algorithmic implementation provided in \cite{Lloyd,Forgy}. Let $\mathbb{V}=\{x_1,\ldots,x_N\}$ be a set of $N\in\mathbb{N}$ real numbers\footnote{The k-means algorithm can more generally be applied to vectors in a $d$-dimensional vector space. Here we shall only focus on the case $d=1$, which is relevant for our work.} that are clustered into $k\leq N$ sets $\{\mathbb{A}_1,\ldots,\mathbb{A}_k\}$, that satisfy similar conditions as in (\ref{DefClusteringGen}).
The goal of the algorithm is to minimise the \emph{within-cluster sum of squares}, \emph{i.e.} to minimise
\begin{align}
&\sum_{a=1}^k\sum_{x\in\mathbb{A}_a}|x-\mu_a|^2\,,&&\text{with} &&\mu_a=\frac{1}{||\mathbb{A}_a||}\sum_{x\in\mathbb{A}_a} x\,,\label{kmeansMinimise}
\end{align}
where $||\mathbb{A}_a||$ is the number of elements in the set $\mathbb{A}_a$. The condition (\ref{kmeansMinimise}) can equivalently be formulated to minimise the differences of elements within any cluster
\begin{align}
\sum_{a=1}^k \frac{1}{||\mathbb{A}_a||} \sum_{x,y\in\mathbb{A}_a}|x-y|^2\,.
\end{align}
For concrete applications, we shall use the implementation of this algorithm within Mathematica, which allows to specify the maximal number of clusters $k$ beforehand.  

\subsection{Genetic Clustering of Sequences}\label{Sect:GeneticClustering}
An algorithm that is designed to cluster spike-protein sequences in a given week $n$, based on their genetic (dis)similarity was developed in \cite{MLvariants}. In this work, we shall use a very similar (but simplified) clustering method to compare with our results (see left panel of Figure~\ref{Fig:OtherClusterVI}). Using the same notation as in Section~\ref{Sect:SeqMethodology} we consider a set of $N$ variants $\mathbb{V}=\{1,\ldots,N\}$ that we cluster into $k$ sets $\{\mathbb{A}_1,\ldots,\mathbb{A}_\ell\}$ satisfying (\ref{DefClusteringGen}). Let
\begin{align}
d_{\text{Lev}}:\,\mathbb{V}\times \mathbb{V}\longrightarrow \mathbb{N}^*\,,
\end{align}
be the \emph{Levenshtein distance} \cite{Levenshtein1965BinaryCC} between two amino acid sequences: we refer to \cite{Levenshtein1965BinaryCC} for the concise definition, and only remark that it corresponds to the minimal number of single-character edits (\emph{i.e.} insertions, deletions or substitutions) to transform one sequence into another. $d_{\text{Lev}}$ is hence also called \emph{edit distance}. Furthermore, for a fixed week $n$, we also define the Ward distance between two clusters as \cite{WardDist}
\begin{align}
\text{dis}(\mathbb{A}_a,\mathbb{A}_b)&=\frac{|\mathbb{A}_a| |\mathbb{A}_b|}{|\mathbb{A}_a|+|\mathbb{A}_b|}\,\bigg[\sum_{{i\in \mathbb{A}_a}\atop{j\in \mathbb{A}_b}}\mathfrak{p}_i(n)\,\mathfrak{p}_j(n)\frac{d_{\text{Lev}}(i,j)^2}{|\mathbb{A}_a|\,|\mathbb{A}_b|}-\sum_{i,j\in \mathbb{A}_a}\mathfrak{p}_i(n)\,\mathfrak{p}_j(n)\,\frac{d_\text{Lev}(i,j)^2}{2|\mathbb{A}_a|^2}\nonumber\\
&\hspace{3cm}-\sum_{i,j\in \mathbb{A}_b}\mathfrak{p}_i(n)\,\mathfrak{p}_j(n)\,\frac{d_\text{Lev}(i,j)^2}{2|\mathbb{A}_b|^2}\bigg]\,.\label{DefWardDistance}
\end{align}
where $|\mathbb{A}_a|$ is defined in (\ref{DefNormsClusters}) and $\mathfrak{p}_i(n)$ are the probabilities introduced in (\ref{DiscDerivative}). Based on these definitions, for each week $n$, the following algorithm was developed in \cite{MLvariants} to group the distinct sequences of $\mathbb{V}$ into clusters 
\begin{enumerate}
\item As a first step, each sequence of $\mathbb{V}$ is assigned into a separate cluster. 
\item Using (\ref{DefWardDistance}), the distances between all clusters are calculated.
\item Among these $N$ clusters $\mathbb{A}_{a=1,\ldots,N}$,  the clusters $\mathbb{A}_a$ and $\mathbb{A}_b$ are combined into a new cluster~if
\begin{align}
&\text{\emph{(i)}}&& 0\leq \text{dis}(\mathbb{A}_a,\mathbb{A}_b)<\text{dis}(\mathbb{A}_{a'},\mathbb{A}_{b'})\,,&&\forall (a',b')\in \{1,\ldots,N\}^2\text{ with }(a',b')\neq (a,b)\neq (b',a')\,,\nonumber\\
&\text{\emph{(ii)}}&&0\leq \text{dis}(\mathbb{A}_a,\mathbb{A}_b)<d_{\text{cutoff}}\,.\label{FusionConditions}
\end{align}
Here $d_{\text{cutoff}}$ is a cutoff distance, which can be chosen as an input for the algorithm.
\item The previous step is repeated for the new set of $N-1$ clusters, until condition \emph{(ii)} is not satisfied anymore.
\item Of the remaining $M\leq N$ clusters, all clusters with $|\mathbb{A}_a|<d_{\text{thresh}}$ are discarded, where the threshold $d_{\text{thresh}}$ is a further input that can be chosen for the algorithm.
\end{enumerate}
As a modification of this algorithm, instead of using a cut-off $d_{\text{cutoff}}$ (and discarding clusters below the threshold $d_{\text{thresh}}$), we shall perform the algorithm until the total number of clusters has reached a pre-determined value $1\leq \ell\leq N$.

\section{Largest Divergences in France (Spring/Summer 2022)}\label{App:TableAlphaDivergences}
In order to document the evolution of the divergences (\ref{AlphaDiv}) for the example period discussed in Section~\ref{Sect:CaseStudyFrance}, we tabulate below all $\mathcal{D}(q_\mu[\mathbb{A}_a,n]||q_\mu[\mathbb{A}_b,n])$ (along with their positions $\mu$ on the spike-protein sequence) that are larger than $0.5$ during the weeks $n\in\{109,\ldots,121\}$. For concreteness, we use the value $\alpha=3/4$ and for further visibility, we have highlighted the divergences in the positions (\ref{Mutations238}) in blue and the clusters that contain the variant BA.5 in brown. A tallying of all divergences is shown in the left panel of Figure~\ref{Fig:FranceDivergencesSummed}, indicating that indeed the largest divergences occur in the positions (\ref{Mutations238}):

\begin{center}
\begin{tabular}{|c||c|c||c|c||c|}\hline
&&&&&\\[-10pt]
{\bf week} & $|\mathbb{A}_a|$ & $-\overline{|\mathbb{A}_a|}$ & $|\mathbb{A}_b|$ & $-\overline{|\mathbb{A}_b|}$ &  largest divergences $\left(\mu,\mathcal{D}(q_\mu[\mathbb{A}_a,n]||q_\mu[\mathbb{A}_b,n])\right)$\\[2pt]\hline\hline
&&&&&\\[-10pt]
109 & 0.75 & 0.054 & 0.0002 & 1.859 & \parbox{9cm}{\footnotesize \hma{(458,5)}, (345, 3.33), (377, 3.33), (379, 3.33), (381, 3.33), (382, 3.33),
(400, 3.33), (1265, 1.04)}\\[8pt]\hline
&&&&&\\[-10pt]
 & & & \hmb{0.004} & \hmb{0.674} & \parbox{9cm}{\footnotesize \hma{(458, 6.52)}, (377, 0.96), (423, 0.61)}\\[2pt]\hline
 &&&&&\\[-10pt]
 & & & 0.019 & 0.247 & \parbox{9cm}{\footnotesize (5, 2.04), (68, 1.88), (64, 1.48), (212, 0.53)}\\[2pt]\hline\hline
 &&&&&\\[-10pt]
 110 & 0.799 & 0.056 & 0.00004 & 1.632 & \parbox{9cm}{\footnotesize \hma{(458, 9.14)}, \hma{(69, 6.34)}, \hma{(70, 6.34)}, \hma{(492, 6.34)}, \hma{(499, 6.34)}, (3, 5.53), (352, 0.7), (710, 0.63)}\\[8pt]\hline
 &&&&&\\[-10pt]
 & & & \hmb{0.0016} & \hmb{0.836} & \parbox{9cm}{\footnotesize (423, 3.78), \hma{(458, 2.62)}, (345, 0.94), (377, 0.94), (379, 0.94), (381, 0.94), (382, 0.94), (400, 0.94), (213, 0.86),  (214, 0.86), (215, 0.86)}\\[12pt]\hline
 &&&&&\\[-10pt]
 & & & 0.0237 & 0.346 &\parbox{9cm}{\footnotesize (68, 1.88), \hma{(458, 1.73)}, (64, 1.59), (212, 0.54)}\\[2pt]\hline\hline
 &&&&&\\[-10pt]
 111 & 0.842 & 0.045 & 0.000027 & 2.00 & \parbox{9cm}{\footnotesize \hma{(458, 9.14)}, (944, 5.27), \hma{(69, 3.08)}, \hma{(70, 3.08)}, \hma{(492, 3.08)}, \hma{(499, 3.08)}, (664, 1.01), (5, 0.57)}\\[8pt]\hline
 &&&&&\\[-10pt]
 & & & \hmb{0.0035} & \hmb{0.649} & \parbox{9cm}{\footnotesize (423, 3.41), \hma{(458, 3.15)}, (345, 0.89), (377, 0.89), (379, 0.89), (381, 0.89), (382, 0.89), (400, 0.89), (213, 0.75), (214, 0.75), (215, 0.75), \hma{(69, 0.51)}, \hma{(70, 0.51)}, \hma{(492, 0.51)}, \hma{(499, 0.51)}}\\[12pt]\hline
 &&&&&\\[-10pt]
 & & & 0.0305 & 0.329 & \parbox{9cm}{\footnotesize \hma{(458, 2.04)}, (68, 1.94), (64, 0.83), (212, 0.57)}\\[2pt]\hline\hline
  &&&&&\\[-10pt]
 112 & 0.855 & 0.022 & 0.00028 & 1.233 &  \parbox{9cm}{\footnotesize \hma{(458, 9.14)}, (944, 2.62), \hma{(69, 1.69), (70, 1.69), (492, 1.69), (499, 1.69)}, (710, 1.42)}\\[8pt]\hline
 &&&&&\\[-10pt]
 & & & \hmb{0.001} & \hmb{0.713} & \parbox{9cm}{\footnotesize \hma{(458, 9.14)}, \hma{(69, 3.51)}, \hma{(70, 3.51)}, \hma{(492, 3.51)}, \hma{(499, 3.51)}, (3, 1.01)}\\[8pt]\hline
  &&&&&\\[-10pt]
 & & & 0.063 & 0.290 & \parbox{9cm}{\footnotesize (1265, 2.73), \hma{(458, 1.5)}, (68, 1.24), (64, 0.51)}\\[2pt]\hline\hline
&&&&&\\[-10pt]
 113 & 0.948  & 0.025 & 0.000039 & 2. & \parbox{9cm}{\footnotesize \hma{(492, 7.05)}, \hma{(499, 7.05)}, \hma{(69, 5.03)}, \hma{(70, 5.03)}, (76, 4.56), (664, 3.76), \hma{(458, 3.7)}}\\[8pt]\hline
 &&&&&\\[-10pt]
 & & & \hmb{0.00129} & \hmb{0.901} & \parbox{9cm}{\footnotesize \hma{(458, 9.14)}, \hma{(492, 3.51)}, \hma{(499, 3.51)}, \hma{(69, 3.51)}, \hma{(70, 3.51)}, (3, 1.01)}\\[8pt]\hline
  &&&&&\\[-10pt]
 & & & 0.00145 & 0.538 & \parbox{9cm}{\footnotesize \hma{(458, 8.62)}, (164, 0.57)}\\[2pt]\hline\hline
&&&&&\\[-10pt]
 114 & 0.844 & -0.014 & \hmb{0.004} & \hmb{0.928} & \parbox{9cm}{\footnotesize \hma{(69, 6.49)}, \hma{(70, 6.49)}, \hma{(492, 6.49)}, \hma{(499, 6.49)}, \hma{(458, 5.75)}, (710, 0.58)}\\[8pt]\hline
 &&&&&\\[-10pt]
 & & & 0.004 & 0.5238  & \parbox{9cm}{\footnotesize \hma{(458, 6.08)}, \hma{(69, 2.04)}, \hma{(70, 2.04)}, \hma{(492, 2.04)}, \hma{(499, 2.04)}, (3, 1.23), (944, 0.88), (64, 0.87), (1270, 0.72), (5, 0.69)}\\[8pt]\hline
 &&&&&\\[-10pt]
 & & & 0.049 & 0.2666 & \parbox{9cm}{\footnotesize (64, 2.56), (68, 2.16), (212, 0.9), (423, 0.8), (352, 0.76)}\\[2pt]\hline\hline
&&&&&\\[-10pt]
 115 & 0.898 & -0.021 & 0.00003 & 2. & \parbox{9cm}{\footnotesize \hma{(69, 6.47)}, \hma{(70, 6.47)}, \hma{(492, 6.47)}, \hma{(499, 6.47)}, \hma{(458, 3.58)}, (6, 1.15), (183, 1.15), (813, 1.15), (19, 0.56)}\\[8pt]\hline
 &&&&&\\[-10pt]
 & & & \hmb{0.012} & \hmb{0.8692}  & \parbox{9cm}{\footnotesize \hma{(69, 4.8)}, \hma{(70, 4.8)}, \hma{(492, 4.8)}, \hma{(499, 4.8)}, \hma{(458, 4.5)}, (3, 0.71)}\\[8pt]\hline

\end{tabular}
\end{center}
\begin{center}
\begin{tabular}{|c||c|c||c|c||c|}\hline
&&&&&\\[-10pt]
{\bf week} & $|\mathbb{A}_a|$ & $-\overline{|\mathbb{A}_a|}$ & $|\mathbb{A}_b|$ & $-\overline{|\mathbb{A}_b|}$ &  largest divergences $\left(\mu,\mathcal{D}(q_\mu[\mathbb{A}_a,n]||q_\mu[\mathbb{A}_b,n])\right)$\\[2pt]\hline\hline
&&&&&\\[-10pt]
 & & & 0.072& 0.23109  & \parbox{9cm}{\footnotesize (64, 2.27), (68, 1.99), (212, 0.78), (423, 0.68)}\\[2pt]\hline
 &&&&&\\[-10pt]
116 & 0.926 & -0.0164 & 0.00022& 1.913 & \parbox{9cm}{\footnotesize \hma{(492, 9.14)}, \hma{(499, 9.14)}, \hma{(69, 4.62)}, \hma{(70, 4.62)}, \hma{(458, 3.51)}, (183, 1.76), (1130, 1.5), (1168, 0.83), (1026, 0.66)}\\[8pt]\hline
&&&&&\\[-10pt]
 & & & \hmb{0.028} & \hmb{0.826}  & \parbox{9cm}{\footnotesize \hma{(492, 7.09)}, \hma{(499, 7.09)}, \hma{(458, 4.41)}, \hma{(69, 3.3)}, \hma{(70, 3.3)}, (3, 0.62)}\\[8pt]\hline
 &&&&&\\[-10pt]
 & & & 0.009 & 0.388  & \parbox{9cm}{\footnotesize \hma{(458, 5.08)}, \hma{(492, 1.68)}, \hma{(499, 1.68)}, (377, 1.03), \hma{(69, 0.54)}, \hma{(70, 0.54)}}\\[8pt]\hline\hline
 &&&&&\\[-10pt]
117 & 0.910 & -0.038 & 0.000117 & 2. & \parbox{9cm}{\footnotesize \hma{(492, 5.)}, \hma{(499, 5.)}, \hma{(69, 4.32)}, \hma{(70, 4.32)}, \hma{(458, 3.43)}, (295, 2.03), (1077, 2.03), (700, 1.74), (1269, 1.16)}\\[8pt]\hline
&&&&&\\[-10pt]
 & & & 0.001 & 1.163  & \parbox{9cm}{\footnotesize \hma{(492, 5.)}, \hma{(499, 5.)}, \hma{(69, 4.32)}, \hma{(70, 4.32)}, \hma{(458, 3.43)}, (183, 1.25), (1265, 1.2), (1168, 1.03), (1026, 0.7)}\\[8pt]\hline
 &&&&&\\[-10pt]
 & & & \hmb{0.064} & \hmb{0.6975}  & \parbox{9cm}{\footnotesize \hma{(492, 3.71)}, \hma{(499, 3.71)}, \hma{(458, 3.43)}, \hma{(69, 3.13)}, \hma{(70, 3.13)}, (3, 0.58)}\\[8pt]\hline\hline
 &&&&&\\[-10pt]
118 & 0.844 & -0.0738 & 0.00008 & 1.615 & \parbox{9cm}{\footnotesize \hma{(492, 4.8)}, \hma{(499, 4.8)}, \hma{(69, 4.16)}, \hma{(70, 4.16)}, \hma{(458, 3.41)}, (664, 3.32), (146, 2.91), (352, 1.37)}\\[8pt]\hline
&&&&&\\[-10pt]
 & & & 0.0069 & 0.847  & \parbox{9cm}{\footnotesize \hma{(492, 4.8)}, \hma{(499, 4.8)}, \hma{(69, 4.16)}, \hma{(70, 4.16)}, \hma{(458, 3.41)}, (1265, 0.7), (446, 0.64), (1168, 0.55)}\\[8pt]\hline
 &&&&&\\[-10pt]
 & & & \hmb{0.118} & \hmb{0.5717}  & \parbox{9cm}{\footnotesize \hma{(492, 3.62)}, \hma{(499, 3.62)}, \hma{(458, 3.36)}, \hma{(69, 3.06)}, \hma{(70, 3.06)}, (710, 1.25), (3, 0.55)}\\[8pt]\hline\hline
 &&&&&\\[-10pt]
119 & 0.766 & -0.123 & 0.00007 & 2. & \parbox{9cm}{\footnotesize \hma{(492, 4.61)}, \hma{(499, 4.61)}, \hma{(69, 4.01)}, \hma{(70, 4.01)}, (352, 3.96), \hma{(458, 3.39)}, (1269, 1.62), (145, 1.08), (664, 1.08), (1266, 1.08)}\\[12pt]\hline
&&&&&\\[-10pt]
 & & & 0.014 & 0.776  & \parbox{9cm}{\footnotesize \hma{(492, 4.61)}, \hma{(499, 4.61)}, \hma{(69, 4.01)}, \hma{(70, 4.01)}, \\\hma{(458, 3.39)}, (1265, 0.88), (446, 0.79), (1168, 0.67)}\\[8pt]\hline
 &&&&&\\[-10pt]
 & & & \hmb{0.198} & \hmb{0.4571}  & \parbox{9cm}{\footnotesize \hma{(492, 3.52)}, \hma{(499, 3.52)}, \hma{(458, 3.31)}, \hma{(69, 3.)}, \hma{(70, 3.)}, (710, 1.15), (3, 0.52)}\\[8pt]\hline\hline
 &&&&&\\[-10pt]
120 & 0.560 & -0.2158 & 0.00002 & 2.& \parbox{9cm}{\footnotesize \hma{(492, 7.97)}, \hma{(499, 6.52)}, \hma{(69, 5.)}, \hma{(70, 5.)}, \hma{(458, 3.37)}, (75, 1.62), (456, 1.62)}\\[8pt]\hline
&&&&&\\[-10pt]
 & & & 0.0009 & 0.858  & \parbox{9cm}{\footnotesize \hma{(492, 6.52)}, \hma{(499, 6.52)}, \hma{(69, 5.)}, \hma{(70, 5.)}, \hma{(458, 3.37)}, (664, 2.91), (352, 2.72), (1026, 1.32), (12, 0.54)}\\[8pt]\hline
 &&&&&\\[-10pt]
 & & & \hmb{0.284} & \hmb{0.386}  & \parbox{9cm}{\footnotesize \hma{(492, 6.34)}, \hma{(499, 6.34)}, \hma{(69, 4.83)}, \hma{(70, 4.83)}, \hma{(458, 3.44)}, (3, 0.51)}\\[8pt]\hline
  &&&&&\\[-10pt]
 & & & 0.155 & 0.0657  & \parbox{9cm}{\footnotesize \hma{(458, 2.49)}, (68, 1.58), (64, 1.53), (423, 0.79), (710, 0.58)}\\[2pt]\hline\hline
 &&&&&\\[-10pt]
121 & 0.439 & -0.273 & 0.00003 & 2. & \parbox{9cm}{\footnotesize \hma{(492, 9.14)}, \hma{(499, 9.14)}, \hma{(69, 5.91)}, \hma{(70, 5.91)}, \hma{(458, 3.39)}, (664, 2.45), (75, 2.45), (423, 1.62), (241, 0.81), (400, 0.81), (450, 0.81), (770, 0.81), (352, 0.53)}\\[16pt]\hline
 &&&&&\\[-10pt]
& & & 0.035 & 0.4997 & \parbox{9cm}{\footnotesize \hma{(492, 9.14)}, \hma{(499, 9.14)}, \hma{(69, 5.91)}, \hma{(70, 5.91)}, \hma{(458, 3.39)}, (446, 1.34), (1026, 0.64)}\\[8pt]\hline
 &&&&&\\[-10pt]
& & & \hmb{0.414} & \hmb{0.2603} & \parbox{9cm}{\footnotesize \hma{(492, 7.76)}, \hma{(499, 7.76)}, \hma{(69, 4.86)}, \hma{(70, 4.86)},\\ \hma{(458, 3.97)}}\\[8pt]\hline
\end{tabular}
\end{center}

\noindent
To further put the divergences above into context, we also tabulate $\mathcal{D}(q_\mu[\mathbb{A}_a,n]||q_\mu[\mathbb{A}_a,n-1])$ in the following, \emph{i.e.} all non-zero divergences of probability distributions between the respectively largest cluster $\mathbb{A}_a$ in consecutive weeks denoted by $n-1$ and $n$
\begin{center}
\begin{tabular}{|c||c|c||c|}\hline
&&&\\[-10pt]
{\bf week} $n$ & $|A_a|$ & $-\overline{|A_a|}$ &  largest divergences $\left(\mu,\mathcal{D}(q_\mu[\mathbb{A}_a,n]||q_\mu[\mathbb{A}_a,n-1])\right)$\\[2pt]\hline\hline
&&&\\[-10pt]
110 & 0.799 & 0.056 & \parbox{12cm}{\footnotesize (5, 0.01)}\\[2pt]\hline
&&&\\[-10pt]
 111 & 0.842 & 0.045 & \parbox{12cm}{\footnotesize (176, 0.01), (504, 0.01), (507, 0.01), (511, 0.01)}\\[2pt]\hline
 &&&\\[-10pt]
112 & 0.855 & 0.022 & \parbox{12cm}{\footnotesize (414, 0.01), (1265, 0.16)}\\[2pt]\hline
 &&&\\[-10pt]
113 & 0.948  & 0.025 & \parbox{12cm}{\footnotesize (64, 0.01), (68, 0.01), (69, 0.01), (70, 0.01), (212, 0.01), (377,
0.01), (458, 0.01), (1265, 0.03)}\\[6pt]\hline
 &&&\\[-10pt]
114 & 0.844 & -0.014  & \parbox{12cm}{\footnotesize (5, 0.06), (64, 0.1), (68, 0.11), (69, 0.01), (70, 0.01), (176,
0.01), (212, 0.03), (213, 0.01), (214, 0.01), (215, 0.01), (254,
0.01), (261, 0.02), (352, 0.03), (377, 0.01), (423, 0.03), (458,0.02), (738, 0.01), (1168, 0.01), (1265, 0.04), (1269, 0.01), (1270,0.03)}\\[20pt]\hline
 &&&\\[-10pt]
 115 & 0.898 & -0.021 & \parbox{12cm}{\footnotesize (5, 0.01), (68, 0.01), (261, 0.01), (345, 0.01), (379, 0.01), (381,
0.01), (382, 0.01), (400, 0.01), (458, 0.01), (1168, 0.01), (1265, 0.03), (1270, 0.01)}\\[8pt]\hline
 &&&\\[-10pt]
116 & 0.926 & -0.0164 & \parbox{12cm}{\footnotesize (64, 0.03), (68, 0.03), (176, 0.01), (212, 0.01), (352, 0.01), (446, 0.01), (1265, 0.27)}\\[8pt]\hline
 &&&\\[-10pt]
117 & 0.910 & -0.038 & \parbox{12cm}{\footnotesize (377, 0.01), (492, 0.01), (499, 0.01)}\\[2pt]\hline
 &&&\\[-10pt]
118 & 0.844 & -0.0738 & \parbox{12cm}{\footnotesize (19, 0.03), (352, 0.01), (377, 0.04), (710, 0.04), (944, 0.01)}\\[2pt]\hline
 &&&\\[-10pt]
119 & 0.766 & -0.123 & \parbox{12cm}{\footnotesize --}\\[2pt]\hline
 &&&\\[-10pt]
120 & 0.560 & -0.2158 & \parbox{12cm}{\footnotesize (19, 0.01), (24, 0.01), (25, 0.01), (26, 0.01), (27, 0.01), (64,0.32), (67, 0.01), (68, 0.32), (95, 0.01), (143, 0.01), (144, 0.01), (145, 0.01), (176, 0.04), (211, 0.01), (212, 0.01), (213, 0.03), (214, 0.02), (215, 0.03), (216, 0.01), (217, 0.01), (251, 0.02), (252, 0.02), (253, 0.02), (254, 0.03), (377, 0.04), (382, 0.01), (411, 0.01), (423, 0.14), (452, 0.01), (458, 0.11), (492, 0.01), (499, 0.01), (502, 0.01), (553, 0.01), (710, 0.01), (738, 0.02), (862, 0.01), (987, 0.01), (1168, 0.02))}\\[30pt]\hline
&&&\\[-10pt]
 121 & 0.439 & -0.273 & \parbox{12cm}{\footnotesize (352, 0.01), (377, 0.01))}\\[2pt]\hline
\end{tabular}
\end{center}
Evidently, these divergences are much smaller than the ones discussed previously. This indicates that the dominant cluster remains relatively stable over a long period of time, with only very small changes in the probability distributions.

\section{Sequences and Mutations}\label{App:SeqMutations}
Throughout Sections~\ref{Sect:FranceSimple} and \ref{Sect:FrancLongTerm} we identify variants of SARS-CoV-2 through the amino acid sequences of their spike proteins\footnote{For simplicity and in order to reduce the amount of data that needs to be handled, we limit ourselves to the spike protein and do not consider the entire genome. Our work can be generalised in a straight-forward manner}. Wherever possible (\emph{i.e.} when a comparison with the recorded spike protein is possible), we name these sequences through the Pango nomenclature \cite{Pango}. Whenever no (simple) identification is possible, we shall label the sequences through consecutive numbering. For concreteness, in this Appendix we describe all sequences that reach a probability of $>0.1$ at any point in time using the same labelling as in Section~\ref{Sect:FrancLongTerm}. We characterise them through the mutations of the spike preotein relative to the original Wuhan variant, which have been determined by the GISAID CovSurver application~\cite{Gisaid1,Gisaid2,Gisaid3}

\begin{center}
\begin{tabular}{|l|l|}\hline
&\\[-12pt]
{\bf seq.} & {\bf point mutations}\\[2pt]\hline\hline
{\footnotesize seq 2} & {\footnotesize D614G} \\\hline
{\footnotesize seq 9} & {\footnotesize S477N, D614G}\\\hline
{\footnotesize seq 10} & {\footnotesize L5F, D614G}\\\hline
{\footnotesize seq 17} & {\footnotesize T19I, L24del, P25del, P26del, A27S, G142D, V213G, G339D, S371F, S373P, S375F, T376A, D405N,}\\
&{\footnotesize R408S, K417N, S477N, T478K, E484A, Q493R, Q498R, N501Y, Y505H, D614G, H655Y, N679K,}\\
&{\footnotesize P681H, N764K, D796Y, Q954H, N969K}\\\hline
{\footnotesize seq 19} & {\footnotesize D614G, S640F}\\\hline
{\footnotesize seq 21} & {\footnotesize D614G, N679K}\\\hline
{\footnotesize seq 23} & {\footnotesize A222V, D614G}\\\hline
{\footnotesize seq 28} & {\footnotesize D614G, T747I}\\\hline
{\footnotesize seq 29} & {\footnotesize D80Y, D614G}\\\hline
{\footnotesize seq 30} & {\footnotesize L176F, D614G}\\\hline
{\footnotesize seq 31} & {\footnotesize K529R, D614G}\\\hline
{\footnotesize seq 32} & {\footnotesize V3G, L5F, D614G}\\\hline
{\footnotesize seq 34} & {\footnotesize D614G, D1084Y}\\\hline
{\footnotesize seq 36} & {\footnotesize N439K, D614G}\\\hline
{\footnotesize BQ.1.1} & {\footnotesize T19I, L24del, P25del, P26del, A27S, H69del, V70del, G142D, V213G, G339D, R346T, S371F, S373P, }\\
&{\footnotesize S375F, T376A, D405N, R408S, K417N, N440K, K444T, L452R, N460K, S477N, T478K, E484A, }\\
&{\footnotesize F486V, Q498R, N501Y, Y505H, D614G, H655Y, N679K, P681H, N764K, D796Y, Q954H, N969K}\\\hline
{\footnotesize B.1.1.7} & {\footnotesize H69del, V70del, Y144del, N501Y, A570D, D614G, P681H, T716I, S982A, D1118H}\\\hline
{\footnotesize BA.1} & {\footnotesize A67V, H69del, V70del, T95I, G142D, V143del, Y144del, Y145del, N211del, L212I, ins214EPE, G339D,}\\
& {\footnotesize  S371L, S373P, S375F, K417N, N440K, G446S, S477N, T478K, E484A, Q493R, G496S, Q498R, N501Y,}\\
& {\footnotesize  Y505H, T547K, D614G, H655Y, N679K, P681H, N764K, D796Y, N856K, Q954H, N969K, L981F}\\\hline
{\footnotesize BA.1.1} & {\footnotesize A67V, H69del, V70del, T95I, G142D, V143del, Y144del, Y145del, N211del, L212I, ins214EPE, G339D,}\\
& {\footnotesize R346K, S371L, S373P, S375F, K417N, N440K, G446S, S477N, T478K, E484A, Q493R, G496S, Q498R,}\\
& {\footnotesize N501Y, Y505H, T547K, D614G, H655Y, N679K, P681H, N764K, D796Y, N856K, Q954H, N969K, L981F}\\\hline
{\footnotesize seq 167} & {\footnotesize T19R, G142D, E156G, F157del, R158del, L452R, T478K, D614G, P681R, D950N}\\\hline
{\footnotesize BA.2} & {\footnotesize T19I, L24del, P25del, P26del, A27S, G142D, V213G, G339D, S371F, S373P, S375F, T376A, D405N,}\\
& {\footnotesize R408S, K417N, N440K, S477N, T478K, E484A, Q493R, Q498R, N501Y, Y505H, D614G, H655Y,}\\
&{\footnotesize N679K, P681H, N764K, D796Y, Q954H, N969K}\\\hline
{\footnotesize seq 186} & {\footnotesize T19R, T95I, G142D, E156G, F157del, R158del, L452R, T478K, D614G, P681R, D950N}\\\hline
{\footnotesize seq 240} & {\footnotesize T19R, E156G, F157del, R158del, L452R, T478K, D614G, P681R, D950N}\\\hline
{\footnotesize seq 530} & {\footnotesize T19R, G142D, E156G, F157del, R158del, T240I, L452R, T478K, D614G, P681R, D950N}\\\hline
{\footnotesize seq 555} & {\footnotesize T19R, G142D, E156G, F157del, R158del, L452R, T478K, D614G, P681R, D950N}\\\hline
{\footnotesize BA.5} & {\footnotesize T19I, L24del, P25del, P26del, A27S, H69del, V70del, G142D, V213G, G339D, S371F, S373P, S375F,}\\
& {\footnotesize T376A, D405N, R408S, K417N, N440K, L452R, S477N, T478K, E484A, F486V, Q498R, N501Y, Y505H,}\\
& {\footnotesize D614G, H655Y, N679K, P681H, N764K, D796Y, Q954H, N969K}\\\hline
{\footnotesize BF.7} & {\footnotesize T19I, L24del, P25del, P26del, A27S, H69del, V70del, G142D, V213G, G339D, R346T, S371F, S373P,}\\
& {\footnotesize  S375F, T376A, D405N, R408S, K417N, N440K, L452R, S477N, T478K, E484A, F486V, Q498R, N501Y,}\\
& {\footnotesize Y505H, D614G, H655Y, N679K, P681H, N764K, D796Y, Q954H, N969K}\\\hline
\end{tabular}
\end{center}

\begin{center}
\begin{tabular}{|l|l|}\hline
&\\[-12pt]
{\bf seq.} & {\bf point mutations}\\[2pt]\hline\hline
{\footnotesize seq 718} & {\footnotesize T19I, L24del, P25del, P26del, A27S, H69del, V70del, G142D, Y144del, V213G, G339D, R346T, S371F,}\\
& {\footnotesize S373P, S375F, T376A, D405N, R408S, K417N, N440K, K444T, L452R, N460K, S477N, T478K, E484A,}\\
& {\footnotesize F486V, Q498R, N501Y, Y505H, D614G, H655Y, N679K, P681H, N764K, D796Y, Q954H, N969K}\\\hline
{\footnotesize seq 726} & {\footnotesize T19I, L24del, P25del, P26del, A27S, H69del, V70del, G142D, V213G, G339D, S371F, S373P, S375F,}\\
& {\footnotesize T376A, D405N, R408S, K417N, L452R, S477N, T478K, E484A, F486V, Q498R, N501Y, Y505H,}\\
& {\footnotesize D614G, H655Y, N679K, P681H, N764K, D796Y, Q954H, N969K}\\\hline
{\footnotesize XBB .1.5} & {\footnotesize T19I, L24del, P25del, P26del, A27S, V83A, G142D, Y144del, H146Q, Q183E, V213E, G252V,}\\
& {\footnotesize G339H, R346T, L368I, S371F, S373P, S375F, T376A, D405N, R408S, K417N, N440K, V445P, G446S,}\\
& {\footnotesize N460K, S477N, T478K, E484A, F486P, F490S, Q498R, N501Y, Y505H, D614G, H655Y, N679K,}\\
& {\footnotesize P681H, N764K, D796Y, Q954H, N969K}\\\hline
{\footnotesize seq 1208} & {\footnotesize T19I, L24del, P25del, P26del, A27S, V83A, G142D, Y144del, H146Q, Q183E, V213E, G252V,}\\
& {\footnotesize G339H, R346T, L368I, S371F, S373P, S375F, T376A, D405N, R408S, K417N, N440K, V445P, G446S,}\\
& {\footnotesize F456L, N460K, S477N, T478K, E484A, F486P, F490S, Q498R, N501Y, Y505H, D614G, H655Y,}\\
& {\footnotesize N679K, P681H, N764K, D796Y, Q954H, N969K}\\\hline
{\footnotesize seq 1218} & {\footnotesize T19I, L24del, P25del, P26del, A27S, V83A, G142D, Y144del, H146Q, E180V, Q183E, V213E, G252V,}\\
& {\footnotesize G339H, R346T, L368I, S371F, S373P, S375F, T376A, D405N, R408S, K417N, N440K, V445P, G446S,}\\
& {\footnotesize N460K, S477N, T478R, E484A, F486P, F490S, Q498R, N501Y, Y505H, D614G, H655Y, N679K,}\\
& {\footnotesize P681H, N764K, D796Y, Q954H, N969K}\\\hline
{\footnotesize seq 1319} & {\footnotesize T19I, L24del, P25del, P26del, A27S, Q52H, V83A, G142D, Y144del, H146Q, Q183E, V213E, G252V,}\\
& {\footnotesize G339H, R346T, L368I, S371F, S373P, S375F, T376A, D405N, R408S, K417N, N440K, V445P, G446S,}\\
& {\footnotesize F456L, N460K, S477N, T478K, E484A, F486P, F490S, Q498R, N501Y, Y505H, D614G, H655Y,}\\
& {\footnotesize N679K, P681H, N764K, D796Y, Q954H, N969K}\\\hline
{\footnotesize seq 1381} & {\footnotesize T19I, L24del, P25del, P26del, A27S, V83A, G142D, Y144del, H146Q, Q183E, V213E, G252V, G339H,}\\
& {\footnotesize R346T, L368I, S371F, S373P, S375F, T376A, D405N, R408S, K417N, N440K, V445P, G446S, L455F,}\\
& {\footnotesize F456L, N460K, A475V, S477N, T478K, E484A, F486P, F490S, Q498R, N501Y, Y505H, D614G,}\\
& {\footnotesize H655Y, N679K, P681H, N764K, D796Y, Q954H, N969K}\\\hline
\end{tabular}
\end{center}

\bibliographystyle{ieeetr}
\bibliography{biblio}

\begin{thebibliography}{100}

\bibitem{PERC20171}
M.~Perc, J.~J. Jordan, D.~G. Rand, Z.~Wang, S.~Boccaletti, and A.~Szolnoki, ``Statistical physics of human cooperation,'' {\em Physics Reports}, vol.~687, pp.~1 -- 51, 2017.

\bibitem{WANG20151}
Z.~Wang, M.~A. Andrews, Z.-X. Wu, L.~Wang, and C.~T. Bauch, ``Coupled disease--behavior dynamics on complex networks: A review,'' {\em Physics of Life Reviews}, vol.~15, pp.~1 -- 29, 2015.

\bibitem{WANG20161}
Z.~Wang, C.~T. Bauch, S.~Bhattacharyya, A.~d'Onofrio, P.~Manfredi, M.~Perc, N.~Perra, M.~Salath\'{e}, and D.~Zhao, ``Statistical physics of vaccination,'' {\em Physics Reports}, vol.~664, pp.~1 -- 113, 2016.

\bibitem{HETHCOTErev}
H.~W. Hethcote, ``The mathematics of infectious diseases,'' {\em SIAM Review}, vol.~42, no.~4, 2000.

\bibitem{BaileyBook}
N.~Bailey, {\em The Mathematical Theory of Infectious Diseases, 2nd ed.}
\newblock Hafner, New York, 1975.

\bibitem{Hamer}
W.~Hamer, ``{Age-incidence in relation with cycles of disease prevalence},'' {\em Trans.~Epidem.~Soc.~London}, vol.~15, pp.~64--77, 1896.

\bibitem{HamerLect1}
W.~Hamer, ``{Epidemic disease in England: The evidence of variability and of persistency of type; Lecture 1},'' {\em Lancet}, pp.~569--574, March 1906.

\bibitem{HamerLect2}
W.~Hamer, ``{Epidemic disease in England: The evidence of variability and of persistency of type; Lecture 2},'' {\em Lancet}, pp.~655--662, March 1906.

\bibitem{HamerLect3}
W.~Hamer, ``{Epidemic disease in England: The evidence of variability and of persistency of type; Lecture 3},'' {\em Lancet}, pp.~733--739, March 1906.

\bibitem{Ross1911}
R.~Ross, ``{The Prevention of Malaria},'' {\em second edition, John Murray, London}, 1911.

\bibitem{Ross1916}
R.~Ross, ``{An application of the theory of probabilities to the study of \emph{a priori} pathometry: Part I},'' {\em Proc.~Roy.~Soc.~Lond.~A}, vol.~92, pp.~204--230, 1916.

\bibitem{RossHudson1916II}
R.~Ross and H.~Hudson, ``{An application of the theory of probabilities to the study of \emph{a priori} pathometry: Part II},'' {\em Proc.~Roy.~Soc.~Lond.~A}, vol.~93, pp.~212--225, 1916.

\bibitem{RossHudson1916III}
R.~Ross and H.~Hudson, ``{An application of the theory of probabilities to the study of \emph{a priori} pathometry: Part III},'' {\em Proc.~Roy.~Soc.~Lond.~A}, vol.~93, pp.~225--240, 1916.

\bibitem{McKendrick1912}
A.~McKendrick, ``{The rise and fall of epidemics},'' {\em Paludism (Transactions of the Committee for the Study of Malaria in India)}, vol.~1, pp.~54--66, 1912.

\bibitem{McKendrick1914}
A.~McKendrick, ``{Studies on the theory of continuous probabilities, with special reference to its bearing on natural phenomena of a progressive nature},'' {\em Proceedings of the London Mathematical Society}, vol.~13, pp.~401--416, 1914.

\bibitem{McKendrick1926}
A.~McKendrick, ``{Applications of mathematics to medical problems},'' {\em Proc.~Edinburgh Math.~Soc.}, vol.~44, pp.~98--130, 1926.

\bibitem{Kermack:1927}
W.~O. Kermack, A.~McKendrick, and G.~T. Walker, ``{A contribution to the mathematical theory of epidemics},'' {\em Proceedings of the Royal Society A}, vol.~115, pp.~700--721, 1927.

\bibitem{DellaMorte:2020wlc}
M.~Della~Morte, D.~Orlando, and F.~Sannino, ``{Renormalization Group Approach to Pandemics: The COVID-19 Case},'' {\em Front. in Phys.}, vol.~8, p.~144, 2020.

\bibitem{DellaMorte:2020qry}
M.~Della~Morte and F.~Sannino, ``{Renormalisation Group approach to pandemics as a time-dependent SIR model},'' {\em Front. in Phys.}, vol.~8, p.~583, 2021.

\bibitem{cacciapaglia2020evidence}
G.~Cacciapaglia and F.~Sannino, ``Evidence for complex fixed points in pandemic data,'' 2020.

\bibitem{MeRG}
G.~Cacciapaglia, C.~Cot, A.~{de Hoffer}, S.~Hohenegger, F.~Sannino, and S.~Vatani, ``Epidemiological theory of virus variants,'' {\em Physica A: Statistical Mechanics and its Applications}, vol.~596, p.~127071, 2022.

\bibitem{HealthPass}
S.~Hohenegger, G.~Cacciapaglia, and F.~Sannino, ``{Effective mathematical modelling of health passes during a pandemic},'' {\em Scientific Reports}, vol.~12, p.~6989, 2022.

\bibitem{Filoche:2024xka}
B.~Filoche, S.~Hohenegger, and F.~Sannino, ``{Information Theory Unification of Epidemiological and Population Dynamics},'' 2 2024.

\bibitem{Cardy_1985}
J.~L. Cardy and P.~Grassberger, ``Epidemic models and percolation,'' {\em Journal of Physics A: Mathematical and General}, vol.~18, pp.~L267--L271, apr 1985.

\bibitem{Grassberger1983}
P.~Grassberger, ``On the critical behavior of the general epidemic process and dynamical percolation,'' {\em Mathematical Biosciences}, vol.~63, no.~2, pp.~157 -- 172, 1983.

\bibitem{Pruessner}
G.~Pruessner, ``Field theory notes, chapter 6,''

\bibitem{Doi1}
M.~Doi, ``Second quantization representation for classical many-particle system,'' {\em J. Phys. A: Math. Gen.}, vol.~9, p.~1465, 1976.

\bibitem{Doi2}
M.~Doi, ``Stochastic theory of diffusion-controlled reaction,'' {\em J. Phys. A: Math. Gen.}, vol.~9, p.~1479, 1976.

\bibitem{Peliti}
L.~Peliti, ``Path integral approach to birth-death processes on a lattice,'' {\em J. Phys. France (Paris)}, vol.~46, pp.~1469--1483, 1985.

\bibitem{Domb}
C.~Domb, ``Fluctuation phenomena and stochastic processes,'' {\em Nature}, vol.~184, pp.~509--12, 1959.

\bibitem{Essam}
J.~W. Essam, ``Percolation theory,'' {\em Rep. Prog. Phys.}, vol.~43, p.~833, 1980.

\bibitem{Stauffer}
D.~Stauffer, ``{Scaling theory of percolation clusters},'' {\em Phys.~Rep.~}, vol.~54, pp.~1--74, 1979.

\bibitem{ABC}
G.~Cacciapaglia, C.~Cot, M.~D. Morte, S.~Hohenegger, F.~Sannino, and S.~Vatani, ``The field theoretical abc of epidemic dynamics,'' 2021.

\bibitem{SHSannino}
S.~Hohenegger and F.~Sannino, ``Renormalisation group methods for effective epidemiological models,'' 2024.

\bibitem{MLvariants}
A.~de~Hoffer, S.~Vatani, C.~Cot, G.~Cacciapaglia, F.~Conventi, A.~Giannini, S.~Hohenegger, and F.~Sannino, ``Variant-driven multi-wave pattern of covid-19 via machine learning clustering of spike protein mutations,'' {\em medRxiv}, 2021.

\bibitem{WasteWaterReview}
F.~Torabi, G.~Li, C.~Mole, G.~Nicholson, B.~Rowlingson, C.~R. Smith, R.~Jersakova, P.~J. Diggle, and M.~Blangiardo, ``Wastewater-based surveillance models for covid-19: A focused review on spatio-temporal models,'' {\em Heliyon}, vol.~9, no.~11, p.~e21734, 2023.

\bibitem{KHAILANY2020100682}
R.~A. Khailany, M.~Safdar, and M.~Ozaslan, ``Genomic characterization of a novel sars-cov-2,'' {\em Gene Reports}, vol.~19, p.~100682, 2020.

\bibitem{10.3389/fmicb.2020.01800}
D.~Mercatelli and F.~M. Giorgi, ``Geographic and genomic distribution of sars-cov-2 mutations,'' {\em Frontiers in Microbiology}, vol.~11, 2020.

\bibitem{ReviewDevelop}
M.~Alessandro, T.~Peacock, L.~Thorne, W.~Harvey, J.~Hughes, T.~Silva, S.~Peacock, W.~Barclay, G.~Towers, and D.~Robertson, ``Sars-cov-2 variant biology: immune escape, transmission and fitness,'' {\em Nature Reviews Microbiology}, vol.~21, pp.~1--16, 01 2023.

\bibitem{SARS1year}
T.~Peacock, R.~Penrice-Randal, J.~Hiscox, and W.~Barclay, ``Sars-cov-2 one year on: evidence for ongoing viral adaptation,'' {\em Journal of General Virology}, vol.~102, 04 2021.

\bibitem{Evolution}
W.~Harvey, M.~Alessandro, B.~Jackson, R.~Gupta, E.~Thomson, E.~Harrison, C.~Ludden, R.~Reeve, A.~Rambaut, S.~Peacock, and D.~Robertson, ``Sars-cov-2 variants, spike mutations and immune escape,'' {\em Nature Reviews Microbiology}, vol.~19, pp.~1--16, 06 2021.

\bibitem{Spectrum}
J.~Bloom, A.~Beichman, R.~Neher, and K.~Harris, ``Evolution of the sars-cov-2 mutational spectrum,'' {\em bioRxiv : the preprint server for biology}, 11 2022.

\bibitem{AntigenicDrift}
R.~Eguia, K.~Crawford, T.~Stevens-Ayers, L.~Kelnhofer-Millevolte, A.~Greninger, J.~Englund, M.~Boeckh, and J.~Bloom, ``A human coronavirus evolves antigenically to escape antibody immunity,'' {\em PLOS Pathogens}, vol.~17, p.~e1009453, 04 2021.

\bibitem{v15010167}
S.~Chatterjee, M.~Bhattacharya, S.~Nag, K.~Dhama, and C.~Chakraborty, ``A detailed overview of sars-cov-2 omicron: Its sub-variants, mutations and pathophysiology, clinical characteristics, immunological landscape, immune escape, and therapies,'' {\em Viruses}, vol.~15, no.~1, 2023.

\bibitem{Serotypes}
E.~Simon-Loriere and O.~Schwartz, ``Towards sars-cov-2 serotypes?,'' {\em Nature Reviews Microbiology}, vol.~20, pp.~1--2, 02 2022.

\bibitem{HU20233003}
S.~Hu, C.~Wu, X.~Wu, X.~Ma, C.~Shu, Q.~Chen, A.~Zheng, H.~Yang, J.~Lu, P.~Du, G.~F. Gao, and Q.~Wang, ``Classification of five sars-cov-2 serotypes based on rbd antigenicities,'' {\em Science Bulletin}, vol.~68, no.~23, pp.~3003--3012, 2023.

\bibitem{microorganisms12030467}
F.~A. Alsuwairi, A.~N. Alsaleh, D.~A. Obeid, A.~A. Al-Qahtani, R.~S. Almaghrabi, B.~M. Alahideb, M.~A. AlAbdulkareem, M.~S. Alsanea, L.~A. Alharbi, S.~I. Althawadi, S.~A. Altamimi, A.~N. Alshukairi, and F.~S. Alhamlan, ``Genomic surveillance and mutation analysis of sars-cov-2 variants among patients in saudi arabia,'' {\em Microorganisms}, vol.~12, no.~3, 2024.

\bibitem{SwedishPortal}
S.~D. Centre, ``Swedish pathogens portal,'' RRID: SCR024866.

\bibitem{dbvar}
I.~Lappalainen, J.~Lopez, L.~Skipper, T.~Hefferon, J.~Spalding, J.~Garner, C.~Chen, M.~Maguire, M.~Corbett, G.~Zhou, J.~Paschall, V.~Ananiev, P.~Flicek, and D.~Church, ``Dbvar and dgva: public archives for genomic structural variation,'' 2013.

\bibitem{Gisaid1}
S.~Khare, C.~Gurry, L.~Freitas, M.~B. Schultz, G.~Bach, A.~Diallo, N.~Akite, J.~Ho, R.~T. Lee, W.~Yeo, G.~C.~C. Team, and S.~Maurer-Stroh, ``Gisaid’s role in pandemic response,'' {\em China CDC Weekly}, vol.~3, p.~1049, 2021.

\bibitem{Gisaid2}
S.~Elbe and G.~Buckland-Merrett, ``Data, disease and diplomacy: Gisaid's innovative contribution to global health,'' {\em Global Challenges}, vol.~1, no.~1, pp.~33--46, 2017.

\bibitem{Gisaid3}
Y.~Shu and J.~McCauley, ``Gisaid: Global initiative on sharing all influenza data – from vision to reality,'' {\em Eurosurveillance}, vol.~22, no.~13, 2017.

\bibitem{NextStrain}
J.~Hadfield, C.~Megill, S.~M. Bell, J.~Huddleston, B.~Potter, C.~Callender, P.~Sagulenko, T.~Bedford, and R.~A. Neher, ``{Nextstrain: real-time tracking of pathogen evolution},'' {\em Bioinformatics}, vol.~34, pp.~4121--4123, 05 2018.

\bibitem{HOFFMANN20212384}
M.~Hoffmann, P.~Arora, R.~Groß, A.~Seidel, B.~F. Hörnich, A.~S. Hahn, N.~Krüger, L.~Graichen, H.~Hofmann-Winkler, A.~Kempf, M.~S. Winkler, S.~Schulz, H.-M. Jäck, B.~Jahrsdörfer, H.~Schrezenmeier, M.~Müller, A.~Kleger, J.~Münch, and S.~Pöhlmann, ``Sars-cov-2 variants b.1.351 and p.1 escape from neutralizing antibodies,'' {\em Cell}, vol.~184, no.~9, pp.~2384--2393.e12, 2021.

\bibitem{BAKHSHANDEH2021104831}
B.~Bakhshandeh, Z.~Jahanafrooz, A.~Abbasi, M.~B. Goli, M.~Sadeghi, M.~S. Mottaqi, and M.~Zamani, ``Mutations in sars-cov-2; consequences in structure, function, and pathogenicity of the virus,'' {\em Microbial Pathogenesis}, vol.~154, p.~104831, 2021.

\bibitem{PHAN2020104260}
T.~Phan, ``Genetic diversity and evolution of sars-cov-2,'' {\em Infection, Genetics and Evolution}, vol.~81, p.~104260, 2020.

\bibitem{Mishra}
T.~Mishra, R.~Dalavi, G.~Joshi, A.~Kumar, P.~Pandey, S.~Shukla, R.~Mishra, and A.~Chande, ``Sars-cov-2 spike e156g/$\delta$157-158 mutations contribute to increased infectivity and immune escape,'' {\em Life Science Alliance}, vol.~5, p.~e202201415, 07 2022.

\bibitem{Mishra2}
T.~Mishra, S.~Mahadev, P.~Ramdas, A.~Sahu, A.~Kumar, and A.~Chande, ``Sars cov-2 nucleoprotein enhances the infectivity of lentiviral spike particles,'' {\em Frontiers in Cellular and Infection Microbiology}, vol.~11, p.~663688, 04 2021.

\bibitem{Fisher}
R.~A. Fisher and E.~J. Russell, ``On the mathematical foundations of theoretical statistics,'' {\em Philosophical Transactions of the Royal Society of London. Series A, Containing Papers of a Mathematical or Physical Character}, vol.~222, no.~594-604, pp.~309--368, 1922.

\bibitem{Hotelling}
H.~Hotelling, ``Spaces of statistical parameters,'' {\em Bulletin of the American Mathematical Society (AMS)}, vol.~36, p.~191, 1930.

\bibitem{Rao}
R.~C. Rao, ``Information and the accuracy attainable in the estimation of statistical parameters,'' {\em Bulletin of the Calcutta Mathematical Society}, vol.~37, pp.~81--91, 1945.

\bibitem{Jeffreys}
H.~Jeffreys, ``An invariant form for the prior probability in estimation problems,'' {\em Proc. R. Soc. Lond. A}, vol.~186(1007), pp.~453--461, 1946.

\bibitem{Lauritzen}
S.~L. Lauritzen, ``Statistical manifolds,'' {\em Differential geometry in statistical inference}, vol.~10, pp.~163--216, 1987.

\bibitem{amari2000methods}
S.~Amari and H.~Nagaoka, {\em Methods of Information Geometry}.
\newblock Translations of mathematical monographs, American Mathematical Society, 2000.

\bibitem{FisherInt}
R.~A. Fisher, ``On the mathematical foundations of theoretical statistics,'' {\em Philosophical Transactions of the Royal Society of London. Series A, Containing Papers of a Mathematical or Physical Character}, vol.~222, pp.~309--368, 1922.

\bibitem{ThomasCover}
T.~M. Cover and J.~A. Thomas, {\em Elements of Information Theory (Wiley Series in Telecommunications and Signal Processing)}.
\newblock USA: Wiley-Interscience, 2006.

\bibitem{Shannon}
C.~E. Shannon, ``A mathematical theory of communication,'' {\em The Bell System Technical Journal}, vol.~27, no.~3, pp.~379--423, 1948.

\bibitem{2004poin.book..119B}
R.~{Balian}, ``{Entropy, a Protean Concept},'' in {\em Poincar{\'e} Seminar 2003}, vol.~38, p.~119, 2004.

\bibitem{lesne_2014}
A.~Lesne, ``Shannon entropy: a rigorous notion at the crossroads between probability, information theory, dynamical systems and statistical physics,'' {\em Mathematical Structures in Computer Science}, vol.~24, no.~3, p.~e240311, 2014.

\bibitem{Khinchin}
A.~Y. Khinchin, ``The concept of entropy in the theory of probability,'' {\em Uspekhi Mat. Nauk}, vol.~8, pp.~3--20, 1953.

\bibitem{Faddeev}
D.~Faddeev, ``{On the concept of entropy of a~finite probabilistic scheme},'' {\em Uspekhi Mat. Nauk}, vol.~11, no.~1(67), pp.~227 -- 231, 1956.

\bibitem{CoverThomas}
T.~M. Cover and J.~A. Thomas, ``Elements of information theory,'' {\em Wiley}, 1991.

\bibitem{Becker}
N.~Becker, ``{The use of epidemic models},'' {\em Biometrics}, vol.~35, pp.~295--305, 1978.

\bibitem{DietzSchenzle}
K.~Dietz and D.~Schenzle, ``{Mathematical models for infectious disease statistics, in: A.~Atkinson (Ed.), A Celebration of Statistics},'' {\em Springer}, pp.~167--204, 1985.

\bibitem{Castillo}
e.~C.~Castillo-Chavez, ``{Mathematical and Statistical Approaches to AIDS Epidemiology, Lecture Notes in Biomath},'' {\em Springer-Verlag, Berlin}, vol.~83, 1989.

\bibitem{Dietz}
K.~Dietz, ``{Epidemics and rumours: A survey},'' {\em J.~Roy.~Statist.~Soc.~Ser. A}, vol.~130, pp.~505--528, 1967.

\bibitem{Dietz2}
K.~Dietz, ``{Density dependence in parasite transmission dynamics},'' {\em Parasit.~Today}, vol.~4, pp.~91--97, 1988.

\bibitem{HethcoteThousand}
H.~Hethcote, ``{A thousand and one epidemic models, in Frontiers in Theoretical Biology, S.A.~Levin, ed., Lecture Notes in Biomath.},'' {\em Springer-Verlag, Berlin}, vol.~100, pp.~504--515, 1994.

\bibitem{OurWorldIndata}
``Our world in data.'' \url{https://ourworldindata.org}.

\bibitem{doi:10.1177/014662168701100401}
G.~W. Milligan and M.~C. Cooper, ``Methodology review: Clustering methods,'' {\em Applied Psychological Measurement}, vol.~11, no.~4, pp.~329--354, 1987.

\bibitem{Hubert1985ComparingP}
L.~J. Hubert and P.~Arabie, ``Comparing partitions,'' {\em Journal of Classification}, vol.~2, pp.~193--218, 1985.

\bibitem{Kaufman}
L.~Kaufman and P.~Rousseeuw, {\em Finding Groups in Data: An Introduction To Cluster Analysis}.
\newblock 01 1990.

\bibitem{Edgar_2004}
R.~C. Edgar, ``Muscle: multiple sequence alignment with high accuracy and high throughput,'' {\em Nucleic Acids Research}, vol.~32, p.~1792–1797, 2004.

\bibitem{Edgar2}
R.~C. Edgar, ``Muscle: multiple sequence alignment method with reduced time and space complexity,'' {\em BMC Bioinformatics}, vol.~5, p.~113, 2004.

\bibitem{KullbackLeibler}
S.~Kullback and R.~A. Leibler, ``{On Information and Sufficiency},'' {\em The Annals of Mathematical Statistics}, vol.~22, no.~1, pp.~79 -- 86, 1951.

\bibitem{Levenshtein1965BinaryCC}
V.~I. Levenshtein, ``Binary codes capable of correcting deletions, insertions, and reversals,'' {\em Soviet physics. Doklady}, vol.~10, pp.~707--710, 1965.

\bibitem{Pango}
A.~Rambaut, E.~Holmes, A.~O’Toole, J.~McCrone, C.~Ruis, L.~du~Plessis, and O.~Pybus, ``A dynamic nomenclature proposal for sars-cov-2 lineages to assist genomic epidemiology,'' {\em Nat Microbiol}, vol.~5, pp.~1403--1407, 2020.

\bibitem{FranceAlpha}
S.~publique France, ``Covid-19, point \'epid\'emiologique hebdomadaire du 28 janvier 2021,'' {\em Point \'epid\'emiologique hebdomadaire}, vol.~28/01/2021, 2021.

\bibitem{Cao}
Y.~Cao, J.~Wang, F.~Jian, T.~Xiao, W.~Song, A.~Yisimayi, W.~Huang, Q.~Li, P.~Wang, R.~An, J.~Wang, Y.~Wang, X.~Niu, S.~Yang, H.~Liang, H.~Sun, T.~Li, Y.~Yu, Q.~Cui, S.~Liu, X.~Yang, S.~Du, Z.~Zhang, X.~Hao, F.~Shao, R.~Jin, X.~Wang, J.~Xiao, Y.~Wang, and X.~S. Xie, ``Omicron escapes the majority of existing sars-cov-2 neutralizing antibodies,'' {\em Nature}, vol.~602, p.~657–663, 2022.

\bibitem{Mohandas}
M.~Sreelekshmy, D.~Y. Pragya, S.~Gajanan, M.~S. Anita, D.~Gururaj, A.~N. Dimpal, P.~Deepak, K.~Manoj, K.~Abhimanyu, M.~Chandrashekhar, and J.~Rajlaxmi, ``Pathogenicity of sars-cov-2 omicron (r346k) variant in syrian hamsters and its cross-neutralization with different variants of concern,'' {\em eBioMedicine}, vol.~79, 2022.

\bibitem{Cele}
S.~Cele, L.~Jackson, K.~Khan, D.~Khoury, T.~Moyo-Gwete, H.~Tegally, C.~Scheepers, D.~Amoako, F.~Karim, M.~Bernstein, G.~Lustig, D.~Archary, M.~Smith, Y.~Ganga, Z.~Jule, K.~Reedoy, J.~E. San, S.-H. Hwa, J.~Giandhari, J.~M. Blackburn, B.~I. Gosnell, S.~A. Karim, W.~Hanekom, A.~von Gottberg, J.~Bhiman, R.~J. Lessells, M.-Y.~S. Moosa, M.~Davenport, T.~de~Oliveira, P.~L. Moore, and A.~Sigal, ``Sars-cov-2 omicron has extensive but incomplete escape of pfizer bnt162b2 elicited neutralization and requires ace2 for infection,'' {\em medRxiv}, 2021.

\bibitem{OmicronFirst}
E.~Callaway, ``Heavily mutated omicron variant puts scientists on alert,'' {\em Nature}, vol.~600, p.~21, 2021.

\bibitem{FranceOmicron}
Reuters, ``France now has 25 omicron covid variant cases - minister,'' {\em 6 December 2021}.

\bibitem{10.1093/ve/veac080}
V.~Hill, L.~Du~Plessis, T.~P. Peacock, D.~Aggarwal, R.~Colquhoun, A.~M. Carabelli, N.~Ellaby, E.~Gallagher, N.~Groves, B.~Jackson, J.~T. McCrone, A.~O'Toole, A.~Price, T.~Sanderson, E.~Scher, J.~Southgate, E.~Volz, W.~S. Barclay, J.~C. Barrett, M.~Chand, T.~Connor, I.~Goodfellow, R.~K. Gupta, E.~M. Harrison, N.~Loman, R.~Myers, D.~L. Robertson, O.~G. Pybus, and T.~C.-. G. U. C.-U.~C. Rambaut, Andrew, ``{The origins and molecular evolution of SARS-CoV-2 lineage B.1.1.7 in the UK},'' {\em Virus Evolution}, vol.~8, pp.~1--13, 08 2022.

\bibitem{Cacciapaglia:2020mjf}
G.~Cacciapaglia and F.~Sannino, ``{Interplay of social distancing and border restrictions for pandemics (COVID-19) via the epidemic Renormalisation Group framework},'' {\em Sci Rep}, vol.~10, p.~15828, 5 2020.

\bibitem{cacciapaglia2020second}
G.~Cacciapaglia, C.~Cot, and F.~Sannino, ``Second wave covid-19 pandemics in europe: A temporal playbook,'' {\em Sci Rep}, vol.~10, p.~15514, 2020.

\bibitem{cacciapaglia2020mining}
G.~Cacciapaglia, C.~Cot, and F.~Sannino, ``Mining google and apple mobility data: Temporal anatomy for covid-19 social distancing,'' {\em Scientific Reports}, vol.~11, p.~4150, 2021.

\bibitem{cacciapaglia2020multiwave}
G.~Cacciapaglia, C.~Cot, and F.~Sannino, ``Multiwave pandemic dynamics explained: How to tame the next wave of infectious diseases,'' {\em Scientific Reports}, vol.~11, p.~6638, 2021.

\bibitem{cacciapaglia2020better}
G.~Cacciapaglia, C.~Cot, A.~S. Islind, M.~{\'O}skarsd{\'o}ttir, and F.~Sannino, ``You better watch out: Us covid-19 wave dynamics versus vaccination strategy,'' 2020.

\bibitem{cacciapaglia2020us}
G.~Cacciapaglia, C.~Cot, A.~S. Islind, M.~{\'O}skarsd{\'o}ttir, and F.~Sannino, ``Impact of us vaccination strategy on covid-19 wave dynamics,'' {\em Scientific Reports}, vol.~11(1), pp.~1--11, 2021.

\bibitem{MR90073}
H.~Steinhaus, ``Sur la division des corps mat\'eriels en parties,'' {\em Bull. Acad. Polon. Sci. Cl. III.}, vol.~4, pp.~801--804, 1956.

\bibitem{MR214227}
J.~MacQueen, ``Some methods for classification and analysis of multivariate observations,'' in {\em Proc. {F}ifth {B}erkeley {S}ympos. {M}ath. {S}tatist. and {P}robability ({B}erkeley, {C}alif., 1965/66), {V}ol. {I}: {S}tatistics}, pp.~281--297, Univ. California Press, Berkeley, CA, 1967.

\bibitem{Lloyd}
S.~Lloyd, ``Least squares quantization in pcm,'' {\em IEEE Transactions on Information Theory}, vol.~28, no.~2, pp.~129--137, 1982.

\bibitem{Forgy}
E.~W. Forgy, ``Cluster analysis of multivariate data: efficiency versus interpretability of classifications,'' {\em Biometrics}, vol.~21, no.~3, pp.~768--769, 1965.

\bibitem{WardDist}
J.~H. Ward, ``Hierarchical grouping to optimize an objective function,'' {\em Journal of the American Statistical Association}, vol.~58, no.~301, pp.~236--244, 1963.

\end{thebibliography}

\end{document}